\documentclass[12pt]{article}
\begin{document}
\input epsf
\def\be{\begin{equation}}
\def\bea{\begin{eqnarray}}
\def\ee{\end{equation}}
\def\eea{\end{eqnarray}}
\def\d{\partial}
\def\la{\lambda}
\def\eps{\epsilon}
\def\l#1{ \label{#1}}
\def\half{ { 1\over 2}}

\def\nono{ { \nonumber}}

%===================================================================
\begin{flushright}
hep-th/0409174\\ PUPT-2136
\end{flushright}
\vskip 0.5cm

\begin{center}
\Large{\bf Bubbling AdS space and 1/2 BPS geometries}

\vspace{ 20mm}

\normalsize{Hai Lin$^1$, Oleg Lunin$^2$ and Juan Maldacena$^2$ }

\vspace{10mm}

\normalsize{\em $^1$ Department of Physics, Princeton University,
Princeton, NJ 08544}

\vspace{0.2cm}

\normalsize{\em $^2$ Institute for Advanced Study, Princeton, NJ 08540}

\vspace{0.2cm}

\end{center}

\vspace{10mm}

\begin{abstract}
\medskip
We consider all 1/2 BPS excitations of $AdS \times S$
configurations in both type IIB string theory and M-theory. In the
dual field theories these excitations are described by free
fermions. Configurations which are dual to arbitrary droplets of
free fermions in phase space correspond to smooth geometries with
no horizons. In fact, the ten dimensional geometry
contains a special two dimensional plane which can be identified
with the phase space of the free fermion system. The topology of
the resulting geometries depends only on the topology of the
collection of droplets on this plane. These solutions also give a
very explicit realization of the geometric transitions between branes
and fluxes. We also describe all 1/2 BPS excitations of plane wave
geometries. The problem of finding the explicit geometries is
reduced to solving a Laplace (or Toda) equation with simple
boundary conditions. We present a large class of explicit
solutions.  In addition, we are led to a rather general class of
$AdS_5$ compactifications of M-theory preserving ${\cal N} =2$
superconformal symmetry. We also find smooth geometries that
correspond to various vacua of the maximally supersymmetric
mass-deformed M2 brane theory. Finally, we present a smooth 1/2
BPS solution of seven dimensional gauged supergravity
corresponding to a condensate of one of the charged scalars.

\end{abstract}

\newpage

\section{Introduction}
\renewcommand{\theequation}{1.\arabic{equation}}
\setcounter{equation}{0}

In this paper we consider a class of 1/2 BPS states that arises
very naturally in the study of the AdS/CFT correspondence for
maximally supersymmetric theories. These states are associated to
chiral primary operators with conformal weight $\Delta =J$, where
$J$ is a particular $U(1)$ charge in the R-symmetry group. For
small excitation energies $J \ll N$ these BPS states correspond to
particular gravity modes propagating in the bulk
\cite{wittenholography}.  As one increases the excitation energy
so that $J \sim N$ one finds that some of the states can be
described as branes   in the internal sphere \cite{giant} or as
branes   in
  AdS   \cite{giantads}. These were called ``giant gravitons".
As we increase the excitation energy to $J \sim N^2 $ we expect to find
new geometries.
The BPS states in question have a simple field theory
description in terms of free fermions \cite{berenstein} (see also \cite{rangoolam}).
In a semiclassical
limit we can characterize these states by giving the regions, or ``droplets",
in phase space
occupied by the fermions. We can also picture the BPS states as
 fermions in a magnetic field
on the lowest Landau level (quantum Hall problem).
In this paper we study the geometries corresponding to
these configurations.
These are smooth geometries that preserve 16 of the original
32 supersymmetries. We are able to give the general form of the solution
in terms of an equation whose boundary conditions are specified on a particular
plane. We can have two types of boundary conditions corresponding to either
of two different spheres shrinking on this plane in an smooth fashion.
This plane, and the corresponding regions are in
direct correspondence with the regions in phase space that were discussed above.
Once the occupied regions are given on this plane, the solution is determined
uniquely and the ten (or eleven) dimensional geometry is non-singular and does not contain
horizons.
\begin{figure}[htb]
\begin{center}
\epsfxsize=4.0in\leavevmode\epsfbox{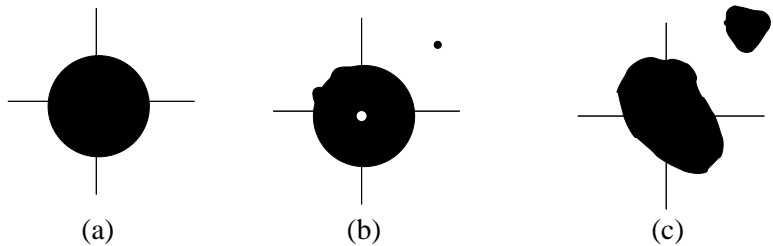}
\end{center}
\caption{ Droplets representing chiral primary states. In the field theory
description these are droplets in phase space occupied by the fermions. In
the gravity picture this is a particular two-plane in ten dimensions which
specifies the solution uniquely. In (a) we see the droplet corresponding
to the $AdS \times S$ ground state. In (b) we see ripples
on the surface corresponding
to gravitons in $AdS \times S$. The separated black region
is a giant graviton brane
which
wraps an $S^3$ in $AdS_5$ and the hole at the center is a giant graviton brane
wrapping an $S^3$ in $S^5$. In (c) we see a more general state.   } \label{droplets}
\end{figure}

The topology  of the solutions is fixed by the topology of the
droplets on the plane. The actual geometry depends on the shape of
the droplets. In fact, this characterization is reminiscent of
toric geometry.  In the type IIB case we simply need to solve a
Laplace equation. A circular droplet gives rise to the $AdS_5
\times S^5$ solution, see figure \ref{droplets}. Small ripples on
the droplet correspond to small fluctuations corresponding to
gravitons in $AdS$. A small droplet far away from the circular one
corresponds to a group of D3 branes wrapping an $S^3$ in $AdS_5$.
A hole inside the circle corresponds to branes wrapping an $S^3$
in $S^5$. In the limit that the droplets become small these
solutions reduce to the giant graviton branes that were discussed
extensively in the literature
\cite{giant,giantads,giantliterature}. Some of our solutions
smoothly interpolate between branes wrapping the sphere and branes
wrapping AdS. We can also have solutions that correspond to new
geometries which cannot be thought of as branes. In other words,
when we put many branes together they back-react on the geometry
and we get new geometries with new topologies that are determined
by geometric transitions. The transition is that the sphere the
branes are wrapping becomes contractible while the transverse
sphere becomes non-contractible and the branes get replaced by
flux.

From the geometrical point of view we can consider this class of
BPS geometries and we can wonder how we quantize them. Of course,
the exact description in terms of fermions is telling us how to do
it. In the type IIB case, a two dimensional plane contained in the
ten dimensional geometry can be identified as the phase space of
free fermions. The quantization of the area in the phase space
amounts to the quantization of fluxes in the geometry. One
interesting lesson is that geometries with very small
topologically non-trivial fluctuations, or spacetime-foam, are
already included when we perform the usual quantization of
ordinary long wavelength gravitons.

These solutions are also interesting because they provide a
relationship between free fermions and string theory which is
rather different than the one we get from the $c=1$ matrix model
(for reviews see \cite{cone}). Here the free fermions arise as the
BPS sector of a ten dimensional string theory. Perhaps we should
not be surprised because integrable systems often lead to  free
fermions and a BPS system is in some sense integrable, so it is
natural to have a free fermion description. It would be nice to
understand better whether there is a reduction of the usual
superstring in AdS \cite{nathan} to a string theory describing
just this 1/2 BPS sector\footnote{See \cite{mcgreevysunny} for a
proposal of a  string theory description of the harmonic
oscillator.}.

We can also describe 1/2 BPS excitations of the plane wave geometry, which
 corresponds to
a half filled plane. In this case the fermion
becomes a relativistic Dirac fermion in 1+1 dimensions. The
light-cone energy of the solution is the same as the usual energy
for a Dirac fermion. Particle-hole duality corresponds to
exchanging the 3-sphere in the first four of the eight transverse
coordinates with a 3-sphere in the last four coordinates.

By performing dualities  we can get solutions which are dual to
the mass deformed M2 brane theory \cite{ib,ibnw}.  This theory is
rather similar to the mass deformed $N=4 $ Yang-Mills theory, or
$N=1^*$ theory, analyzed by Polchinski and Strassler \cite{jpms}.
The mass term preserves $SO(4) \times SO(4)$ symmetry in $SO(8)$
and the theory has vacua that contain M5 branes wrapping an $S^3$
in the first four coordinates or an $S^3$ in the last four
coordinates. Our solutions are non-singular and describe all
possible vacua of this theory. By changing the fluxes on the
various spheres, we can smoothly interpolate between the solutions
with the M5 branes wrapping the first $S^3$ and the solutions with
those wrapping the second $S^3$. This system has also been
recently analyzed in \cite{ibnw}, in terms of slightly different
variables. Our approach leads to a simple way of constructing {\it
non-singular} geometries.

We have also performed a similar analysis for  the M-theory case,
which corresponds to giant gravitons in $AdS_4 \times S^7$ or
$AdS_7 \times S^4$. In this case we have similar droplets, and the
11 dimensional geometry is obtained after solving a three
dimensional Toda equation.  In this case we could only solve the
equations explicitly in very simple examples.
 We also consider the M-theory plane wave. In this way we could find
geometries that are dual to the BMN matrix model \cite{bmn}. In particular, we
find more evidence that the M5 brane
emerges as a state of the BMN matrix model \cite{mfivefromBMN}.

By performing a Wick rotation of the above analysis we are led to
a characterization of all M-theory compactifications to $AdS_5$
that preserve ${\cal N}=2$ four dimensional supersymmetry. These
are again given by solutions of the Toda equation but with
slightly different boundary conditions. Indeed, we fit the
previously known solutions \cite{maldanunez} into this class. This
constitutes an extension of the analysis in \cite{Gauntlett, Gauntlett2} which characterized M-theory compactifications
to $AdS_5$  preserving ${\cal N} =1$ four dimensional
supersymmetry.

This paper is organized as follows. In section 2 we discuss the
geometries associated to 1/2 BPS states in $AdS_5 \times S^5$ or
the type IIB pp-wave. In section 3 we discuss the 1/2 BPS
geometries describing states in $AdS_{7}\times S^{4}$, $AdS_{4}\times S^{7}$, or
M theory pp-wave. In various appendices we give more
technical details.

\begin{figure}[htb]
\begin{center}
\epsfxsize=2.0in\leavevmode\epsfbox{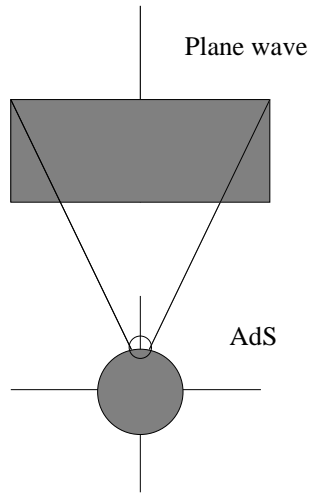}
\end{center}
\caption{Plane wave configurations correspond to filling the
lower half plane. This can be understood from the fact that the plane wave
solution is a limit of the $AdS \times S$ solution.  }
 \label{pplimit}
\end{figure}

\section{1/2 BPS geometries in type IIB string theory}
\renewcommand{\theequation}{2.\arabic{equation}}
\setcounter{equation}{0}

\subsection{1/2 BPS states in the field theory}

We consider ${\cal N}=4 $ super Yang Mills on $S^3 \times R$. We
are interested in the class of states that preserves one half of
the supersymmetries. These are the states associated to chiral
primary operators that are built by taking products of traces of
powers of a single chiral scalar field of $N=4$ Yang Mills.
Denoting by $\phi^i$ the six scalars, we are interested in the
field $Z = \phi^1 + i \phi^2$, and the operators $\prod_i ( Tr
Z^{n_i})^{r_i}$. These BPS states can be described in a variety of
ways. The one that will be most useful for our purposes will be
the description in terms of free fermions discussed  in
\cite{berenstein} (see also \cite{BKRuch,silva}). These free fermions
arise in the following way. We are interested in states with
$\Delta -J=0$. The only such state is the lowest Kaluza-Klein mode
of the field $Z$ on the $S^3$. This mode has a harmonic oscillator
potential which arises from its conformal coupling to the
curvature of $S^3$ \cite{wittenholography}. So we are interested
in the gauge invariant states of a matrix $Z$ in a harmonic
oscillator potential. Standard arguments for matrix quantum
mechanics \cite{brezin} imply that the system reduces to $N$
fermions in a harmonic oscillator potential. We can think of these
fermions as forming droplets in phase space. The ground state
corresponds to a circular droplet. Equivalently, we can say that
we have a quantum hall fluid. We can form the new Hamiltonian $H'
= H-J= \Delta - J$, where $J$ is the angular momentum in the 12
plane. In terms of this new Hamiltonian we have a Landau level
problem. The 1/2 BPS states are the ground states of $H'$ and
correspond to the lowest Landau level. The AdS ground state
corresponds to a circular droplet. The conformal dimension
$\Delta=J$ of any excitation is given by the angular momentum on
the Hall plane, or the energy of the harmonic oscillator, above
the ground state corresponding to the circular droplet. It is also
interesting to take the plane wave limit of these configurations.
In terms of the droplets this amounts to zooming in on the edge of
a droplet, as shown in figure \ref{pplimit}.
 So the plane wave can be thought of as
a Hall configuration where we fill the lower half  plane ($x_2
<0$). BPS excitations correspond to particles and/or hole
excitations. These look like the states of a relativistic fermion.
In fact, the lightcone energy of the states, $-p_- \sim J$, is
indeed given by the expression of the energy for a relativistic
fermion.

These BPS states preserve 16 non-trivial supersymmetries as well as
$ SO(4)\times SO(4) \times R$ bosonic symmetries, where
$R$ corresponds to the Hamiltonian $H' = H-J$. This generator
commutes with the preserved supercharges.

\subsection{1/2 BPS geometries in type IIB supergravity}

We now look for the most general
type IIB geometry that is invariant under $SO(4) \times SO(4)
\times R$. This implies that the geometry will contain two
three-spheres and a Killing vector. We only expect the five--form
field strength to be excited.  So we assume we have a geometry of
the form \bea
ds^2&=&g_{\mu\nu}dx^\mu dx^\nu +e^{H+G}d\Omega_3^2+e^{H-G}d{\tilde\Omega}_3^2\\
F_{(5)}&=&F_{\mu\nu}dx^\mu\wedge dx^\nu\wedge d\Omega_3+ {\tilde
F}_{\mu\nu}dx^\mu\wedge dx^\nu\wedge d{\tilde\Omega}_3
 \l{fiveform} \eea
where $\mu,\nu = 0, \cdots, 3$. In addition, we assume that the
dilaton and axion are constant and that the three-form field strengths
are zero. The self duality condition on the five-form field
strength implies that
 $F_{\mu\nu}$ and $\tilde F_{\mu\nu}$ are dual to each other in four
 dimensions:
 \be \label{fequation}
 F = e^{3 G} *_4 {\tilde F} ~,~~~~~ F = d B, ~~~~ \tilde F = d \tilde B
 \ee
We now demand that this geometry preserves the Killing spinor,
i.e. we require that there are solutions to the equations \bea
\nabla_M\eta+\frac{i}{480}\Gamma^{M_1M_2M_3M_4M_5}
F^{(5)}_{M_1M_2M_3M_4M_5}\Gamma_M \eta=0 \eea This equation is
analyzed using techniques similar to the ones developed in
\cite{solvingsusyequations,Gauntlett,Gauntlett2}. One first writes
the ten dimensional spinor as a product of four dimensional
spinors and spinors on the spheres. Due to the spherical symmetry
the problem reduces to a four dimensional problem involving a four
dimensional spinor. One then constructs various forms by using
spinor bilinears. These forms have interesting properties. For
example, we can construct a Killing vector, which we assume to be
non-zero. This is the translation generator, $\Delta - J$.
There is another interesting form which is a closed one form. This
can be used to define a local coordinate $y$. This coordinate $y$
is rather special since one can show that $y$ is the product of
the radii of the two $S^3$s. By analyzing the  Killing spinor
equations one can relate the various functions appearing in the
metric to a single function. This function ends up obeying a
simple differential equation. We present the details of this
analysis in appendix A. The end result is:
 \bea ds^2 &=& - h^{-2}
(dt + V_i dx^i)^2 + h^2 (dy^2 + dx^idx^i) + y e^{G
 } d\Omega_3^2 + y e^{ - G} d \tilde \Omega_3^2 \l{solmetric}
\\
h^{-2} &=& 2 y \cosh G , \l{solmetric2}
\\
 y \partial_y V_i &=& \epsilon_{ij} \partial_j z,\qquad
 y (\partial_i V_j-\partial_j V_i) = \epsilon_{ij} \partial_y z
  \l{solmetric3}
 \\
 z &=&{ 1 \over 2} \tanh G   \l{solmetric4}
 \\
F &=& dB_t \wedge (dt + V) + B_t dV + d \hat B ~,~~~~~~ \nonumber
\\
 \tilde F &=&
d\tilde B_t \wedge (dt + V) + \tilde B_t dV + d \hat { \tilde B}
\l{4dgf2}
 \\
 B_t &=& - {1\over 4} y^2 e^{2 G
 } ,~~~~~~~~~~~~~~~~~~~~~~~~~~\tilde B_t = - { 1 \over 4}  y^2 e^{- 2 G}
\\
 d \hat B &=&  - { 1 \over 4} y^3 *_3 d ( { z + \half \over y^2 }) ~,~~~~~~~~~~~~~~~~
d \hat {\tilde B} = - { 1 \over 4} y^3 *_3 d ( { z - \half \over y^2 }) ~~ \l{4dgf}
 \eea where $i=1,2$ and $*_3$ is the flat space epsilon
symbol in the three dimensions parameterized by $y,x_1,x_2$. We
see that the full solution is determined in terms of a single
function $z$. This function obeys the linear equation \be
 \partial_i \partial_i z + y \partial_y ( { \partial_y z \over y} ) =0
\label{zequation} \ee Since the product of the radii of the two
3-spheres is $y$,  we would have singularities at $y=0$ unless $z$
has a special behavior. It turns out that the solution is
non-singular as long as $z = \pm \half $ on the $y=0$ plane spanned
by $x_1,x_2$. Let us consider the case $z=\half$ at $y=0$.  Then
we see that $z$ will have an expansion $z \sim
 \half - e^{-2 G} =\half - y^2 f(x) + \cdots$, where
$f(x)$ will be positive with our boundary conditions.
From this we find that $e^{-G} \sim y c(x)$. So we see that the metric in the
$y$ direction and the second 3-sphere directions becomes
\be
 h^2 dy^2 + y e^{- G} d\tilde \Omega_3^2  \sim {  c(x)} ( dy^2 + y^2
 d\tilde \Omega_3^2 )
 \ee
 In addition we see that $h$ remains finite and the radius of the first sphere
 also remains finite. One can also show that $V$ remains finite by using the
 explicit expression we write below. When $z=- \half $ the discussion is similar.
 In fact the transformation $z\to -z$ and an exchange of the two
 three--spheres is a symmetry of the equations. This corresponds to a particle
 hole transformation in the fermion system. This will not be a symmetry
 of the solutions if the fermion configuration itself is not particle-hole
 symmetric, or the asymptotic boundary conditions are not
 particle-hole symmetric (as in the $AdS_5 \times S^5$ case).
We will explain below that  the solution is non-singular at the
boundary of the two regions.   So in order to determine the
solution we need to specify regions in the $x_1,x_2$ plane where
$z= \pm \half $. These two signs corresponds to the fermions and
the holes, and the $x_1,x_2$ plane corresponds to the phase space.
After defining $\Phi = z/y^2$ the equation (\ref{zequation})
becomes the Laplace equation in six dimensions for $\Phi$ with
spherical symmetry in four of the dimensions, $y$ is then the
radial variable in these four dimensions.
 The boundary
values of $z$ on the $y=0$ plane are charge sources for this
equation in six dimensions.  It is then  straightforward to write
the general solution once we specify the boundary values.  We find
\bea
z(x_1,x_2,y)&=&\frac{y^2}{\pi}\int_{\cal D} \frac{z(x_1',x'_2,0)
dx_1'dx_2'}{[({\bf x}-{\bf x}')^2+y^2]^2} \l{zepres} =
-\frac{1}{2\pi}\int_{\d {\cal D}} dl~ n_i'\frac{x_i-x_i'}{[({\bf x}-{\bf x}')^2+y^2]}
+ \sigma
\\
V_i(x_1,x_2,y) &=&  { \epsilon_{ij} \over \pi} \int_{\cal D}
\frac{z(x_1',x'_2,0) (x_j - x'_j) dx_1'dx_2'}{[({\bf x}-{\bf
x}')^2+y^2]^2} =  {\epsilon_{ij}  \over 2 \pi} \oint_{\d {\cal D}} {
dx'_j \over ({\bf x}-{\bf x}')^2+y^2  } \l{vexpres} \eea where in
the second expressions for $z ,V_i$ we have used that
 $z(x_1', x_2', 0)$ is locally constant
and we have integrated by parts to convert integrals over droplets ${\cal D}$ into
the integrals
over the boundary of the droplets $\d{\cal D}$. In these expressions
 $n_i$ is the unit normal vector to the droplet pointing towards the $ z =\half$ regions,
 $\sigma  $ is a contribution from infinity which arises in the
 case that $z$ is constant  outside a circle of very large radius (asymptotically
 $AdS_5 \times S^5$ geometries). $\sigma = \pm \half$ when we have $z = \pm \half$
 asymptotically. The contour integral in (\ref{vexpres}) is oriented in such a way
 that the $z = - \half$ region is to the left.
We see from the second expression for $V$ in (\ref{vexpres}) that
$V$ is finite as $y\to 0$ in the interior of a droplet. We also
see from (\ref{vexpres}) that $V$ is a globally well defined
vector field.\footnote{ In the cases that we consider, where at
most the $x_1$ coordinate is compact, there are no
compact two cycles   in the $x_1,x_2,y$ space. So  we do
not have any compact two cycles on which  we could find a
non-zero integral of $dV$.  }
 This is important
since we want the time direction parameterized by $t$ to be well defined
 (i.e.
we do not want NUT charge).

\subsection{Examples}

Let us now consider a simple solution which is associated to the half
filled plane.
We have the boundary conditions
\bea
z(x_1',x_2',0)=\frac{1}{2}\mbox{sign}~x'_2
\eea
From this data we can compute the entire function $z(x_2,y)$ using (\ref{zepres}),
(\ref{vexpres})
\bea
z(x_2,y) &=& \frac{1}{2}\frac{x_2}{\sqrt{x_2^2+y^2}} \l{zpp}
\\
V_{1} &=& \half { 1 \over \sqrt{ x_2^2 + y^2} } ~,~~~~~~V_2 =0 \l{vpp}
\eea
Inserting this into the general ansatz (\ref{solmetric})
and performing the change of coordinates
\bea
y & = & r_1 r_2
\\
x_2 & = & \half ( r_1^2 - r_2^2)
\eea
we obtain the usual form of the metric for the plane wave \cite{figueroa}
\be \label{ppwavemetric}
ds^2  = - 2 dt  dx_1  - (r^2_1 + r_2^2) dt^2 + d{\vec r}_1^{\, 2} + d
{\vec r}_2^{ \, 2}
\ee
We see that the final solution is smooth,
 despite the fact that on the $y=0$ plane  $V$ diverges  at the boundary
between two regions ($x_2=0$ in this case).
In fact, this computation shows that, in general, the boundary between two regions
is smooth. The reason is that locally the boundary between two regions looks
like the plane wave and therefore we will get a non-singular metric.

Let us now recover the familiar $AdS_5 \times S^5$ geometry.
In this case it is convenient to introduce a function $\tilde z = z -\half$.
The Laplace equation for $\tilde z/y^2$ has sources on a disk of radius
$r_0$. We choose polar coordinates $r, \phi$ in the $x_1,x_2$ plane.
We obtain
\bea
{\tilde z}(r,y)&=&-\frac{y^2}{\pi}\int_{\rm Disk}
\frac{ r' dr' d\phi}{[r^2+ r'^2-2r r'\cos\phi+y^2]^2}
\nonumber\\
{\tilde z}(r,y;r_0) &\equiv&\left.
\frac{r^2- r_0^2+y^2}{2\sqrt{ (r^2+r_0^2+y^2)^2-4r^2r_0^2  }
 }\right. - \half  \l{ads1}
\\
V_\phi&=&-r\sin\phi V_1+r\cos\phi V_2=-\frac{1}{2\pi}\int_{\d {\cal D}}
\frac{rr'\cos\phi' d\phi'}{r^2+r'^2+y^2-2rr'\cos\phi'}\nonumber\\
V_\phi(r,y;r_0)&\equiv&-\frac{1}{2}\left.\left(
\frac{r^2+y^2+r_0^2}{\sqrt{(r^2+ r_0^2+y^2)^2-4 r^2r_0^2}}-1
\right)\right. \l{ads2}
\eea
Inserting this into the general ansatz  and
performing the change of coordinates
\bea
y &=& r_0 \sinh \rho \sin \theta
\\
r &= &r_0 \cosh\rho \cos \theta
\\
\tilde \phi &= & \phi - t
\eea
we see that we get
the standard $AdS_5 \times S^5$ metric
\be \l{ourads}
ds^2 = r_0 [ - \cosh^2 \rho d t^2 + d\rho^2+ \sinh^2 \rho d\Omega_3^2 +
d \theta^2 + \cos^2 \theta d \tilde \phi^2 + \sin^2 \theta d\tilde \Omega_3^2 ]
\ee
So we see that $r_0 =R^2_{AdS}=R_{S^5}^2$.
 In fact, under an overall scaling of the
coordinates $ (x_i , y) \to \lambda ( x_i, y)$ the metric scales
by a factor $\lambda$. This is what we expect since   the total area
of the droplets is equal to the number of branes, a fact which
we will demonstrate later.
 By comparing the value of the $AdS$ radius
we obtained in (\ref{ourads}) and the standard answer, $R^4_{AdS} = 4 \pi l_p^4N$,
we can write the precise quantization condition on the area of the droplets in
the 12 plane as\footnote{ We define $l_p = g^{1\over 4} \sqrt{\alpha'}$.}
\be \l{hbardef}
  ({\rm Area}) =  4 \pi^2 l_p^4 N ~,~~~~~~~~~~~{\rm or}~~~\hbar = 2 \pi l_p^4
  \ee
where $N$ is an integer, and we have defined an effective $\hbar$ in the $x_1,x_2$
plane, where we think of the $x_1,x_2$ plane as phase space.

\begin{figure}[htb]
\begin{center}
\epsfxsize=4.0in\leavevmode\epsfbox{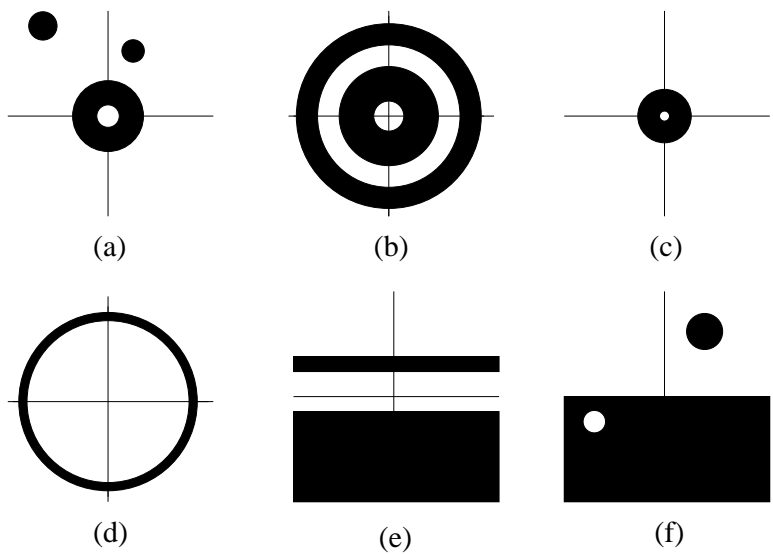}
\end{center}
\caption{We see various configurations whose solutions can be easily constructed
as superpositions of the $AdS_5 \times S^5$ solution and the plane wave solution.
In (a) we see an example of the type of configurations that can be obtained
by superimposing the circular solution (\ref{ads1}). In (b) we see generic
configurations that lead to solutions which have two Killing vectors and
lead to static configurations in $AdS$. In (c) we see the solution corresponding
to a superposition of D3 branes wrapping the $\tilde S^3$ in $S^5$. In (d) we see the
configuration resulting from many such branes, which can be thought of as a
superposition of branes on the $S^3$ of $AdS_5$ uniformly distributed along
the angular coordinate $\tilde \phi$ of $S^5$. In (e) we see a configuration
that can be viewed as an excitation of a plane wave with constant energy
density. In (f) we see a plane wave excitation with finite energy.
 } \label{rings}
\end{figure}

Now that we have constructed the solution for a circular droplet, we can
construct in a trivial way the solutions that are superpositions of circles,
see figure \ref{rings}(a) \footnote{Note that, even though the figure depicts
black rings, these solutions are not related in any obvious way to the ``black rings''
discussed recently  \cite{blackrings}. The solutions in \cite{blackrings} contain horizons,
while ours do not. They also preserve a different number of supersymmetries.}.
Among these the ones corresponding to
concentric circles have an extra Killing vector.
These lead to time independent configurations in $AdS$. All other
solutions will depend on $\phi=t + \tilde \phi$ where $t$ is the time
in $AdS$ and $\tilde \phi$ is an angle on the asymptotic $S^5$, see (\ref{ourads}).
The solutions corresponding to concentric circles are therefore superpositions
of (\ref{ads1}) and (\ref{ads2})
\be
\tilde z = \sum_i (-1)^{i+1} \tilde z(r, y;r^{(i)}_0),\qquad
V_\phi = \sum_i (-1)^{i+1} V_\phi(r, y;r^{(i)}_0)
\ee
Here $r_0^{(1)}$ is the radius of the outermost circle, $r_0^{(2)}$ the next one,
etc (see figure \ref{rings}(b)).
Let us discuss the
solution corresponding to a single black ring \ref{rings}(c).
When the white hole in the
center is very small, this can be viewed as branes wrapping a maximal $\tilde S^3$ in
$S^5$. When the area of this hole, $N_h$, is smaller than the original area, $N$,
 of the
droplet ($N_h \ll N$),
the solution will locally look like an $AdS_5 \times S^5$ solution near
the hole. When we increase the number of branes  wrapped on $\tilde S^3$ in $S^5$ the
area of the holes becomes very large and in the limit we get a rather thin
ring, which could be viewed as a superposition of D3 branes wrapping
an $S^3$ in $AdS_5$ \footnote{ Note that this seems to disagree with a
proposal in \cite{Balasubramanian:2001dx} for realizing a larger radius $AdS$ space
inside a smaller radius $AdS$ space.}, see figure \ref{rings}(d).

\begin{figure}[htb]
\begin{center}
\epsfxsize=3.0in\leavevmode\epsfbox{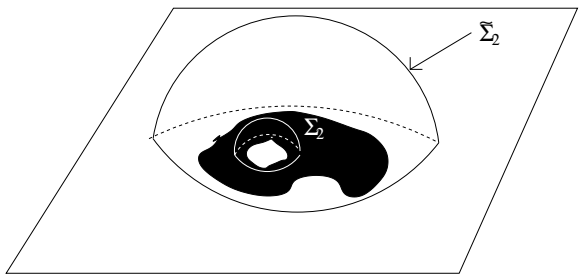}
\end{center}
\caption{ We can construct a five-manifold by adding the sphere
$\tilde S^3$ fibered over the surface $\tilde \Sigma_2$. This is a
smooth manifold since at the boundary of $\tilde \Sigma_2$ on the
$y=0$ plane the sphere $\tilde S^3$ is shrinking to zero. The flux
of $F_5$ is proportional to the area of the black region inside
$\tilde \Sigma_2$. Another five manifold can be constructed by
taking $\Sigma_2$ and adding the other three--sphere $S^3$. The
flux is proportional to the area of the white region contained
inside $\Sigma_2$.    } \label{cups}
\end{figure}

\begin{figure}[htb]
\begin{center}
\epsfxsize=2.0in\leavevmode\epsfbox{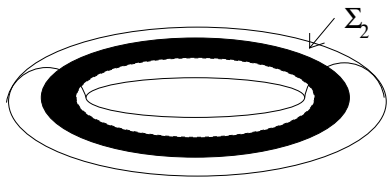}
\end{center}
\caption{ We see here an example of a two dimensional surface, $ \Sigma_2$,
 that is
surrounding a ring. If we add the three--sphere $\tilde S^3$
fibered over $ \Sigma_2$ we get a five manifold with the topology
of $S^4 \times S^1$.} \label{ringcup}
\end{figure}

\subsection{Topology and charges of the solutions.}

Let us explore the topology of the solutions. This analysis is
somewhat similar to that used in toric geometry. As long as
$y\not =0$ we have two $S^3$s. Let us denote these two spheres as
$S_3$ and $\tilde S_3$. At the $y=0$ plane the first sphere shrinks in
a non-singular fashion if $z= -\frac{1}{2}$ while the second sphere, $\tilde S^3$,
 shrinks if
$z=\frac{1}{2}$. Both spheres shrink at the boundary of the two regions.
In fact there is a shrinking $S^7$ at these points, since the
geometry is locally the same as that of a pp-wave. For example, in
the $AdS_5 \times S^5$ solution the second sphere, $\tilde S_3$,
shrinks at $y=0$ outside the circle, this is the three--sphere
contained in $S^5$. On the other hand the three--sphere
contained in $AdS_5$ shrinks at $y=0$ inside the circular droplet.
Consider a  surface $\tilde \Sigma_2$ on the $(y,\vec x)$ space
that ends at $y=0$ on a closed, non-intersecting curve lying in a
region with $z=\frac{1}{2}$ see figure \ref{cups} . We can construct a
smooth five dimensional manifold by fibering the second three
sphere, $\tilde S^3$, on $\tilde \Sigma_2$. This is a smooth manifold which is
topologically a five--sphere. We can now measure the flux of the
five-form field strength $F_5$ on this five-sphere. Looking at the
expressions for the field strength (\ref{fiveform}) in terms of
the four dimensional gauge field (\ref{4dgf}), (\ref{4dgf2}) we
find that the spatial components are given by $\tilde F|_{spatial}
= d( \tilde B_t V) + d \hat {\tilde B}$. Since $B_t V$ is a
globally well defined vector field the flux is given by \be
 \tilde N =- { 1 \over 2 \pi^2 l_p^4 } \int d \hat{ \tilde B}
 = { 1 \over 8 \pi^2 l_p^4 }\int_{\tilde \Sigma_2} y^3*_3 d
 \left({z - \half \over y^2} \right) =  { ({\rm Area})_{z=-\half} \over
 4 \pi^2 l_p^4 }
 \l{areaflux} \ee
where $\tilde \Sigma_2$ is the two surface in the three
dimensional space spanned by $y,x_1,x_2$. This expression gives
the total charge inside this region for the Laplace equation,
which in turn is equal to the total area with $z=-\half$ contained
within the contour on which $\tilde \Sigma_2$ ends at $y=0$, see
figure \ref{cups}. Note that (\ref{areaflux}) leads to the
quantization of area, (\ref{hbardef}). In the $AdS_5 \times S^5$
case there is only one non-trivial five--sphere and this integral
gives the total flux. This flux is quantized in the quantum
theory.

We can consider an alternative five--sphere by considering a
surface that ends on the $y=0$ plane on a region with $z
=-\frac{1}{2}$ (see figure \ref{cups}). The flux over this
five--manifold is given by \be N =  { 1 \over 2 \pi^2 l_p^4 } \int
d \hat{  B}
 = - { 1 \over 8 \pi^2 l_p^4 }\int_{\Sigma_2} y^3*_3 d
 \left({z + \half \over y^2} \right) =  { ({\rm Area})_{z=\half} \over
 4 \pi^2 l_p^4 }
\ee and it measures the total area of the other type, with $z=\frac{1}{2}$, contained in
this region. If these fluxes are non-zero, then
these spheres are not contractible. So if we have a large number
of droplets, we have a complicated topology for the solution. In
addition we can construct other 5-manifolds which are not
five--spheres by considering more complicated surfaces. For
example we get the five-manifold with topology  $S^4 \times S^1$
from the surface depicted in figure \ref{ringcup}.

Another  interesting property of the solutions is their energy or
their angular momentum $J$. These are equal to each other due to
the BPS condition $\Delta = J$. As explained in \cite{berenstein},
this energy is the energy of the fermions in a harmonic oscillator
potential minus the energy of the ground state of $N$
fermions\footnote{ Equivalently we can express it as the angular
momentum of the quantum Hall problem.}. From the gravity solution
it is easier to read off the angular momentum. This involves
computing the leading terms in the $g_{\phi+t,t}$
components of the metric. The details of this computation are
given in appendix \ref{AppEnergy}. The final expression is
\bea
\Delta &=& J =
{ 1\over 16 \pi^3 l_p^8 } \left[ \int_{\cal D} d^2 x (x_1^2 + x_2^2) - {
1 \over 2 \pi}
 \left( \int_{\cal D} d^2 x \right)^2 \right] \nonumber
 \\
 &=&  \int_{\cal D} { d^2 x \over 2 \pi \hbar }  { \half (x_1^2 + x_2^2) \over
 \hbar } - \half \left( \int_{\cal D} { d^2 x  \over 2 \pi \hbar } \right)^2
 \l{energyf}
\eea
where $\cal D$ is the domain where $z = - \half$, which is the domain where the fermions are.
Using the definition of $\hbar$ in (\ref{hbardef}) we see that this is the quantum
energy of the fermions minus the energy of the ground state.

None of the solutions described here has a horizon and they are
all regular solutions. A singular solution was considered in
\cite{myerstat}. That solution was obtained as the extremal limit
of a charged black hole in gauged supergravity \cite{5dgbh1,5dgbh2}. Since it is a BPS solution it  obeys our equations.
We find that the boundary conditions on the $y=0$ plane are such
that we have a disk, similar to the one we have in $AdS$ but the
boundary value of $2z$ is  not  $-1$  but $-1/(1+q)$ where $q$ is
the charge parameter of the singular solution. Of course, the
solution is singular because it violates our boundary condition,
but it could be viewed as an approximation to the situation where
we dilute the fermions, or we consider a uniform gas of holes in
the disk, which agrees with the picture in \cite{myerstat}.

\begin{figure}[htb]
\begin{center}
\epsfxsize=2.0in\leavevmode\epsfbox{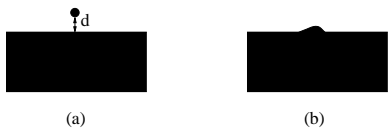}
\end{center}
\caption{In (a) we see a small circular droplet of area $\hbar$ that is at some distance
$d$ from a bigger droplet which is described by a smooth gravity solution.
If the distance $d$ is larger than $\sqrt{\hbar}$ this is a topologically non-trivial
excitation  with
an energy of the order of $M_{pl}$ above the ground state. As $d \to 0$ the
energy decreases, but its curvature increases and, in the fermion picture, such an
excitation is better described
in terms of ripples of the Fermi surface, as in  (b). These are
 gravitons propagating
in the smooth original geometry. Notice that the geometries corresponding
to pictures (a) and (b) have different topologies.    } \label{wave}
\end{figure}

Note that any droplet which is far away from other droplets will
look locally near the droplet like $AdS_5  \times S^5$. In
particular if we have a fermion droplet with $z =-\half$
 surrounded by a sufficiently large region with $z=\half$,
then in the $z = \half$ region we see that $S^3$ is not contractible\footnote{ In the asymptotic region, $S^3$ is in $AdS_5$ and $\tilde S^3$ is in
$S^5$. }. The droplet can be viewed as branes
wrapped on $S^3$. On the droplet itself, the $S^3$ is contractible
but now there is a new $S^5$ that is not contractible, so we have
a geometric transition. The $S^5$ is constructed by fibering the
three--sphere $\tilde S^3$ on a surface $\tilde \Sigma_2$ which
surrounds the droplet  as in figure \ref{cups}. If the amount of
flux is small, these geometries can be viewed as branes in an
background geometry, but as the flux becomes large they can be
viewed as smooth geometries with fluxes. Note that even a single
brane, can be viewed as a highly curved smooth geometry with flux,
but this geometrical description is misleading for some aspects of
the physics. In fact we expect that curvatures will become high if
the dimensions of the droplets become of order one. More
precisely, we expect that curvatures will become high if the
linear dimensions of the droplet become of order $\sqrt{\hbar}$,
or if two droplets come close together at distances smaller than
$\sqrt{\hbar}$. Note that if we have a small circular droplet of
area $\hbar$ and we bring it to within a distance of order
$\sqrt{\hbar}$ from the big circular droplet corresponding to the
$AdS$ ground state, then the configuration has an energy of order
$M_{pl}$ in ten dimensions. If we bring this small droplet to a
distance $d \ll \sqrt{\hbar}$ from the big circular droplet we
will get a highly curved geometry that formally has very small
energy. On the other hand low energy excitations  described by
gravity modes correspond to small long wavelength fluctuations of
the big circular droplet. It is clear from the fermion picture
that a fermion very close to the Fermi surface is well described
by the boson characterizing long wavelength excitations of the
fermion fluid. So we conclude that highly curved topologically
non-trivial excitations with very small energies are already
included as gravity modes, see figure \ref{wave}.
 Here we have always discussed the curvature of the solution in
Planck units. If the string coupling is small, the geometry can be rendered
invalid by stringy corrections at a smaller curvature scale.

The solutions we are discussing here are somewhat reminiscent of
the Coulomb branch solutions that arise when we consider $D3$
branes on $R^{1+3}$. In fact, the $SO(4)$ invariant subset of the
latter can be obtained from the solutions in this paper by taking
appropriate limits (see appendix \ref{dilute}).

Distributions of droplets in a compact region
of the 12 plane lead to solutions with
  $AdS_5 \times S^5$ asymptotics. Solutions which
correspond to finite deformations of the half filled plane are
asymptotic to the pp-wave geometry. Let us discuss the latter
solutions a bit more. Solutions with a small droplet or a small
hole, see figure \ref{rings} (f), correspond to branes wrapping
the $S^3$ or $\tilde S^3$. Large size droplets correspond to new
geometries with fluxes. One can also consider solutions that are
translation invariant along $x_1$. These are solutions
corresponding to empty and occupied bands see figure \ref{rings}
(e). These solutions have infinite energy, but finite energy
density. We can also compactify the direction $x_1$. This is
really a DLCQ compactification, since the solution asymptotes to a
pp-wave where $x_1 = x^-$, see (\ref{ppwavemetric}). These
solutions correspond to the DLCQ of the pp-wave. The momentum
$-p_-$ is the energy of the fermion configuration after we take
the Fermi surface to be in a position such that the total number
of particles and holes is the same.

\subsection{ M2 brane theory with a mass deformation}

In this section we consider geometries that are dual to the M2 brane theory
with a mass deformation \cite{cpnw,ibnw}.
Starting with the usual theory on coincident M2 branes,
it is possible to introduce a mass deformation
that preserves 16 supercharges. This deformation preserves an $SO(4) \times SO(4)$
subgroup of the $SO(8)$ R-symmetry group of the conformal M2 brane theory.
One interesting aspect of this theory is that its features are rather
similar to those of ${\cal N}=4$ SYM with a mass deformation. Namely, the mass
deformed M2 brane theory also has vacua that are given by dielectric branes \cite{myers}. In
this case these are M5 branes that are wrapping a 3-sphere in the first four of the
eight transverse coordinates or a 3-sphere in the last four of the eight transverse
coordinates.
We can obtain these solutions by U-dualizing some of the solutions discussed
above.
The authors of \cite{ibnw} managed to
reduce the problem to finding a solution of a harmonic equation.
The relation
between their function and ours is given in appendix \ref{benawarner}.
Our parametrization of the ansatz has the advantage that it is
very simple to select out the non-singular solutions.
\begin{figure}[htb]
\begin{center}
\epsfxsize=2.0in\leavevmode\epsfbox{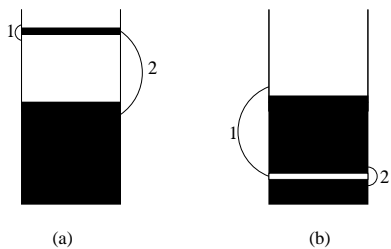}
\end{center}
\caption{ We see the configurations corresponding to two of the
vacua of the mass deformed M2 brane theory. The vacuum in (a) can
be viewed as  dielectric M5 branes wrapping the $S^3$ in the first
four coordinates of the eight transverse coordinates. The
configuration in (b) corresponds to a vacuum with  M5 branes
wrapping the $S^3$ in the second four coordinates. The two
geometries have the same topology. Consider arcs in the $x_2,y$
plane that enclose the fermions or the holes and end at $y=0$. We
can construct four spheres by taking one of these arcs and
tensoring the $S^3$ that shrinks to zero at the tip of the arcs.
The two different $S^3$ are denoted in the figure by indices $1$
and $2$. The flux of $F_4$ over these four spheres is equal to the
number of particles or holes enclosed by the arcs. Note that the
horizontal line in this figure does not correspond to a coordinate
in the final M-theory geometry. } \label{twovacua}
\end{figure}

This system is intimately related to the type IIB solutions that
we considered above. One way to see the connection is the
following. It was argued in \cite{banks} that the DLCQ of type IIB
string theory with $N$ units of DLCQ momentum is the same as the
theory on $N$ M2 branes on a torus. We can now consider the DLCQ
of the maximally supersymmetric pp-wave, where we periodically
identify along the lightlike Killing direction, $x^- \sim x^- + 2
\pi R$ in (\ref{ppwavemetric}). The sector with $N$ units of
momentum $-p_- = N/R$ is given by the mass deformed M2 brane
theory on a torus \footnote{ There has been another proposal for
the DLCQ limit of this theory in \cite{jabbari}, which involves a
rather different theory.}.
% Our proposal gives a clear description
%of BPS D1 and F1 branes wrapped along the DLCQ direction in terms of states with
%momentum along the two directions of the torus. It is not clear how to
%describe such states from the point of view of \cite{jabbari}.}.
From the pp-wave point of view it is clear that there can be
supersymmetric vacua that correspond to D3 branes wrapping either
of the $S^3$s. On the M-theory side, these map into vacua of the
M2 brane theory where the M2 branes form an M5 brane wrapping
either of the two $S^3$s. There is a large number of vacua that
are in one to one correspondence with the partitions of $N$.
Perhaps the simplest way to  count these vacua is to recall yet
another description of this DLCQ theory in terms of a limit of a
gauge theory in \cite{mukhi}. According to the description in
\cite{mukhi} the vacua are given in terms of chiral primary
operators of a particular large $N$ limit of an orbifold theory.
It is a simple matter to count those and notice that they are
equivalent to partitions of $N$. This is of course related in a
simple manner to the fermion fluid picture for the pp wave. Once
we compactify $x^-$ we have fermions on a cylinder, where we fill
half the cylinder. The asymptotic conditions automatically imply
that we are only interested in states with zero $U(1)$ charge. The
$U(1)$ charge is related to the position of the Fermi level. We
always choose it such that the total number of particles and holes
is zero. The energy of the fermions is the same as the number $N$
 of M2 branes. These are relativistic fermions which can be bosonized and the number
 of states with energy $E= N$ is indeed given by the partitions of $N$.
States which contain  highly energetic holes or particles, as
shown in figure \ref{twovacua}, correspond to M5 branes wrapping
one or the other $S^3$. Configurations in between are better
thought of as smooth geometries with fluxes. An interesting fact
is that the geometry corresponding to a highly energetic fermion,
as in figure \ref{twovacua}(a), and the geometry corresponding to
a highly energetic hole, as in figure \ref{twovacua} (b), are
topologically the same. The reason is that the  geometry contains
two distinct $S^4$s through which we have a non-vanishing flux.
Consider for example the configuration in figure \ref{twovacua}
(a), which can be interpreted as M5 branes wrapping one of the
$S^3$s. One $S^4$ is the obvious one that is transverse to these
branes. The other $S^4$ arises in an interesting way. Consider the
three--sphere that these branes are wrapping. At the center of the
space, where one normally imagines the $M2$ branes, this
three--sphere is contractible. As we start  going radially
outwards we encounter the M5 branes, the backreaction of the
branes on the geometry will make the $S^3$ on their worldvolumes
contractible. So the end result is that the $S^3$ contracts to
zero on both end points of the interval that goes between the
origin and the branes. This produces another $S^4$. Through this
$S^4$ we have a large flux, which we might choose to view as part
of the background flux, the flux that was there before we put in
the M2 branes, the flux which is responsible for the mass term on
the M2 brane theory. A configuration with highly energetic holes
corresponds to M5 branes wrapping the second $S^3$. This is
topologically the same as the configuration with highly energetic
fermions. In other words, the two configurations in figure
\ref{twovacua} have the same topology. They only differ in the
amount of four form flux over the two $S^4$s.

In this problem there is a precise duality under the interchange of the two
three--spheres, which maps solutions into each other. Some special solutions will
be invariant under the duality. This is particle hole duality in the fermion
picture.

Finally, let us give the explicit form of the  solutions
\bea
ds^2_{11} &=& e^{ 4\Phi \over 3} ( - dt^2 + dw_1^2 + dw_2^2)  \nonumber
\\
& & ~~~~~~~~~~+
e^ { - 2 \Phi \over 3} \left[  h^2 ( dy^2 + dx_2^2) + y e^{G} d\Omega_3^2 +
y e^{-G} d\tilde \Omega_3^2  \right]
 \l{mtwo1}\\
e^{2 \Phi} &=& { 1 \over h^2 - h^{-2} V_1^2} \l{mtwo2}
\\
F_{4} &=& - d(e^{ 2\Phi} h^{-2} V_1)\wedge dt \wedge dw_1\wedge
dw_2  \nonumber
\\
 && ~~~~~~~~~~ - {1 \over 4} e^{-2 \Phi}
 [ e^{-3 G} *_2 d(  y^2 e^{2 G} ) \wedge d\tilde \Omega_3 + e^{3 G}
*_2 d(  y^2 e^{-2G}) \wedge d\Omega_3 ] \l{mtwo3} \eea where $*_2$
is the flat epsilon symbol in the coordinates $y,x_2$ and $h,G$
are given by the expressions we had above
(\ref{solmetric2})--(\ref{solmetric4}), (\ref{zequation}). These
functions are determined by considering boundary conditions
corresponding to strips that are translation invariant along
$x_1$, see figure \ref{rings}(e) and equations (\ref{zpp}),
(\ref{vpp}). Note that since we had translation symmetry along
$x_1$ in the original IIB solution, only the component $V_1$ is
nonzero. The coordinate $x_1$ does not appear in this M-theory
solution because it was U-dualized. So  $z$, $V_1$ are given by
superpositions of solutions of the form (\ref{zpp})--(\ref{vpp}).
In other words \be  \l{solstrips} z(x_2,y) = \sum_i (-1)^{i+1}
z^{pp}(x_2-x_2^i , y),\qquad V_1(x_2,y) = \sum_i (-1)^{i+1}
V_1^{pp}(x_2-x_2^i , y) \ee where $z^{pp},V_1^{pp}$ are the
functions in (\ref{zpp}), (\ref{vpp}), and $x_2^i$ is the position
of the $i$th boundary starting from the bottom of the Fermi
sea\footnote{ For odd $i$ the boundary changes from black to white
while for even $i$ the boundary changes from white to black. See
figure \ref{rings} (e).}. The relation between our parametrization
of the solution, (\ref{mtwo1})-(\ref{mtwo3}), and the
parametrization in \cite{ibnw} is given in appendix
\ref{benawarner}.

\begin{figure}[htb]
\begin{center}
\epsfxsize=2.0in\leavevmode\epsfbox{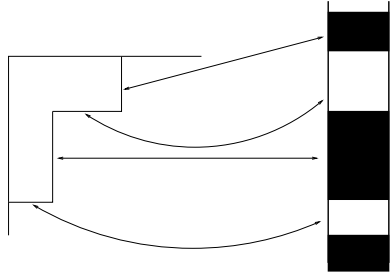}
\end{center}
\caption{  Correspondence between the Young diagrams and the states
of free fermions. We start from the bottom left of the Young diagram,
each time we move right by $n$ boxes  we add $n$ holes and each time we
move up by $n$ boxes we add $n$ fermions. The energy of the configuration
is equal to the total number of boxes of the Young diagram.
} \label{young}
\end{figure}

These solutions can be related to Young diagrams in a simple way which is
pictorially represented in figure \ref{young}. We
start at the bottom of the Young diagram and we move along the boundary. Each
time we move up we add as many fermions as boxes, each time we move right we
add holes. The Fermi level
is set so that the total number of holes is equal to the total number of fermions.
Then the energy of the fermion system is equal to the number of boxes, and in
our case this is the number of M2 branes.
Of course, small curvature solutions are only those where the Young diagram
has a small number of corners and a large number of boxes.
 This is in contrast to the situation encountered
in other cases \cite{douglaskazakov,vafacrystal}
where smooth Young diagrams
correspond to smooth macroscopic configurations.
In our case, a Young diagram which
contains edges separated by few boxes  leads to
solutions with Planck scale curvature.

Using Young diagrams we can describe in a similar way the
circularly symmetric configurations in the $x_1, x_2$ plane.
Solutions that are not circularly symmetric are given by
superpositions of these diagrams. In other words, the Young
diagrams are in direct correspondence with the momentum basis for
the fermions. Generic states that are not invariant under
translations (or rotations) are given, in the Hilbert space, by a
superposition of these. In the gravity description the only states
that lead to smooth geometries are those which form well defined
droplets in the Fermi sea.

\begin{figure}[htb]
\begin{center}
\epsfxsize=2.0in\leavevmode\epsfbox{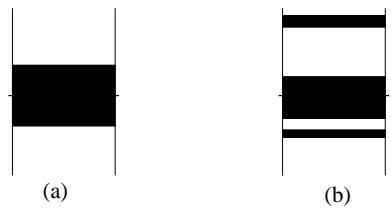}
\end{center}
\caption{We see fermion configurations corresponding to a single
isolated strip, or set of strips. These fermions are the same as
the ones that appear in 2d QCD on a cylinder, or $SU(N)$ group
quantum mechanics. In (a) we display the ground state and in (b)
we display an excited state.
From the point of view of D4 brane on $S^3\times S^1\times R$ these are
all supersymmetric ground states.
} \label{singlestrip}
\end{figure}

It is also interesting to consider fermion distributions that are
not asymptotic to the distributions for $AdS \times S$ or plane
waves. For example, we can consider a single isolated strip of
fermions, as in figure \ref{singlestrip}. If we compactify the
$x_1$ coordinate then the fermion configuration is the same as the
one we have in two dimensional QCD on a cylinder. In fact, the
dual field theory configuration for a single isolated strip (or
single collection of strips) is $N$ M-fivebranes wrapped on $S^3
\times T^2 \times R$, where $N$ is given by  the area of the
strip. We can also think of this as Yang Mills theory on $S^3
\times S^1 \times R $. The reduction on $S^3$ leaves us with a
gauge theory in two dimensions, which has BPS vacua that are in
correspondence with the states of 2d Yang-Mills theory on a circle
\cite{2dqcd}\footnote{Recently an interesting connection between
2d Yang-Mills theory and topological strings was proposed in
\cite{vafa2dQCD}.}. We discuss this a bit more in appendix
\ref{AppD4brane}.   There are other asymptotic configurations that
could be explored, such as wedges in the $12$ plane, etc.

In the  ${\cal N}=1^*$ theory considered in
\cite{jpms} we expect a similar situation, where geometries will be non-singular
but could have large curvatures when some of the fluxes become small.

\subsection{ Analytic continuation to $AdS_3 \times S^3 \times S^1$ }
\label{iibcontinuation}

If we want to describe  solutions with $AdS_3\times S^3$ factors,
 rather than $S^3\times S^3$, then the
following minor changes should be made from
(\ref{solmetric})-(\ref{4dgf2}):
 \be y = i y'
~,~~~~~~~~G = G'+ { i \pi \over 2} ~,~~~~~~~~x_j = i x'_j \ee \be
d\Omega_3^2 = - d s^2_{AdS_3} \ee Then we find that \be \l{analce}
h^{-2} = - y'( e^{G'} -e^{-G'} )~,~~~~~ V_j = -i V_j', ~~~~~ z = { 1
\over 2 \tanh G'} \ee After we insert these expressions in the
ansatz (\ref{solmetric})-(\ref{4dgf2}) and we take all primed
quantities to be real we get a real solution with an $AdS_3 \times
S^3$ factors. The coordinate parameterized by $t$ is now spacelike,
so we can take it to be compact. We can set $t = \chi$
\footnote{Note that we do not Wick rotate $t$.}. It would be nice
to see if there are any solutions with a compact internal
manifold. Of course we have the well known $AdS_3 \times S^3
\times T^4$ or $AdS_3 \times S^3 \times K3$ solutions. But our
ansatz does not cover these, because in our case the $\chi$
translation generator appears in the right hand side of the
supersymmetry algebra (as a $U(1)$ central charge).
 We did not manage to find any solutions with
a compact internal manifold\footnote{
It is clear from the expression for $z$ in (\ref{analce}) that at $y=0$ we
can only have one type of boundary condition since $z$ cannot continuously
change from positive to negative, with real $G'$.}.

We find an interesting limit of these analytically continued
solutions, under which the $AdS_3\times S^3$ factor becomes a
6-dimensional flat space and the remaining 4-dimensional
transverse space turns out to be a Hyper-Kahler manifold with a
translational Killing vector \cite{GH}. Let us look for a solution
of the equation, \be
\partial_{i'} \partial_{i'} z + y' \partial_{y'} ({ 1 \over y'} \partial_{y'} z )
 = 0
 \ee
We will consider the solution in the range $ | y' - y_0| \ll y_0 $
and rewrite $z$ as $z \sim y_0 g$, then $g$ satisfies the 3d
Laplace equation \be \l{eforg}
\partial_{i'} \partial_{i'} g + \partial_{y'}^2 g =0 \ee
We could consider solutions where $g$ is asymptotically a constant
if wanted. We will now take the limit $y_0 \to \infty$ and $z \to
\infty$ and from (\ref{analce}) we see \be z \sim { 1 \over 2
G'},~~~~~~~h^{-2} \sim - y_0 2 G' \sim - { 1 \over g}
\label{iiblimit} \ee We now insert this and (\ref{analce}) into
the general ansatz (\ref{solmetric}) for the ten dimensional
metric. We find that the radii of $AdS_3 \times S^3$ go to
infinity and we recover six dimensional flat space. The remaining
four dimensional manifold becomes \bea
 ds_4^2 &=& {1 \over g}(d\chi + V'_i dx'^i)^2 + g (dy'^2 + dx'^idx'^i) \\
d V' &=& { 1 \over y'} *_3 dz \sim *_3 d g \label{dv_iib} \eea
This is a metric of the Gibbons-Hawking form, which is the general
form for a 4d hyper-Kahler manifold with one translational Killing
vector \cite{GH} (see also \cite{todahkahler,bakas}). In
particular note that $y_0$ scales out of the definition of $V'$ in
(\ref{dv_iib}).  In order to have interesting solutions we should
allow delta functions in the right hand side of (\ref{eforg}) with
appropriate coefficients.

\section{1/2 BPS geometries in M-theory}
\renewcommand{\theequation}{3.\arabic{equation}}
\setcounter{equation}{0}

In this section we perform an analogous analysis for M-theory
solutions associated to 1/2 BPS geometries  in $AdS_{4,7}\times
S^{7,4}$. We first describe how to reduce the problem to the Toda
equation. We then discuss some interesting dualities and Wick
rotations.

Let us consider 1/2 BPS geometries in $AdS_7 \times S^4$. These are
associated to the chiral primaries of the $(0,2)$ theory.
The chiral primaries of the $(0,2)$ theory can also be described in
terms of Young diagrams with at most $N$ rows \cite{ofer}, as in  four
dimensional ${\cal N} =4$ SYM. So in terms of labeling of states
we also have free fermions on a plane. We expect a similar picture if
we study chiral primaries of the 2+1 superconformal field theory
related to $AdS_4 \times S^7$.

These states preserve 16 supercharges which
transform under $SO(3) \times SO(6) \times R$. In this case
we expect that the $R$ translation generator does not leave the
spinor invariant, rather the spinor has non-zero energy
under this generator. This $R$ generator should leave the geometries
invariant.

We look for  supersymmetric solutions of 11D supergravity which have
$SO(6)\times SO(3)$ symmetry
\bea \l{11d_general}
ds_{11}^2&=&e^{2\la}\left(4 d\Omega_5^2+e^{2A}d{\tilde\Omega}_2^2+
ds_4^2\right)\\
G_{(4)}&=&G_{\mu_1\mu_2\mu_3\mu_4}dx^{\mu_1}\wedge
dx^{\mu_2}\wedge d x^{\mu_3}\wedge
dx^{\mu_4}+F_{\mu_1\mu_2}dx^{\mu_1}\wedge dx^{\mu_2} \wedge
d^2{\tilde\Omega} \eea where $d \Omega_5^2$ and $d{\tilde
\Omega}_2^2$  are the metrics on unit radius spheres\footnote{The
factor of $4$ in front of the five--sphere metric was inserted for
later convenience, and it corresponds to setting the parameter $m$
in appendix \ref{AppMderiv} to $m=\half$. } and $\mu_i =0,
\cdots,3$. Using the equations for the field strength, one can
show that \be \l{ione} G_{\mu_1\mu_2\mu_3\mu_4}=I_1
e^{-3\la-2A}\eps_{\mu_1\mu_2\mu_3\mu_4} \ee with constant $I_1$.
In the solutions related to chiral primaries on $AdS \times S$ or
pp-waves the $S^2$ or the $S^5$ can shrink, at least in the
asymptotic regions. These spheres cannot shrink in a
non-singular manner if the flux $I_1$ were non-vanishing. The
reason is that the flux density would diverge at the points where
the spheres shrink. So from now on we set $I_1=0$. In order to
continue constraining the metric we decompose the Killing spinor
in terms of a four dimensional Killing spinor and spinors on $S^2$
and $S^5$. So we have an effective problem in four dimensions with
a four dimensional gauge field $B_\mu$ and two scalars
$A,\lambda$. A closely related problem was analyzed  in
\cite{Gauntlett}, where general supersymmetric M-theory solutions
with $SO(2,4) \times U(1)$ symmetry were considered. Our solutions
preserve more supersymmetries, but after a suitable Wick rotation
they are particular examples of the general situation considered
in \cite{Gauntlett} so we can use some of their methods. After a
rather long analysis, which can be found in appendix
\ref{AppMderiv}, the end result is:
 \bea \l{4d_general} ds_{11}^2&=&-{ 4e^{2\la} ( 1
+ y^2e^{-6\la} ) } (dt+V_idx^i)^2 + \frac{ e^{-4\la} }{1+y^2e^{-6\la}}
[ dy^2 + e^{ D} (dx_1^2 + dx_2^2) ]\nonumber\\
&&+4e^{2\la} d\Omega_5^2+y^2e^{-4\la}d{\tilde\Omega}_2^2
\\
G_{(4)}&=&F\wedge d^2{\tilde\Omega} \nonumber\\
 e^{-6\la} &=& {\partial_y  D \over y(1 -  y \partial_y D) }\nonumber
\\
V_i &=& \half \epsilon_{ij} \partial_j D ~~~~~~~~{\rm or}~~~~~~~
dV = \half *_3 [ d ( \partial_y D) + ( \partial_y D)^2 dy]
\\
F & = & dB_t \wedge (dt + V) + B_t d V + d \hat B
\\
B_t &=& - 4 y^3 e^{-6\la}\nonumber
\\
d \hat B & =& 2  *_3  [ ( y  \partial_y^2 D + y (\partial_y D)^2
 - \partial_y D ) dy + y \partial_i \partial_y D dx^i ] \nonumber
 \\
& =&  2 \tilde *_3 [  y^2 ( \partial_y { 1 \over y} \partial_y e^{D} ) dy
+ y dx^i \partial_i
 \partial_y D ]  \l{bspatial}
\eea
where $i,j=1,2$, and $  *_3$ is the epsilon symbol of the three
dimensional metric $ dy^2 + e^{ D} dx_i^2 $, and $\tilde *_3$ is the flat
space $\epsilon$ symbol.
The function $D$ which determines the solution obeys the equation
\be \l{toda}
(\partial_1^2 + \partial_2^2 )D + \partial_y^2 e^{D} =0
\ee
This is the 3 dimensional continuous version of the Toda equation.
Note that (\ref{toda}) implies that the expression for $d \hat B$ in
(\ref{bspatial}) is closed.
Notice that the form of the ansatz is preserved under $y$ independent
conformal transformations of the $12$ plane if we shift $D$ appropriately.
Namely
\be
x_1 + i x_2 \to f(x_1 + i x_2)  ~,~~~~~~D \to D - \log |\partial f|^2
\ee
Note that the coordinate $y$ is given in terms of the radii of
five--sphere and the two--sphere by $y = R_2 R_5^2/4 = e^{ 2\lambda} e^{\lambda + A}$.
This implies that the 2--sphere or 5--sphere shrinks to zero size at
$y=0$.
Let us first understand what happens when the two--sphere
shrinks to zero and the five--sphere remains with constant radius. From the
condition that $\lambda $ remains constant as $y \to 0$ we find that
$e^{D} $ is an $x$ dependent constant at
$y=0$ and in addition we find that $\partial_y D =0$ at $y=0$.
These conditions ensure that the $y$ coordinate combines with the sphere
coordinates in a non-singular fashion. We now can consider the case
where the five-sphere shrinks. In this case $R_2$ is a constant, so
that $e^{2 \lambda} \sim y $. This happens when
$D \sim  \log y $ as $y\to 0$.
In this case
we see that the geometry is non-singular.  After redefining the coordinate $y= u^2$,
we see that the $y$ and 5-sphere components of the metric become locally the
metric of $R^6$.
In summary, we have the following two possible boundary conditions at
$y=0$
\bea
 \partial_y D &=&0~,~~D = { \rm finite} ~,~~~~~~~~~~~~~~
   S^2 ~{\rm shrinks} \l{stwoshr}
\\
 D &\sim& \log y  ~,~~~~~~~~~~~~~~~~~~~~~~~~~ ~~   S^5 ~{\rm shrinks}
 \l{sfiveshr} \eea

We can also separate the $12$ plane into droplets where we have
one or the other boundary condition above. We can now consider
four cycles obtained by fibering the two--sphere over
 a two-surface $\Sigma_2$ on
the $y,x_1,x_2$ space which ends at $y=0$ in a region where the
$S^2$ shrinks, see figure \ref{cups}. This is a non-singular
four-cycle\footnote{ This four cycle has the topology of a sphere $S^4$
if
  $\Sigma_2$   is
topologically a disk ending at $y=0$.}. Since $B_t V$ is a
globally well defined vector field, we find that the flux of the
four form over this four cycle is given by computing the integral
\be \l{fivebrch} N_5 \sim -\frac{1}{\mbox{vol}_{S^2}}\int_{\Sigma_4}  G_{(4)} =
-\int_{\Sigma_2} d \hat B =  \int_{\cal D}
 dx_1 dx_2 2
(  y^{-1} e^{D})|_{y=0} \ee where ${\cal D}$ is the region in the
$x_1,x_2$ plane with the $S^5$ shrinking boundary condition,
(\ref{sfiveshr}),
 which lies inside the surface $\Sigma_2$.
So the area of this region measures the number of 5-branes in this
region.

We can similarly measure the number of two branes by considering
the flux of electric field. Namely we consider now a seven cycle
which is given by fibering the  five--sphere over a two surface $\Sigma'_2$
which ends on the $y=0$ in a region where the five--sphere shrinks.
Then the electric flux is given by
\bea N_2 &\sim& \frac{1}{\mbox{vol}_{S^5}}\int_{\Sigma_7}
*_{11} G_{(4)} = \int_{\Sigma'_2} [ \Phi dV + g_0^{-1} e^{ 3
\lambda - 2 A} *_3 d B_t ]\nonumber\\
 &=& \int_{\cal D} 2 \tilde *_3 [ y^3 \partial_y^2 ( y^{-1} e^{D}) dy
 + y^2 \partial_i \partial_y D dx^i ]
= \int_{ \cal D}dx_1dx_2 \,  2 e^{  D}|_{y=0} \l{twobrch} \eea
where $\tilde *_3$ is the flat space $\epsilon$ symbol and $\cal
D$ is the region in the $12$ plane where the $S^2$ shrinks which
is inside the original $\Sigma'_2$ surface. This integral counts
the number of two branes. If the five--branes were fermions the
two-branes are holes. The equation (\ref{toda}) implies that the
two form we are integrating in (\ref{twobrch}) is closed.

Notice that in both cases the fluxes are given by the area
measured with a metric constructed form $D$. So we first have to
solve the Toda equation, (\ref{toda}), find $D$, and only then can
we know the number of $M2$ and $M5$ branes associated to the
droplets. Note also that any two droplets which differ by a
conformal transformation seem to give us the same answer. In fact,
if we consider circular droplets of different sizes, then a
conformal transformation would map them all into a circular
droplet of a specific size. The point is that the boundary
conditions (\ref{stwoshr}), (\ref{sfiveshr}) do not fix the
solution uniquely. Given a solution $D(x_i,y)$,  the function
$D(x_i, y \lambda) - 2 \log \lambda $ is also a solution with the
same boundary conditions. We see from (\ref{fivebrch}),
(\ref{twobrch}) that this change rescales the charges. We expect
that this is the only freedom left in determining the solution,
but we did not prove this. In other words, we expect that the
solution is completely determined by specifying the shapes of the
droplets. As opposed to the IIB case, we do not know the
correspondence between the precise shape of the droplets in phase
space and the shape of the droplets in the $y=0$ plane\footnote{
In particular, notice that the densities, given by
(\ref{fivebrch}) and (\ref{twobrch}), are not constant. }.
 But we expect that their
topologies  are the same.

Let us now discuss some examples. The simplest example is the pp-wave solution.
In this case $x_1$ is an isometry direction. The necessary change of variables
is
\bea
y & = & { 1 \over 4} r_5^2 r_2 \nono
\\
x_2& = & { r_5^2 \over 4 } - {r_2^2 \over 2} \nono
\\
e^{D} &=& {r_5^2 \over 4} \l{mthpp} \eea where $r_5$ and $r_2$ are
the radial coordinates in the first six transverse dimensions and
the last three transverse dimensions respectively. In
(\ref{mthpp}) we could in principle find $D$ in terms of $x_2,y$,
this involves solving a cubic equation.  One can check that $D$
defined in this fashion obeys the Toda equation. It is also easy
to see that this expression obeys the appropriate boundary
conditions for $x_2>0$ and $x_2<0$ which represent a half filled
plane, and corresponds to the Dirac sea.

Another example is given by the $AdS_7 \times S^4$
solution\footnote{In equations (\ref{ads7_exp}) and
(\ref{ads4_exp}) we use the polar coordinates in $x_1, x_2$ plane:
$ds_2^2=dx^2+x^2 d\psi^2$.}: \bea e^{D}=\frac{r^2 L^{-6}}{4+
r^2},\qquad x=(1+{ r^2 \over 4} )\cos\theta,\quad 4 y=
L^{-3}r^2\sin\theta \label{ads7_exp} \eea Where $\theta$ is a
usual angle on $S^4$ and $r$ is the radial coordinate in $AdS_7$
and $L$ is the radius of $S^4$. Notice that in this case the
solution asymptotes to $D \sim 0$ at large distances. So we expect
that any solution with $AdS_7 \times S^4$ asymptotics can be
obtained by solving Toda equation (\ref{toda}) with the boundary
conditions (\ref{stwoshr}), (\ref{sfiveshr}) and $D \sim 0$ at
infinity.

We can similarly describe the solution for $AdS_4 \times S^7$:
\bea\label{ads4_exp}
e^{D}=4L^{-6}\sqrt{1+\frac{r^2}{4}}\sin^2\theta,\qquad
x=\left(1+\frac{r^2}{4}\right)^{1/4}\cos\theta,\qquad
2y=L^{-3}r\sin^2\theta \eea Here $L$ is the radius of $AdS_4$.
Notice that in both $AdS \times S$ cases we have circular
droplets.

Unfortunately the Toda equation is not as easy to solve as the Laplace equation,
so it is harder to find new solutions.
There are a few other known solutions. There are two singular solutions that
correspond to extremal limits of charged
 black holes in 7d and 4d gauged supergravity\footnote{i.e. the
 $AdS_7$ black hole in \cite{liuminasian,gubser} and the
 $AdS_4$ black hole in, for example, \cite {duff}.}.
 They obey our equations but not the boundary conditions.
In section \ref{gsugrasection} we will construct a new non-singular solution of
7d gauged supergravity. This solution is
 associated to a droplet of elliptical shape.

In the case of solutions with an extra Killing vector we can reduce the
problem to a Laplace equation using \cite{ward}. Let us consider a translational
Killing vector. In the case that the solution is independent of $x_1$ the
equation (\ref{toda}) reduces to the two dimensional Toda equation
\be \l{twodtoda}
 \partial_2^2 D + \partial_y^2 e^{ D} =0
 \ee
 This equation can be transformed to a Laplace equation by the
 change of variables
 \be \l{trickw}
 e^{D} = \rho^2 ~,~~~~~~~~~~y=\rho \partial_\rho V  ~,~~~~~~~~ x_2=\partial_\eta V
 \ee
 Then the equation (\ref{twodtoda}) becomes the cylindrically symmetric Laplace
 equation in three dimensions
 \be
 {1 \over \rho } \partial_\rho ( \rho \partial_\rho V) + \partial^2_\eta V =0
 \ee
 It would be nice to understand more precisely the boundary conditions in terms of these
 new variables in order
 to find interesting solutions.
 The pp-wave solution (\ref{mthpp}) can be expressed as
 \bea
V=\rho^2\eta-\frac{2}{3}\eta^3 \eea Note that only the region
$\eta>0$ is meaningful. In fact, at $y=0$,  half of the $x_2$ line
is mapped to $\rho=0$ and the other half to $\eta =0$. As we
consider other solutions of the Laplace equation we expect to find
more complicated boundaries. It would be nice to analyze this
further. By solving this equation one can obtain solutions with
pp-wave asymptotics that represent particles with nonzero $-p_-$,
which are translationally invariant along $ x^- $ (this can happen
at the level of classical solutions). One can then compactify
$x^-$ and reduce to type IIA. In this way we obtain non-singular
geometries that are the gravity duals of the BMN matrix model
\cite{bmn}. These solutions were explored in \cite{hai} in the
Polchinski-Strassler approximation. By using the methods of this
paper it is possible, in principle (and probably also in
practice), to obtain non-singular solutions corresponding to
interesting vacua of the BMN matrix model. The Young diagrams
representing different vacua of the BMN matrix model are directly
mapped to strips in $y=0$ plane, just like in the case of M2 brane
theory with mass deformation (see figure \ref{young}). In
particular, our solutions make it clear that it will be possible
to find configurations that correspond to D0 branes that grow into
NS5 (or M5) branes, as discussed in \cite{mfivefromBMN}. Such a
solution would come from boundary conditions on the $y=0$ plane
for the Toda equation as displayed in figure \ref{twovacua}(a).
The solution where the D0 branes grow into D2 branes on $S^2$ is
then related to a boundary condition of the form shown in figure
\ref{twovacua}(b). Note that, despite appearances, the topology of
these two solutions would be the same. In the type IIA language
both solutions would contain a non-contractible $S^3$ and a
non-contractible $S^6$. These spheres are constructed from the
arcs displayed in figure \ref{twovacua}, together with either
$S^2$ or $S^5$.

In this paper we considered plane wave excitations with
$p_+=0$ and $p_- \not =0$. Solutions that correspond to a
plane wave plus particles with $p_-=0$ but $p_+ \not =0$ were discussed in
\cite{columbiaguy}, together with their matrix model interpretation.

It is also possible to use this trick, (\ref{trickw})  when we
consider solutions that are rotationally symmetric in the
$x_1,x_2$ plane. The reason is that the plane and the cylinder can
be mapped into each other by a conformal transformation, and
conformal transformations are a symmetry of the Toda equation
(\ref{toda}). More explicitly, if we write two dimensional metric
as $dr^2+r^2d\phi^2$ and look for solutions which do not depend on
$\phi$, then three dimensional Toda equation reduces to
(\ref{twodtoda}) with following replacement: $x_2\rightarrow \ln
r$, $D\rightarrow D+2\ln r$.

Note that the Toda equation (\ref{toda}) has also appeared in the related
problem of finding four dimensional hyper-Kahler manifolds with a so called
 ``rotational" Killing vector \cite{todahkahler} (see also \cite{bakas}).
In fact the expression for the hyper-Kahler  manifold in terms of
the solution of the Toda equation can arise as a limit of our
expression for the four dimensional part of the metric
(\ref{4d_general}), which will be discussed in subsection
(\ref{n2})\footnote{ This is similar to the limit of the ansatz
and equations for the four dimensional part of the type IIB
geometries (\ref{solmetric}) to the expressions one finds for
hyper-Kahler metrics with ``translational" Killing vectors, as
discussed in subsection (\ref{iibcontinuation}).}. This equation
also arises in the large $N$ limit of the 2d Toda theory based on
the group $SU(N)$ in the $N \to \infty$ limit, the $SU(N)$ Dynkin
diagram becomes a continuous line parameterized by $y$
\cite{todasun}.

\subsection{ $ { \cal N } = 2 $ superconformal field theories from M-theory}
\label{n2}
 We consider now solutions of 11 dimensional supergravity
that contain an $AdS_5$ factor and a six dimensional compact
manifold. These solutions can be interpreted as conformal field
theories in four dimensions. In \cite{Gauntlett} all solutions
with ${\cal N} =1$ superconformal symmetry were characterized. In
this subsection we consider all solutions with ${\cal N} =2$
superconformal symmetry in four dimensions. So we are interested
in solutions preserving 16 supercharges. The superconformal
algebra implies that we have an extra $SU(2) \times U(1)$ bosonic
symmetry. In fact the full superalgebra is a Wick rotated version
of the superalgebra preserved by the M-theory 1/2 BPS geometries
we discussed above.

Our analysis of the supersymmetry conditions was a local analysis, so
after a simple Wick rotation we use the same general solution to
describe the M-theory
geometries dual to ${\cal N} =2 $ superconformal field theories.

In order to find the metric it is convenient to note that after an
analytic continuation of coordinates we have
\bea\label{AnalCont52In} \psi \rightarrow \tau && \alpha
\rightarrow i \rho\nonumber
\\
\cos^2 \alpha d\psi^2 + d \alpha^2 + \sin^2 \alpha d\Omega_3^2 & \rightarrow
& - ( - \cosh^2 \rho d\tau^2 + d \rho^2 + \sinh^2 \rho d \Omega_3^2 )
\\
d\Omega_5^2 & \rightarrow & -ds_{AdS_5}^2 \nonumber \eea
In addition we perform the analytic continuation
\be\label{AnalCont52Out} \lambda  = \tilde \lambda + i { \pi \over
2} \ee
with the
rest of the coordinates remaining the same. If we now take real
$\tilde \lambda$ we get appropriate minus signs
that produce a metric with the correct signature. Note
that $t$ is now a spacelike coordinate. We denote it by $\chi = t$ and take
it to be compact.
 Just to be more explicit, we rewrite the full ansatz
for the geometry\footnote{One can also consider an analytic
continuation to $AdS_2\times S^5$ reduction: $y=iy'$, $x_i=ix'_i$.
The resulting metric is $ds_{11}^2=e^{2\tilde \la}\left( 4
ds_{S^5}^2+y^2e^{-6{\tilde\la}} ds^2_{AdS_2}- ds_4^2\right)$ where
$ds_4^2$ and $e^{-6{\tilde \la}}$ are given by (\ref{m4dconti})
and (\ref{m4dcontLa}). To have a metric with correct signature one
should consider a region where $y^2e^{-6 \tilde \la}>1$. } \bea
ds_{11}^2&=&e^{2\tilde \la}\left( 4
ds_{AdS_5}^2+y^2e^{-6{\tilde\la}} d{\tilde\Omega}_2^2+
ds_4^2\right) \nonumber\\
ds_4^2&=& { 4  ( 1 - y^2e^{-6 \tilde \la} ) } (d\chi+V_idx^i)^2 +
\frac{ e^{-6\tilde\la} }{1 -   y^2e^{-6\tilde \la}} [ dy^2 + e^{D}
(dx_1^2 + dx_2^2) ]   \label{m4dconti}
\\
G_{(4)}&=&F\wedge d^2{\tilde\Omega}
\nonumber\\
\label{m4dcontLa}
 e^{-6\tilde \la}&=& - {\partial_y D \over y(1 -  y \partial_y D) }
\\
V_i &=& \half \epsilon_{ij} \partial_j D ~~~~~~~~{\rm or}~~~~~~~
dV = \half *_3 [ d ( \partial_y D) + ( \partial_y D)^2 dy]
\\
F & = & dB_{\chi} \wedge (d\chi + V) + B_{\chi} d V + d \hat
B,\qquad B_{\chi}= 4 y^3 e^{-6\tilde \la}\nonumber
\\
\label{m4dcontEnd}
d \hat B & =& 2  *_3  [ ( y  \partial_y^2 D + y (\partial_y D)^2
 - \partial_y D ) dy + y \partial_i \partial_y D dx^i ]
 \\
& =&  2 \tilde *_3 [  y^2 ( \partial_y { 1 \over y} \partial_y e^{D} ) dy
+ y dx^i \partial_i
 \partial_y D ]\nonumber
\eea
where $D$ obeys the same equation (\ref{toda}). Of course we cannot take
the same solutions we had before since, for example, they
would lead to negative values of $e^{ 2 \tilde \la}$. We now need to solve
these equations with other boundary conditions. Before we discuss the
boundary conditions let us look at particular
solutions which were  found by different methods in \cite{maldanunez}.
The solutions in \cite{maldanunez} correspond to the the following solution
of (\ref{toda}):
\be \l{solmn}
 e^{D} = { 1 \over x_2^2} ( { 1 \over 4} - y^2 )
 \ee
Now the metric in the two dimensional space parameterized by $x_1
, x_2$ becomes the metric of two dimensional hyperbolic space. As
in \cite{maldanunez} we can perform a quotient to produce a
Riemann surface of genus $g>1$.  We note that, at $y=1/2$, $e^{D}$
becomes zero, this means that the circle parameterized by $\chi$
is shrinking. One can check in general that if $e^{D}$ becomes
zero linearly as $ e^{D} = {\rm const}(y_c-y)$, with $y_c \not
=0$, then the circle $\chi$ shrinks in a smooth manner, as long as
$\chi \sim \chi + 2 \pi $. Note that this periodicity is
reasonable given that the $\chi$ dependence of the Killing spinor
is $e^{ i \half \chi}$ (see appendix \ref{AppMderiv}). So we see
that the spinor behaves in a smooth way too. Another new feature
of these solutions is that the $\chi$ circle is non-trivially
fibered over the Riemann surface. In fact we can compute the flux
of $dV$ and find that it is equal to \be \l{fluxv} \int_{\Sigma_2
= H_2/\Gamma} d V = g ( 4 \pi ) \ee where $g$ is the genus of the
Riemann surface obtained as a quotient of hyperbolic space, $H_2$,
by the discrete group $\Gamma$.
 Since $V$ is a gauge field for the KK reduction on the circle
parameterized by $\chi$, and since the spinor carries $1/2$ a unit
of charge\footnote{ All bosonic KK modes carry integer units of
charge.}, we see that the flux (\ref{fluxv}) is correctly
quantized. Note that the solutions associated to 1/2 BPS chiral
primaries discussed above did not have any topologically
non-trivial closed two manifold on which we could integrate $dV$
in order to find a non-trivial flux. This is good since in that
case $\chi=t$ is a non-compact time-like direction.

It would be nice to produce other non-trivial solutions of these equations.
It might be  possible to give a complete classifications of
all the solutions.  These geometries could arise as special regions of a
warped compactification to four dimensions, so it would be nice to understand
them better. For example, we would like to know
what the moduli space of deformations is.
For the solution in  (\ref{solmn}) we see that the boundary conditions on $D$
are that
$\partial_y D =0$ at $y=0$, which is telling us that the $S^2$ is
shrinking at $y=0$, and $ D \sim  \log( y_c - y)$ at $y = y_c$.

In order to show that equations (\ref{m4dconti})--(\ref{m4dcontEnd}) give the most
general solution we need to
show a couple of things that we assumed when we derived the previous solutions.
The first is that the $U(1)$ circle, parameterized by $\chi$,
 is trivially fibered over the $S^2$. This was natural in the Lorentzian
 context since the $\chi$ direction is timelike.
 The algebra implies that the $U(1)$ charge of the spinor is nonzero.
If the $\chi$ circle were non-trivially fibered on $S^2$, then the effective
spin of the spinor on $S^2$ would be shifted and the solutions of the Killing
spinor equations would not transform in the appropriate representation of
$SO(3)$. See appendix \ref{AppMderiv} for more details.
Another assumption we made was that $I_1$ defined in (\ref{ione}) vanishes.
If $I_1$ is non-zero, then neither  $S^2$ nor $S^5$ can shrink.
 We show in appendix \ref{AppMderiv}
that if we start with a solution of our equations
and we try to add $I_1$ infinitesimally, then
 we break supersymmetry. We do not know if there are any  supersymmetric
 solutions with
 non-zero $I_1$ \footnote{It is easy to construct
 non-supersymmetric solutions
 with the required isometries and non-vanishing $I_1$, e.g. consider
 $AdS_5 \times H_2/{\Gamma} \times S^2 \times S^2$ and
 choose appropriate fluxes.}.
We have set  $I_1=0$ above.

Similar to the limit discussed in subsection
\ref{iibcontinuation}, we find an interesting limit under which
the $AdS_5 \times S^2$ factor becomes a 7-dimensional flat space
and the remaining 4-dimensional manifold turns into a hyper-Kahler
manifold with one rotational Killing vector \cite{todahkahler}. In this
case we focus on a region of the solution near
$y_0$ with $y_0 < 0$ and very large. We look for a solution of the Toda
equation of the following form \be
 D(x_1,x_2, y) = \tilde D(x_1,x_2, C y) - 2 \log C~, ~~~~~~~ \tilde y = C y
 \ee
where $C$ is a constant. We find that $\tilde D$ obeys the Toda
equation \be
 (\partial_1^2+\partial_2^2) {\tilde D} + \partial_{\tilde y}^2 e^{\tilde D} =0
 \ee
We now assume that $y\partial_y D$ is very large, so we have
 \be
 y \partial_y D \sim  y_0 \partial_{y} D \to \infty ~,~~~~~~~e^{ - 6 \tilde \lambda}
  \sim { 1 \over y_0^2} ( 1 + { 1 \over y_0 \partial_y D} )
 \ee
We then insert this into the ansatz (\ref{m4dconti}) and we find
that the $S^2$ and the $AdS_5$ become flat space after a rescaling
of the coordinates. The remaining four dimensional part of the
metric becomes
\bea
  e^{2 \tilde \lambda}ds^2_4 &=&  { e^{2 \tilde \lambda} \over C |y_0|} \left[
{ 4 \over \partial_{\tilde y} \tilde D} ( d \chi + V_i dx^i)^2 +
\partial_{\tilde y} D ( d\tilde y^2 + e^{ \tilde D} dx^{i}dx^{i} ) \right] \\
V_i &=& { 1 \over 2} \epsilon_{ij} \partial_j \tilde D \eea So we
see that we recover the ansatz of the general 4d hyper-Kahler
manifold with a rotational Killing vector in \cite{todahkahler}
(see also \cite{bakas}) if we take $C = |y_0|^{-1/3}$.

It is also interesting to note that the solutions in \cite{Gauntlett} can
be analytically continued in a similar fashion to give solutions corresponding
to 1/4 BPS states in $AdS_7$ which have two non-zero angular momenta $J_{12}$ and
$J_{34}$ in $SO(5)$. For example, let us look at the metric for a general $AdS_5$ reduction
which was derived in \cite{Gauntlett} (see equations (2.1), (2.51) and (2.55) in that paper):
\bea
&&ds_{11}^2=e^{2\la}\left[ds^2_{AdS_5}+e^{-6\la}\left(g_{ij}dx^i dx^j+\frac{dy^2}{\cos^2\zeta}
\right)+\frac{\cos^2\zeta}{9m^2}(d\psi+\rho)^2\right]\\
&&\cos^2\zeta=1-4m^2y^2e^{-6\la}\nonumber
\eea
We now can perform an analytic continuation analogous to (\ref{AnalCont52In}),
(\ref{AnalCont52Out}), which amounts to replacing $ds_{AdS_5}^2$ by $-ds_{S^5}^2$, $\la$ by
${\tilde\la}+\frac{i\pi}{2}$ and $\zeta$ by $i{\tilde\zeta}$. Then we end up with the metric
which has $SO(6)$ symmetry:
\bea
&&d{\tilde s}_{11}^2=e^{2{\tilde\la}}
\left[ds^2_{S^5}+e^{-6{\tilde\la}}\left(g_{ij}dx^i dx^j+\frac{dy^2}{\cosh^2{\tilde\zeta}}
\right)-\frac{\cosh^2{\tilde\zeta}}{9m^2}(d\psi+\rho)^2\right]\\
&&\cosh^2{\tilde\zeta}=1+4m^2y^2e^{-6\tilde\la}\nonumber
\eea

\subsection{ Gauged supergravity solution}
\label{gsugrasection}

In this section we describe the solution to the Toda equation
that corresponds to an elliptical droplet.
This solution can be obtained by using gauged supergravity.
We consider the truncation of $AdS_7 \times S^4$ to the modes
contained in 7-dimensional gauged supergravity \cite{gaugedsugra}.
 We can find a solution
of 7-dimensional gauged supergravity that corresponds to   a
chiral primary. In this seven dimensional gauged supergravity solution there is
only one field that is $SO(3)$ invariant and charged under the
$U(1)$ generator that we considered above. The $U(1)$ symmetry
has an associated gauged field in gauged supergravity. So we will
be interested in a solution that preserves $SO(3)$ symmetry and
carries charge under this $U(1)$ gauge field. The only fields that
are excited are the charged scalar, a neutral scalar and the gauge
field. In addition we demand that the solution preserves an
$SO(6)$ symmetry in the $AdS_7$ directions. From the 7-dimensional
point of view this is a BPS gauged Q-ball \cite{Coleman}. The
final solution is completely smooth and has no horizons. If one
sets the charged scalar to zero it is possible to find a singular
solution which is the extremal limit of charged black hole
solutions of 7-d gauged supergravity \cite{liuminasian, gubser}. Once we have found the seven dimensional solution,
we can uplift it to 11 dimensions using \cite{nastase}.

The bosonic fields of 7--dimensional gauged supergravity \cite{gaugedsugra}
contain the graviton,  $SO(5)$ gauge fields $A_\mu$ and 14 scalars which
form a coset $SL(5,R)/SO(5)$. We write a Lagrangian and supersymmetry
transformations for this theory in the Appendix \ref{AppGauSUGRA}.
After we impose $SO(3)$
symmetry in the internal space, then the gauge field and coset have the
form\footnote{There should be similar ansatz to (\ref{SO3Ans}) in 4d and 5d gauged sugergravity, where the neutral scalar, the $U(1)$ gauge field and the charged scalar under $U(1)$ gauge field are excited, and the spherical symmetry is preserved in $AdS$ directions. They are particular examples of the general smooth 1/2 BPS geometries.}
\bea\label{SO3Ans}
&&{V_I}^i=\left[\begin{array}{cc}
e^{-3\chi}g&{\bf 0}_{2\times 3}\\
{\bf 0}_{3\times 2}&e^{2\chi}{\bf 1}_{3\times 3}\\
\end{array}
\right],\qquad
{A_{\mu I}}^J=\left[\begin{array}{ccc}
iA_\mu\sigma_2&{\bf 0}_{2\times 3}\\
{\bf 0}_{3\times 2}&{\bf 0}_{3\times 3}\\
\end{array}
\right].
\eea
where $g$ is an element of $SL(2,R)/U(1)$ coset, which corresponds to the
charged scalar field we described above.  We  parameterize $g$
in terms of two functions $\rho$ and $\theta$
\be
g=\exp(i\theta\sigma_2)\exp(-\rho\sigma_3)
\ee
Under $U(1)$ gauge transformations, $\theta$ transforms as
$\theta \to \theta + \epsilon$, so we can work in the gauge where
$\theta =0$.

Taking a metric with $U(1)\times SO(6)$ symmetry and solving the Killing
spinor equations we can express everything in terms of a single function
$H(x)$ \footnote{In gauged supergravity the natural time coordinate is
${\tilde t}$ which is related to $t$ as ${\tilde t}=\frac{t}{m}$.}
\bea
&&ds_7^2=-fH^{-4/5}d{\tilde t}^2+H^{1/5}\frac{dr^2}{f}+
H^{1/5}r^2 d\Omega_5^2\nonumber\\
&&A_{\tilde t}=
\frac{1}{2}H^{-1},\quad e^{\chi}=H^{-\frac{1}{10}},\quad \nonumber
\\
&& f=1+m^2 r^2 H,\quad \cosh 2\rho=\d_x(x H),\quad x \equiv 4m^4
r^4  \l{gsbulk} \eea The function $F(x) = x H(x)$ obeys a
nonlinear differential equation \footnote{ Notice that the black
hole constructed in \cite{liuminasian} corresponds to a particular
solution of this equation $H_{BH}=1+\frac{Q}{x}$ but this solution
is singular at $x=0$, which doesn't obey our boundary condition.}
\bea\label{TheEquation} (2\sqrt{x}+F)F''=(1-(F')^2) \eea
Non-singular solutions obey the boundary condition \be
 F(x=0) =0 ~,~~~~~~~~~~~{\rm or}~~~~~ H(x=0) = C\ge 1  \l{bco}
\ee
 It is clear
that (\ref{TheEquation}) admits a family of solutions parametrized
by the constant $C$ in (\ref{bco}), which in turn determines the
charge or mass of the solution (see appendix \ref{AppCharge}). This
equation is further discussed in appendix \ref{AppGauSUGRA}. It is
possible to show that in the limit of very large charge the
solutions go over to Coulomb branch solutions, similar to
\cite{larsen}, but with branes distributed over a thin ellipse.
Notice that any solution of gauged supergravity can be lifted to a
solution of eleven dimensional supergravity using the general
formalism of \cite{nastase}, and for our solution the resulting
metric is \bea\label{GaugSGRsln11}
ds_{11}^2&=&\Delta^{1/3}ds_7^2+\frac{1}{4m^2}\Delta^{-2/3}\left\{
e^{4\chi}(\cos^2\theta d\theta^2+\sin^2\theta d\Omega_2^2)\right.
\nonumber\\
&+&
e^{-6\chi-2\rho}\left[d(\cos\theta\cos\phi)+
2mH^{-1}\cos\theta\sin\phi d{\tilde t}\right]^2\nonumber\\
&+&\left. e^{-6\chi+2\rho}\left[d(\cos\theta\sin\phi)-
2mH^{-1}\cos\theta\cos\phi d{\tilde t}\right]^2 \right\} \eea
where \bea
\Delta=\cos^2\theta(e^{6\chi+2\rho}\cos^2\phi+e^{6\chi-2\rho}\sin^2\phi)+
e^{-4\chi}\sin^2\theta \eea After some work (see appendix
\ref{todagauged}) it is possible to rewrite this in the form of
our ansatz (\ref{4d_general}) by making the changes of variables
\bea
 x_1 + i x_2 & = & ( e^{i \phi} \cosh \rho - e^{- i \phi} \sinh \rho )
 { \cos \theta \over \sqrt{ \sinh 2 \rho} }
 \\
 y &=& m^2 r^2 \sin \theta
 \\
 e^D &=& { m^2 r^2 f  \sinh 2 \rho }
\eea
where $\rho$ and $f$ are defined in (\ref{gsbulk}). It is possible to see from
this expression that at $y=0$ the region with boundary condition (\ref{sfiveshr}),
which corresponds to shrinking $S^5$, is confined to an ellipse in the $12$ plane.
This ellipse becomes more squeezed as the constant $C$ in (\ref{bco}) becomes
larger (see appendix \ref{todagauged}).

\section{ Conclusions and discussion}

In this paper we have considered smooth geometries that correspond
to half BPS states in the field theory.  The most convenient field
theory description of the half BPS states is in terms of free
fermions. These fermions form an incompressible fluid in phase
space. For each configuration of droplets of this fluid we have
assigned a geometry in the dual gravity theory. The phase space
and the configuration of droplets are mapped very nicely into a two
dimensional plane in the gravity solution. This is the plane at
$y=0$, spanned by $x_1,x_2$. This plane contains droplets that are
defined as the regions where either one or the other of the two spheres
is shrinking in a smooth fashion. In the type IIB case we have two
3-spheres and one or the other shrinks at $y=0$, and in the
M-theory case we have a 2-sphere and a 5-sphere and one or the
other shrinks at $y=0$. The full geometry is determined by one function
which obeys a partial differential equation in the three dimensions
$y,~ x_1,~ x_2$. Regularity of the solution requires either of two kinds
of  specific boundary
conditions on the  $y=0$ plane. These two types of boundary conditions
determine the different droplets in the $x_1,~x_2$ plane.
 From the gravity point of view, the
quantization of the area of the droplets comes from the
quantization of the fluxes.
If we only knew the construction of
the gravity solutions we would need to know how to quantize them.
The free fermion description is telling us how to do it. It would
be nice to see how to derive this prescription directly from the
gravity point of view. In particular, it would be nice to understand
in what sense the $x_1,x_2$ plane becomes non-commutative.

For M-theory solutions the correspondence between phase space and
the $y=0$ plane is not as straightforward as in the IIB case. In particular, the
flux density is not constant. Nevertheless, we expect that, just as in the IIB case,
the full solutions
are determined by specifying the shapes of the boundaries between the two regions.

It is possible to define a path integral such that only
these geometries contribute. This can be done by defining a suitable index
and taking appropriate limits, \cite{mms}.
One interesting aspect of the quantization of these geometries
is that
topologically non-trivial fluctuations, which are highly curved
and have small energies (quantum foam),
such as the ones arising from small
droplets close to the Fermi sea, are already included in the path
integral when we consider ordinary low energy gravity and do not
need to be considered separately. The fact that the geometric and
field theory descriptions match so nicely can be viewed as a test
of $AdS/CFT$ in the half BPS sector. This system displays very
clearly the geometric transitions \cite{ks,transitions} that arise
when one adds branes to a system. Adding  a droplet of fermions to
an empty region corresponds to  adding branes that are wrapped on
the sphere that originally did not shrink  at $y=0$. Once we have
a new  droplet this sphere will now shrink in the interior of   the
droplet. In the process we have also created a new
non-contractible $S^5$ that consists of the sphere that was
shrinking before and the two dimensional manifolds (or ``cups")
depicted in figure \ref{cups}.

The well known $c=1$ string theories are also dual to free
fermions. But in the $c=1$ case the string theory side does not
have a simple geometric description since there are fields with
string scale gradients.  In our case we can describe the
corresponding geometries in a very simple way. In the $c=1$ matrix
model there is a special operator, called the ``macroscopic loop"
operator or FZZT brane \cite{fzzt}. It would be nice to find whether there is
an analogous brane in our case. It seems that this system is going
to be very useful for understanding some aspects of quantum
gravity. Our correspondence to free fermions is reminiscent to the
free fermions that appear in the topological string context
\cite{aganagic}, which also arises in the study of a BPS sector of
the full string theory.

One natural question is if we have a similar characterization of
chiral primaries in $AdS_3 \times S^3$. Smooth geometries
corresponding to chiral primaries in   $AdS_3 \times S^3 \times
K3$ or  $AdS_3 \times S^3\times T^4$ were constructed in
\cite{lmm}. These were constructed by specifying the profiles of
several scalars on a circle. Of course, we could  get fermions by
fermionizing these scalars, and it would be nice to see if these
fermions are useful in describing the geometries.

It would be nice to find a relationship between the free fermions
we discussed here in the M-theory context and the free fermions
that are closely connected to the Toda equation \cite{todafermions}.
 In particular,
one would like to solve the Toda equation in some cases. In the
case that we have an extra Killing vector this seems to be
possible. All that is needed is a better understanding of the
boundary conditions for the new variables that linearize the Toda
equation (\ref{trickw}). A particularly interesting set of
solutions would be those corresponding to translation invariant
excitations of the M-theory pp-wave. Upon compactification, these
would lead to geometries that are dual to the different vacua of
the BMN matrix model.

It seems natural to think that smooth solutions corresponding to
the ${\cal N}=1^*$ theory, studied by  Polchinski and Strassler
\cite{jpms},  will be similar in spirit to those described in this
paper. In particular, the solutions associated to the mass
deformed M2 brane theory have taught us that there is no
fundamental topological distinction between the solutions where
the M2 grows into a dielectric M5 brane wrapped on an $S^3$ in the
first four coordinates and a dielectric brane wrapped on  an $S^3$
in the second four coordinates.

It would be nice to see if there are simple generalizations of these
geometries to those preserving 1/4 or  1/8 supersymmetries which can
be viewed as general chiral primaries of an ${\cal N}=1$ subalgebra
of ${\cal N}=4$. Geometries with 1/16 supersymmetries are expected to
be even richer since they include supersymmetric black holes
\cite{gut}.

It would be nice to study further the $AdS_5$ compactifications of
M-theory we discussed above. In particular, it would be nice to see
whether there are any solutions with non-zero $I_1$. It seems possible
that one could classify completely all such geometries.

Our geometries have $\Delta - J =0$. It would be nice to find
the energy gap to the next non-BPS
excitation. In other words, the simplest non-BPS excitation will
have $\Delta -J = E_{gap}$.  In
the weakly coupled field theory, we have that $E_{gap} = 1$.
For excitations  around a weakly curved
 $AdS_5 \times S^5$ ground state we also have $E_{gap} =1$.
It might be that for some chiral primaries the
gap is smaller, as was found in \cite{lmm} for the $AdS_3 \times S^3$ case.

\section*{Acknowledgments}

We would like to thank S. Cherkis for discussions.
This work was supported in part by NSF grant PHY-0070928  and DOE grant
DE-FG02-90ER40542.
% HL also thanks Princeton University for Assistantship in Instruction.
JM would like to thank the hospitality of the IHES and CERN where
some of this work was done.

\appendix

\section{Type IIB solutions}
\label{AppIIB}
\renewcommand{\theequation}{A.\arabic{equation}}
\setcounter{equation}{0}

In this appendix we derive the solution (\ref{solmetric})--(\ref{4dgf2})
that we described
in the main text\footnote{
Our conventions for normalizing differential forms are as follows.
We will take
$
A^{(k)}=\frac{1}{k!}\sum A_{i_1\dots i_k} dx^{i_1}\wedge\dots\wedge dx^{i_k}
$.
For example, $
A^{(1)}=A_i dx^i,
$,
$F=dA^{(1)}=\d_j A_i dx^j\wedge dx^i=
\frac{1}{2}F_{ij}dx^i\wedge dx^j $.
The dual $B=^\star F$ is defined by $
F_{ij}=\eps_{ijk} B_k$, $B_k=\frac{1}{2}\eps_{ijk}F_{ij} $.}.

We start with the following $SO(4) \times SO(4)$ symmetric ansatz in type IIB
supergravity
\bea
ds^2&=&g_{\mu\nu}dx^\mu dx^\nu +e^{H+G}d\Omega_3^2+e^{H-G}d{\tilde\Omega}_3^2\\
F_{(5)}&=&F_{\mu\nu}dx^\mu\wedge dx^\nu\wedge d\Omega+
{\tilde F}_{\mu\nu}dx^\mu\wedge dx^\nu\wedge d{\tilde\Omega}
\label{tendmetric}
\eea
where $d\Omega_3^2, ~ d\tilde \Omega_3^2$
denote the metric on 3--spheres with unit radius.
The self duality condition for the ten dimensional gauge field implies that we
have only one independent gauge field in four dimensions
\be
F = e^{3 G} *_4 \tilde F , ~~~~~~\tilde F = - e^{-3 G} *_4 F~,~~~~~~~~~
F = d B ~,~~~~~~~~\tilde F = d \tilde B
\l{fequationapp}
\ee
Then we get only one nontrivial equation for the Killing spinor
\bea \label{kspin}
\nabla_M\eta+\frac{i}{480}\Gamma^{M_1M_2M_3M_4M_5} F^{(5)}_{M_1M_2M_3M_4M_5}\Gamma_M
\eta=0
\eea
We choose a basis of gamma matrices
\bea
\Gamma_\mu=\gamma_\mu\otimes 1\otimes 1\otimes 1,\quad
\Gamma_a=\gamma_5\otimes \sigma_a\otimes 1\otimes \hat \sigma_1,\quad
\Gamma_{\tilde a}=\gamma_5\otimes 1\otimes \tilde \sigma_a\otimes \hat \sigma_2,
\eea
where $\sigma_a , ~\tilde \sigma_a , ~ \hat \sigma_a $ are ordinary Pauli matrices.
In this basis
\bea
\Gamma_{11}=\Gamma_0\dots \Gamma_3\prod\Gamma_a\prod\Gamma_{\tilde a}
= \gamma^5 {\hat\sigma}^3~,~~~~~~~ \gamma^5 = i \Gamma_0 \Gamma_1 \Gamma_2 \Gamma_3
\eea
The spinor obeys the chirality condition
\bea  \l{chicond}
\Gamma_{11}\eta=\gamma^5{\hat\sigma}_3\eta=\eta.
\eea

We begin with the spinor equation on the sphere.
Suppose we have a spinor on a unit radius 3-sphere.
We consider spinors obeying the equation
\be
\nabla_c \chi = a { i \over 2}   \gamma_c \chi  \label{spheresp} ~,~~~~~a = \pm 1
\ee
The solutions of this equation transform in the spinor representation under the
$SO(4)$ isometries of the sphere. The chirality of the $SO(4)$ spinor representation
is correlated with the sign of $a$ \footnote{
Chirality $+ $ or $-$  under $SO(4)$ means that the spinor transforms in
the $(1,2)$ or $(2,1)$ under $SU(2) \times SU(2) = SO(4)$. }.
Now let us consider the full metric (\ref{tendmetric}),
the warp factors lead to the following
spin connections in the sphere directions
\bea \label{spincon}
\nabla_a={\nabla'}_a-\frac{1}{4}{\Gamma^\mu}_a\d_\mu (H+G),\qquad
\nabla_{\tilde a}={\nabla'}_{\tilde a}-
\frac{1}{4}{\Gamma^\mu}_{\tilde a}\d_\mu (H-G),
\eea
where $\nabla'$ contains the spin connection on a unit sphere.
We now decompose the ten dimensional spinor as
\be
 \eta = \epsilon_{a,b} \otimes \chi_a  \otimes \tilde \chi_b
\ee
where $\chi_{a} , \tilde \chi_b$ obey equation (\ref{spheresp}) with overall
signs $a, b = \pm 1$. The spinor $\epsilon_{ab}$ is acted on by the four
dimensional $\gamma$ matrices and the matrices $\hat \sigma $.
For simplicity we now drop the indices $a,b$ on the spinor $\epsilon$.
We are interested in geometries that are asymptotically $AdS_5 \times S^5$ or
 plane waves which preserve
a half of the original supersymmetries.
Since the original supersymmetries have correlated chiralities under
$SO(2,4)$ and $SO(6)$, and we are looking at supercharges with
$H'= \Delta - J=0$, we   expect them to
have chiralities $++$ or $--$ under the
$SO(4)\times SO(4)$ generators.

The expression  involving the five--form becomes
\bea
 M & \equiv&  { i \over 480} \Gamma^{M_1M_2M_3M_4M_5} F^{(5)}_{M_1M_2M_3M_4M_5}  \\
 M& =& { i \over 48} (e^{-{3 \over 2} (H+G)}\Gamma^{\mu\nu} F_{\mu\nu}\eps_{abc}\Gamma^{abc}-
e^{- {3 \over 2} (H-G)}
\Gamma^{\mu\nu} {\tilde F}_{\mu\nu}\eps_{{\tilde a}{\tilde b}{\tilde c}}
\Gamma^{{\tilde a}{\tilde b}{\tilde c}})
\\
M &=& - { 1 \over 4}  e^{-{ 3\over 2} (H+G) } \gamma^{\mu \nu} F_{\mu \nu} \gamma^5 {\hat\sigma}^1
\eea
where we used  (\ref{fequationapp}) and
the fact that $M$ acts on spinors with negative ten-dimensional
chirality.
Equation (\ref{kspin}) then  becomes the system
\bea \label{simplif1}
 &&(iae^{- \half (H+G)}\gamma_5{\hat\sigma}_1+ \half  \gamma^\mu\d_\mu (H+G))\eps+
 2 M \eps=0\\
&&(ibe^{- \half(H-G)}\gamma_5{\hat\sigma}_2+ \half \gamma^\mu\d_\mu (H-G))\eps-
 2 M \eps=0   \l{simplif2} \\
&&\nabla_\mu\eps+
M\gamma_\mu\eps=0 \l{simplif3}
\eea
 The system
(\ref{simplif1})--(\ref{simplif3})  describes our reduction.
They describe equations that are
effectively four dimensional. The hatted sigma matrices could be removed by
using (\ref{chicond}) and redefining the four dimensional gamma matrices by
$\gamma^\mu \to \hat \sigma^1 \gamma^\mu$.
We chose to leave them in this form
to preserve more explicitly the duality symmetry between
the two three--spheres. The four dimensional system involves the four dimensional
metric, one gauge field and two scalar fields.

\subsection{ Spinor bilinears}

It is now convenient to construct some spinor bilinears.
An interesting set of  spinor
bilinears is
\bea \l{defbilinears}
&&K_\mu= -\bar \eps \gamma_\mu \eps,\qquad
L_\mu= \bar \eps \gamma^5\gamma_\mu \eps ~,~~~~~~~~~\bar \epsilon =
\epsilon^\dagger \Gamma^0
\nonumber\\
&&f_1=i \bar \eps {\hat\sigma}_1\eps,\quad
f_2=i \bar \eps  {\hat\sigma}_2\eps,\qquad
Y_{\mu\nu}=\bar \eps  \gamma_{\mu\nu}{\hat\sigma}_1\eps
\eea
where $\bar \epsilon = \epsilon^\dagger \gamma^0$.
Using (\ref{simplif3}) one can show that
\bea
\label{33Eqnf1}
\nabla_\mu f_1&=&- e^{-\frac{3}{2}(H-G)} \tilde F_{\mu\nu} K^\nu =
+\frac{e^{-\frac{3}{2}(H+G)}}{2}\eps_{\mu\nu\la\rho}F^{\la\rho} K^\nu\\
\label{33Eqnf2}
\nabla_\mu f_2&=& - e^{-\frac{3}{2}(H+G)}F_{\mu\nu} K^\nu\\
\label{33Eqnf3}
\nabla_\nu K_\mu&=&- e^{-\frac{3}{2}(H+G)}\left[F_{\mu\nu}f_2-
\frac{1}{2}\eps_{\mu\nu\la\rho}F^{\la\rho}f_1\right]\nonumber\\
&=& -e^{-\frac{3}{2}(H+G)}F_{\mu\nu} f_2 -
e^{-\frac{3}{2}(H-G)} \tilde F_{\mu \nu} f_1  \\
\label{33Eqnf4}
%\nabla_\nu L_\mu&=& e^{-\frac{3}{2}(H+G)} \left[
%- \half g_{\mu \nu} F_{\lambda \rho} Y^{\lambda \rho}
% + {F_\mu }^{\rho }Y_{\rho\nu } + {F_\nu }^{\rho }Y_{\rho\mu } \right] CHECK\nonumber\\
\nabla_\nu L_\mu&=& e^{-\frac{3}{2}(H+G)} \left[
- \half g_{\mu \nu} F_{\lambda \rho} Y^{\lambda \rho}
- {F_\mu }^{\rho }Y_{\rho\nu } - {F_\nu }^{\rho }Y_{\rho\mu } \right]
% -e^{-3A}{F_\mu }^{\rho }Y_{\rho\nu}
% \\
% \nabla_\sigma Y_{\mu \nu} &=& e^{-3 A}[ - F_{\mu\nu} L_\sigma -
% \half F_{\mu \sigma} L_\nu + \half F_{ \nu \sigma} L_\mu] \l{yequat}
\eea
 Another interesting set of spinor bilinears involves taking
the the spinor and its transpose. We are going to be particularly
interested in a one-form which obeys a useful equation \bea
\l{oneform} \omega_\mu  &=&\epsilon^t \Gamma^2 \gamma_\mu \epsilon
~,
\\
d \omega &=& 0 \l{omegaequation}
\eea
where in our conventions\footnote{
Our conventions are such that
 $\Gamma_0, ~\Gamma_3, ~\Gamma_1$ are symmetric and $\Gamma_2$ is antisymmetric.}
$\Gamma^2 \gamma_\mu^t \Gamma^2 = - \gamma_\mu$,  and
the last equation says that the exterior derivative vanishes.

By Fierz rearrangement identities we can show\footnote{We found a useful summary
of these identities in \cite{fierz}.}
\be \l{normkill}
 K\cdot L =0 ~,~~~~~~L^2 = - K^2 = f_1^2 + f_2^2
\ee
We now use all these facts to constrain the metric and the gauge fields.

\subsection{Implications of the equations for the bilinears.}

First we observe that $K^\mu$ is a Killing vector and $L_\mu dx^\mu$ is
a (locally) exact form.
We begin by  choosing a coordinate $y$ through
\bea
 \gamma dy=L_\mu dx^\mu ~,~~~~~~~~\gamma = \pm 1
\eea
We will later determine the sign of $\gamma$.
We choose the other three coordinates in the subspace orthogonal to $y$
\bea
ds^2=h^2 dy^2+{\hat g}_{\alpha\beta}dx^\alpha dx^\beta
\eea
Let us now look at the vector $K^\mu$. Using the relation
\bea
0=K^\mu L_\mu =  K^y L_y= \gamma K^y
\eea
 we find that $K^\alpha$
is a vector in three dimensional space spanned by $x^\alpha$.
Choosing one of the coordinates along $K^\alpha$ (we will call it $t$),
we find the  metric
\bea
ds^2=-h^{-2}(dt+V_i dx^i)^2+h^2( dy^2+\tilde h_{ij}dx^i dx^j)
\eea
were $i,j$ take values $1,2$. We have used the equation $K^2 = - L^2$
to link the $g_{tt}$ and the $g^{yy}$ coefficients of the metric.
We also pulled out a factor of $h^2$ out of the remaining two dimensions
for later convenience.

Now we look at  equation (\ref{33Eqnf2}). Since $K^\mu$ has only one
component $K^t=1$ (we can always choose this normalization of $t$), and
$B_i$ is independent of $t$, that equation becomes
\bea
\d_\mu f_2=- e^{-{3 \over 2} (H+G)}\d_\mu B_t, \quad \mbox{i.e.}\quad
df_2=- e^{- { 3 \over 2}(H+G)}d B_t
\eea
We now compute
\bea  \l{fcomm}
\d_\mu B_t= F_{\mu\nu} K^\nu= - F_{\mu\nu} \bar \eps \gamma^\nu \eps=
- \frac{1}{4} \bar \eps [\gamma_\mu,{\not F}] \eps
\eea
Now we recall the equation coming from the sphere (\ref{simplif1}) and its adjoint
\bea
&&\frac{1}{2}e^{-{ 3 \over 2}(H+G)}{\not F}\eps=(iae^{- \half (H+G)}+
\half \gamma_5{\not\d}(H+G)
{\hat\sigma}_1)\eps,
\nonumber\\
&&\frac{1}{2}e^{-{3 \over 2} (H+G)}\bar  \eps {\not F}=
\bar \eps (iae^{-\half (H+G)}+ \half \gamma_5{\not\d} (H+G){\hat\sigma}_1)\nonumber
\eea
Using this in (\ref{fcomm}) we obtain
\bea
\d_\mu B_t= e^{{ 3 \over 2}( H+G)} \half \d_\mu (H+G)~ \eps^\dagger\Gamma^0\gamma_5{\hat\sigma}_1\eps=
 -  e^{{3 \over 2} (H+G)} \half \d_\mu (H+G)~ f_2  \l{gfequ}
\eea
We now get an equation which involves only $f_2$ and $H+G$
\bea
\d_\mu f_2= \half f_2 \d_\mu (H+G),
\eea
which can be easily solved
\bea   \l{gaugefield}
f_2= 4 \alpha  e^{ \half (H+G) } ,\qquad B_t=- \alpha e^{2 (H+G)}
\eea
In the same way, starting from equations (\ref{33Eqnf1}), (\ref{simplif2}),
we can prove that
\bea \l{tigauge}
f_1 = 4 \beta e^{\half (H-G)}, \qquad \tilde B_t=- \beta e^{2 (H-G)} ~,~~~~~~~~4 \beta=1
\eea
Here we have set $ 4 \beta =1$ by choosing the overall sign of the five--form field
strength
and an  appropriate rescaling of the Killing
spinor\footnote{Note that the previous conditions, $K^t=1$
and $L_y = \gamma$, only determine the normalizations of  $t$ and $y$
 but do not determine the normalization
of the Killing spinor.}. We will fix $\alpha$ below.

We will now show that $H$ has a simple coordinate dependence.
 We begin with the equation coming  from the sum of
(\ref{simplif1}) plus (\ref{simplif2}) and its adjoint
\bea\label{SphereEqnPlus}
 {\hat\sigma}_1{\not\d}H\eps&=&(-iae^{- \half (H+G)}\gamma_5+be^{- \half (H-G)})
 \eps\\
 \bar \eps {\hat\sigma}_1{\not\d}H&=&
- \bar \eps (-iae^{-\half (H+G) }\gamma_5+be^{- \half (H-G)})
\eea
We find
\bea
\d_\mu H f_1&=&i\d_\mu H \bar \eps {\hat\sigma}_1\eps=
{ i \over 2} \bar \eps  [\gamma_\mu,( -iae^{- \half (H+G)}\gamma_5+be^{- \half (H-G)})]\eps=
\\
&=&
- a e^{-\half (H+G)}\bar  \eps \gamma_5\gamma_\mu \eps=- ae^{- \half (H+G) }L_\mu\nonumber
\eea
so $H$ is a function of $y$ only. Using (\ref{tigauge})
we can determine this function
\bea \l{yexpr}
e^{H}= -  a \gamma y = y ~,~~~~~~~~~~~~\gamma = - a
\eea
where we have fixed the sign of  $\gamma$.
We now fix  $\alpha$ by
multiplying (\ref{SphereEqnPlus}) by $\bar \epsilon \gamma^5 \hat \sigma^1$
 which gives
\bea
%% h^{-2} \partial_y Z L_y &=& - a f_1 e^{-A} - b f_2 e^{-B}
%%\\
h^{-2}  \gamma \partial_y e^{H} &=& - a h^{-2} =  -{ a  } (
f_1^2 + { b a   \over 4 \alpha } f_2^2 ) \l{fixconst}
\eea
By comparing with (\ref{normkill}) we see that we need to have
$ a b 4 \alpha =  1 $.
We can now choose $4 \alpha = 4 \beta $ \footnote{
The  sign choice in $\alpha = \pm \beta$
corresponds to whether we look at chiral primaries with $\Delta  \mp J =0$.}.
Note that with these choices only
supersymmetries with $b=a$ are preserved, but still we have both choices of
sign for $a$ \footnote{ This is true for the generic solution. Special backgrounds like
$AdS_5 \times S^5$ or the plane wave background preserve more supersymmetries.}.
We now go back to
(\ref{SphereEqnPlus}), and we also
recall that $g_{yy}=h^2$. Then we find
\bea
\left( \frac{1}{hy}{\hat\sigma}_1{\Gamma}^3+iae^{-\half (H+G)}\gamma_5-be^{- \half (H-G)}
\right)\eps=0
\eea
Using (\ref{normkill}), (\ref{gaugefield}), (\ref{tigauge}), this reduces to the
projector
\bea
\label{TheProject}
\left( \sqrt{ 1 +e^{-2G}} {\hat\sigma}_1{\Gamma}^3
+  a ie^{-G}\gamma_5  - a
\right)\eps=0
\eea

The definitions (\ref{defbilinears}) and
the equations $K^t=1$,
$L_y = -a$
imply that $\epsilon^\dagger \epsilon =1$ and $\epsilon^\dagger \Gamma^0 \Gamma^5 \Gamma^3
\epsilon = -a$. Since  $\Gamma^0 \Gamma^5 \Gamma^3$ is a unitary operator we conclude that
we must also have the following projection condition
\bea
\left[1+ a\Gamma^0\Gamma^5\Gamma_3\right]\eps=0,\quad\mbox{or}\quad
\left[1+ ai\Gamma_1\Gamma_2\right]\eps=0  \l{resproj}
\eea
The two projectors (\ref{TheProject}) and (\ref{resproj}) imply that the Killing spinor
has the form
\bea\l{epsilonone}
\epsilon &=& e^{ i \delta \gamma^5 \Gamma^3 \hat \sigma^1 } \epsilon_1 ~,~~~~~
\Gamma^3 \hat \sigma^1 \epsilon_1 = a \epsilon_1,\qquad
\sinh 2 \delta = a e^{ - G}
\eea
We can fix the scale of $\epsilon_1$ by inserting (\ref{epsilonone}) in the expression
for $f_2$ which gives
\be \l{epsilonzero}
\epsilon_1 = e^{ { 1 \over 4} (H+G) } \epsilon_0 ~,~~~~~~~ \epsilon_0^\dagger \epsilon_0 =1
\ee
We can  set the phase of $\epsilon_0$ to zero by performing a local Lorentz rotation
in the $12$ plane. Then $\epsilon_0$ is a constant spinor.

We can now insert this expression for the Killing spinor
in the definition of the one form
(\ref{oneform}) to find that
\bea
\omega_{\hat 2} &=&  \epsilon^t \Gamma^2 \Gamma^2 \epsilon =  e^{ \half (H+G)}
\cosh 2 \delta  \, \epsilon^t_0
\epsilon_0 = h^{-1} \epsilon^t_0
\epsilon_0
\nonumber\\
\omega_{\hat 1} & = & \epsilon^t \Gamma^2 \Gamma^1 \epsilon = - i a  h^{-1} \epsilon^t_0
\epsilon_0
\\
\omega_\mu & = & \omega_{\hat c} e^{\hat c}_\mu dx^\mu =  ({\rm constant})
({\tilde e}^{\hat 1}_i+ i a {\tilde e}^{\hat 2}_i) dx^i\nonumber
\eea
Where $\tilde e^{\hat c}_i $ is the
vielbein of the metric $\tilde h_{ij} = {\tilde e}^{\hat c}_i
{\tilde e}^{\hat c}_{j} $ and $e^{\hat i}_i = h {\tilde e}^{\hat i}_j$ is the
full vielbein for the
four dimensional metric in the directions 1,2. Equation (\ref{omegaequation})
implies that these vielbeins are
independent of $y$ and that the two dimensional metric is flat. So we  choose
coordinates such that $\tilde h_{ij} = \delta_{ij}$.

We now use equation (\ref{fequationapp}) to write an expression
for the gauge field
\bea
 B &=& B_t (dt + V) +  \hat B ~,~~~~~~
\nonumber \\
d \hat B + B_t dV &=& -h^2 e^{3G}~_3^\star d\tilde B_{t} \l{gfone}
\\
\tilde B &=& \tilde B_t (dt + V) +  \hat {\tilde B}\nonumber
\\
d \hat{ \tilde B} + \tilde B_t d V  &=&  h^2 e^{-3G} ~_3^\star d  B_{t}  \l{gftwo}
\eea
where $\hat B$, $\hat {\tilde B}$  have no components
along the time direction and
$*_3$ it the flat space epsilon symbol in the directions $y,x_1,x_2$.
It is now possible to obtain an expression for the  vector $V$.
We start from the antisymmetric part of the equation for the Killing spinor
(\ref{33Eqnf3})
\bea
-{ 1\over 2} d[h^{-2}(dt + V) ] &=& \half dK =   e^{ - (H+G)} F  + e^{- (H-G)} \tilde F
\eea
This equation splits into two equations, one gives no new information,
the equation giving new information is
\bea
{ 1\over 2} h^{-2} dV  &=& - e^{- (H+G) }(   d \hat B + B_t d V) - e^{- (H-G)}
(   d  \hat {\tilde B} +  \tilde B_t d V )\nonumber\\
&=& h^2 ( e^{ - H  + 2 G} *_3 d \tilde B_t - e^{-H - 2 G} *_3 d B_t )
\\
\l{finexpv}
d V &=&  2 h^4 y *_3 d G={1 \over y}  *_3 d z,\qquad
z\equiv { 1 \over 2} \tanh G
\eea
where in the last couple of equations we used (\ref{normkill}) written as
\be
h^{-2} = y (e^{G} + e^{-G} )
\ee
The consistency
condition $d(dV)=0$
gives the equation
\be
 { 1 \over y}  \partial_i^2 z + \partial_y ( { 1 \over y} \partial_y z) =0
\ee

 {} From equations (\ref{gfone}), (\ref{gftwo})  and (\ref{finexpv})
 we can  determine the gauge fields
\bea
d \hat B &=& - { 1 \over 4} y^3 *_3 d ( { z + \half \over y^2 } )
\\
d \hat {\tilde B} &=& - { 1 \over 4} y^3 *_3 d ( { z - \half \over y^2 } )
\eea

In summary, we have derived the full form of the metric and gauge fields
described in
(\ref{solmetric})--(\ref{4dgf2}). In addition we found the expression (\ref{epsilonone}),
(\ref{epsilonzero}) for the Killing spinor. It is possible to show that this killing
spinor obeys all other equations, so that we have a consistent solution.

\section{Dilute gas approximation and Coulomb branch.}
\label{dilute}
\renewcommand{\theequation}{B.\arabic{equation}}
\setcounter{equation}{0}

\begin{figure}[htb]
\begin{center}
\epsfxsize=2.0in\leavevmode\epsfbox{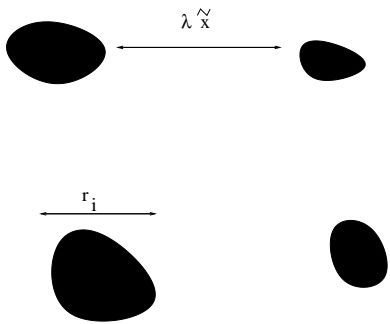}
\end{center}
\caption{Dilute gas approximation for the three branes: we scale
$\la$ to infinity while keeping ${\tilde x}$ and typical size of
the droplets $r_i\sim \sqrt{A_i}$ fixed. } \label{FigGas}
\end{figure}
Let us consider an interesting limit of the general solution
(\ref{solmetric})--(\ref{4dgf2}) which leads to the Coulomb branch
of D3 branes \cite{larsen}. We begin with a boundary condition in
$y=0$ plane which corresponds to droplets with areas $A_i$ distributed
in this plane (see figure \ref{FigGas}), and we
take a dilute gas approximation by increasing the distance between
the droplets while keeping $A_i$ fixed. More explicitly, we make
the rescaling
\bea x\rightarrow \la {\tilde x},\qquad
x'\rightarrow \la {\tilde x'},\qquad y\rightarrow \la {\tilde
y},\qquad A_i ~-&\mbox{fixed}
\eea
in equations (\ref{solmetric}),
(\ref{zepres}). We will be interested in the
$\la\rightarrow\infty$ limit of the solution, while positions of
the droplets ${\tilde{\bf x}}_i$ in the rescaled coordinates are
kept fixed. Using (\ref{zepres}), (\ref{solmetric2}),
(\ref{solmetric4}), we get the approximate expressions
\bea
e^{2G}=\frac{{\tilde y}^2}{\pi\la^2}\sum_i\frac{A_i}{ [(\tilde{\bf
x}-\tilde{\bf x}'_i)^2+{\tilde y}^2]^2} \equiv \frac{{\tilde
y}^2}{\la^2}H,\qquad h^{-2}=ye^{-G}=\la^2 H^{-1/2}
\eea
and we
observe that the terms containing $V_i dx^i\sim \la^{-2}$ are
subleading as $\la$ goes to infinity.
In the limit the metric
(\ref{solmetric}) becomes
\bea\label{Coulomb}
ds^2=H^{-1/2}\left[-d{\tilde t}^2+dw_idw_i\right] +H^{1/2} (d{\tilde y}^2
+{\tilde y}^2d\Omega_3^2+ dx^idx^i) \eea Here ${\tilde t}=\la t$
and three dimensional flat space parameterized by $(w_1,w_2,w_3)$
arose from $d{\tilde\Omega}_3^2$ in the large radius limit: $\la^2
d{\tilde\Omega}_3^2 \sim dw_idw_i$.

We see that in the limit $\la\rightarrow\infty$ we obtain a simple
multi--center solution (\ref{Coulomb}) which is parameterized by
one harmonic function \bea H=\frac{1}{\pi}\sum_i\frac{A_i}{
[(\tilde{\bf x}-\tilde{\bf x}'_i)^2+{\tilde y}^2]^2} \eea
This
solution corresponds to the SO(4) invariant sector of the Coulomb
branch of the SYM theory \cite{larsen}.

\section{Relation to the solutions of Bena and Warner \cite{ibnw}.}
\label{benawarner}
\renewcommand{\theequation}{C.\arabic{equation}}
\setcounter{equation}{0}

In this appendix we compare the solution (\ref{mtwo1})--(\ref{mtwo3}) to the
solution in \cite{ibnw}.
 First we look at the
metric written in that paper:
\bea
ds_{11}^2&=&16L^4e^{2B_0}\left[-dt^2+dz^2+dx_{11}^2\right]+e^{2B_1-B_0}(du^2+dv^2)
\nonumber\\
&+&2 u^2 e^{2B_3-B_0}d\Omega^2+2 v^2 e^{-2B_3-B_0}d{\tilde\Omega}^2
\eea
To compare with our solution we should make
identifications:
\bea
&&e^{4\Phi/3}=16L^4e^{2B_0},\qquad e^{H+G}=8L^2  u^2 e^{2B_3},\quad
e^{H-G}=8L^2  v^2 e^{-2B_3},\nonumber\\
&&h^2(dx^2+dy^2)=4L^2 e^{2B_1}(du^2+dv^2)
\eea
Combining the warp factors and using orthogonality of the coordinates, we find
\bea
y= 8L^2 uv,\qquad x=4 L^2  (u^2-v^2)
\eea
then we get relations for $h$ and $G$:
\bea
h=\frac{e^{B_1}}{4L\sqrt{u^2+v^2}},\qquad e^{G}=\frac{u}{v}e^{2B_3}
\eea
As a cross check we look at the relation between $h$ and $y,G$:
\bea
y(e^G + e^{-G})&=&8L^2 (u^2 e^{2B_3}+v^2 e^{-2B_3})=8L^2 \left(
\frac{uv M_u}{M_v}+\frac{uv M_v}{M_u}\right)\nonumber\\
&=&\frac{8L^2  uv}{M_u M_v}4L^2(u^2+v^2)=16 L^2 e^{-2B_1}(u^2+v^2)\nonumber\\
h^{-2}&=&16 L^2 e^{-2B_1}(u^2+v^2)
\eea
where we used formulas of \cite{ibnw} as well as definitions
\bea
M_u=\sqrt{2L^2 u^2+vg^{(0,1)}-ug^{(1,0)}},\qquad
M_v=\sqrt{2L^2 v^2-vg^{(0,1)}+ug^{(1,0)}}
\eea
Now we want to relate their  function $g$ with  our  function
$z$
\bea
z=\frac{1}{2}\frac{e^{ 2 G}-1}{e^{ 2 G}+1}=
\frac{1}{2}\frac{u^2e^{4B_3}-v^2}{u^2e^{4B_3}+v^2}=
\frac{1}{2}\frac{M_u^2-M_v^2}{M_u^2+M_v^2}=
\frac{(v\d_v-u\d_u)g}{2L^2(u^2+v^2)}=- 4 \d_x g
\eea
We could now translate the conditions we found for non-singular solutions into
conditions for sources for their harmonic function $h_{bw} = g/(uv)^2 $.
We end up having sources at $y=0$ which correspond to $u=0$ or $v=0$. At $u=0$
we have uniform spherically symmetric charge distributions in the coordinates
parameterized by $v$ with positive or
negative charge and we have a similar situation at $v=0$, $u \not =0$.
This charges have a specific coefficient. Since we have already
given the full form of the solutions in our parametrization,
we will not fill in the details.

\section{ Isolated strips and 2d QCD. }
\label{AppD4brane}
\renewcommand{\theequation}{D.\arabic{equation}}
\setcounter{equation}{0}

Here we analyze the gravity solutions corresponding to a single
strip (or strips) as in figure \ref{singlestrip}.  It is simply a limit of the
configurations we considered above when we discussed the theory
related to mass-deformed M2 branes.
 The
resulting asymptotic geometry corresponds to a set of $M5$ branes
wrapped on $S^3 \times R^{1+2}$.
 If we compactify one of the
dimensions this becomes, at low energies, a 4+1 Yang Mills theory
on an $S^3 \times R^{1+1}$ where four of the five transverse
scalars have a mass given by the inverse radius of the sphere and
the fifth scalar, call it $Y$, does not have a mass term but it has
a coupling of the form $Tr( Y F_{01})$ where $01$ are the
directions in $R^{1+1}$. The number of D4 branes is the total
width of the strip (or strips). The M-theory form of the solutions
is as in (\ref{mtwo1})-(\ref{mtwo3}) with $z$ and $V$ given by
(\ref{zstrip})-(\ref{vstrip}). Here we just give the form of the
solution in IIA notation.  This solution is a simple
 U-dualization of (\ref{mtwo1})--(\ref{mtwo3})
\bea
ds^2_{IIA}&=&e^{2 \Phi} (-dt^2+dw^2)
+h^2(dy^2+dx^2)+y e^{G}d\Omega_3^2+
y e^{-G}d{\tilde\Omega}_3^2\nonumber\\
e^{2\Phi}&=&\frac{1}{h^2-h^{-2}V^2},\qquad
{\cal B}^{NS}=-\frac{h^{-2}V}{h^2-h^{-2}V^2} dt\wedge dw\nonumber\\
\nonumber\\
F_{4} &=& -{1\over 4} e^{-2 \Phi} ( e^{-3 G} *_2 d(y^2 e^{2 G} )
\wedge d\tilde \Omega_3 + e^{3 G} *_2 d( y^2 e^{-2G})\wedge
d\Omega_3 ) \l{twoa3} \eea where $*_2$ is a flat 2D epsilon
symbol. The solution that corresponds to D4 branes on $S^3 \times
R^2$ (or M5 branes on $R^2 \times S^1 \times S^3$) is this
solution where we consider metrics given by a single strip. We
take a source in a form of a strip \bea {\tilde
z}(y=0)=-\theta(x)\theta(1-x) \eea Using the general formula
(\ref{solstrips}) we find the solution \bea \l{zstrip} {\tilde
z}(x,y)
&=&\left.\frac{1}{2}\frac{x-x'}{\sqrt{({x}-{x}')^2+y^2}}\right|^R_L
=-\frac{1}{2}\left[\frac{x}{\sqrt{x^2+y^2}}-
\frac{x-1}{\sqrt{(x-1)^2+y^2}}\right]\\
 \l{vstrip}
 V&=&
-\left.\frac{1}{2}\frac{1}{\sqrt{({x}-{x}')^2+y^2}}\right|^R_L=
\frac{1}{2}\left[\frac{1}{\sqrt{x^2+y^2}}-\frac{1}{\sqrt{(x-1)^2+y^2}}\right]
\eea
In the leading order we find
\bea
{\tilde z}=-\frac{y^2}{2(x^2+y^2)^{3/2}},\quad
V=-\frac{x}{2(x^2+y^2)^{3/2}},\quad
h^{-2}=\sqrt{2}(x^2+y^2)^{3/4}
\eea
Then introducing polar coordinates in the $x,y$ plane, we find an asymptotic form
of the metric
\bea
ds_{IIA}^2=\sqrt{2}r^{3/2}\left[-dt^2+dw^2+d\Omega_3^2\right]+
\frac{1}{\sqrt{2}r^{3/2}}\left[dr^2+r^2 d\theta^2+r^2\sin^2\theta
d{\tilde\Omega}_3^2\right]
\eea
which asymptotes at large $r$ to the metric of the D4 brane (or M5 brane
when uplifted to 11 dimensions).

Let us look at the metric for M5 branes from (\ref{mtwo1}) by
imposing boundary condition corresponding to a superposition of
isolated strips. We observe that the first few terms in the large
$r$ expansion are described by 7d gauged supergravity\footnote{For
completeness we also give the relation between coordinates $x_2,y$
in (\ref{mtwo1}) and $r,\theta$ which we use here:
$x_2=r'\cos\theta'$, $y=r'\sin\theta'$,
$r'=r+\left(\frac{3P_0}{10}+\frac{P_1\cos\theta}{2P_0}\right)+
\frac{3P_0^4-5P_1^2+15P_0P_2-25\cos
2\theta(P_1^2-P_0P_2)}{100rP_0^2}
$,\\
$\theta'=\theta-\frac{P_1\sin\theta}{2rP_0}+
\frac{6P_0^2P_1+5\cos\theta(5P_1^2-4P_0P_2)\sin\theta}{40r^2P_0^2}$
}:
\bea
ds_{11}^2&=&\left(\frac{2}{P_0}\right)^{1/3}\Delta^{1/3}\left[
\left(r+\frac{P_0}{2}+\frac{P^2_0}{10r}
\right)(-dt^2+d{\bf w}^2)
+rd{\Omega}_3^2\right.\nonumber\\
&&\qquad\left.
+P_0\frac{dr^2}{2r^2}\left(1-\frac{9P_0}{10r}+\frac{39P_0^2}{100r^2}
\right)
\right]\nonumber\\
&+&\left(\frac{P_0}{2}\right)^{2/3}\Delta^{-2/3}(T^{-1})_{IJ}dY^I dY^J
\eea
Here we parameterized a deformed $S^4$ by a five dimensional unit vector
$Y_I$ ($Y_I Y_I=1$) such that $Y_5=\cos\theta$ and $Y_i=\sin\theta\mu_i$,
where four dimensional unit vector $\mu_i$ parameterises ${\tilde S}^3$.
The matrix
$T_{IJ}$ has the form
\bea
T_{IJ}=\mbox{diag}(T,T,T,T,T^{-4}),\quad
T=1-\frac{P_0}{10r}+\frac{P_0^4-15P_1^2+20P_0P_2}{100r^2P_0^2}
\eea
and
\bea
\Delta\equiv Y_I T_{IJ}Y_J&=&
1+\frac{(3+5\cos 2\theta)P_0}{20r}+\frac{2\cos^2\theta}{r^2}
\left(\frac{P_0^2}{20}-\frac{P_0^4-15P_1^2+20P_0P_2}{50P_0^2}\right)
\nonumber\\
&+&\frac{P_0^4-15P_1^2+20P_0P_2}{100P_0^2 r^2}\sin^2\theta
\eea
The solution is specified by the moments of the distribution:
\bea
P_n\equiv (n+1)\int_{\cal D} dx x^{n}
\eea
By making a shift in coordinate $x$, we can go to the frame where
$P_1=0$.
In the next order in $1/r$ expansion a generic metric is not described by
the ansatz of the gauged supergravity, however if the charges of the
solution satisfy the relation
\bea
P_3=-\frac{P_1^3-2P_0P_1P_2}{P_0^2},
\eea
then even in the next order in $1/r$ we excite only fields from the gauged
SUGRA.

In the field theory, we have a 2d Yang Mills reduced from 4+1 Yang
Mills theory on $S^3 \times R^{1+1}$. The lagrangian is of the
schematic form \bea
 S \sim \int_{2d} Tr [ -{1 \over 2}F^2 + (DY)^2 + Y F + \cdots]
\eea
We have supersymmetric ground states
such
that the electric field
$E = { \partial L  \over \partial \dot A_1}= F_{01} + Y  =Y$, since $F_{01}=0$
on these ground states.
It
is interesting that these states are characterized by the value of
$Tr[ E^2 ]\sim Tr[Y^2]$, which in the fermion picture
corresponds to the energy of non-relativistic fermions
\be \l{enerqcd}
E_{NR} = \int_{Strips} dx \half x^2
\ee
In the gravity picture this quantity appears as the leading (angular dependent)
deviation from
the metric we described above. It is a quantity similar to a dipole moment.
This quantity is well defined for the BPS solutions we are considering.
It would be nice to know if there is a quantity that is conserved, and it is defined
in the full interacting theory, which would reduce to (\ref{enerqcd}) on the
supersymmetric ground states.

\section{Asymptotic form of the metric}
\label{AppEnergy}
\renewcommand{\theequation}{E.\arabic{equation}}
\setcounter{equation}{0}

We now derive expressions for the energy and angular momentum of
the solution. To do this it is convenient to use radii of the
spheres $u,v$ as independent coordinates and rewrite metric in the
form \bea ds^2&=&-(u^2+v^2)(dt+V_i
dx^i)^2+\frac{1}{u^2+v^2}(dy^2+d{\bf x}^2)+u^2 d\Omega_3^2+ v^2
d{\tilde\Omega}_3^2 \eea and $x_1$, $x_2$ can be expressed in
terms of $u,v$ using the relation:
\bea {\tilde z}\equiv
-\frac{y^2}{\pi}\int_{\cal D}\frac{d^2x'}{[({\bf x}-{\bf x}')^2+u^2
v^2]^2} =-\frac{y^2}{v^2(v^2+u^2)} \eea
Here ${\cal D}$ is a
region of $y=0$ plane where ${\tilde z}=-1$. Let us assume that
such regions are contained inside a circle with sufficiently big
radius in $x_1$--$x_2$ plane. For $r\equiv |x|$ and $y$ larger than this
radius we
 can perform a multipole expansion in ${\tilde z}$ and $V_i$
\bea {\tilde z}&=&-
\frac{y^2}{\pi}\left[\frac{1}{(r^2+y^2)^2}\int_{\cal D} d^2 x'-
\frac{2}{(r^2+y^2)^3}\int_{\cal D} d^2 x'\left\{({\bf x}')^2-\frac{6
r^2 ({\bf x}'{\bf n})^2}{r^2+y^2} \right\} \right] \l{expanz}
\\ \l{expanv}
V_i &=&-\frac{1}{2\pi}\frac{2r}{(r^2+y^2)^2}\int_{\d {\cal D}}
dx'_i({\bf n}{\bf x'}) \eea where we have chosen the origin so
that the dipole vanishes, $\int_{\cal D} d^2 x' {\bf x}_i'=0 $.

We now define \bea Q^2 &=& { 1 \over \pi} \int_{\cal D} d^2x'
\l{qexp}
\\
W &=& Q^{-2}{ 1 \over \pi} \int_{\cal D} d^2 x' |x'|^2  \l{wexp}
\\
W_{\bf n} &=& Q^{-2} { 1 \over \pi} \int_{\cal D} d^2 x' n_i n_j
x'_i x'_j \equiv \half W + \half \tilde W \cos \phi \eea
where in the
last line $n_i $ is a unit vector in the direction $x^i$ and we
have defined the origin of the angle $\phi$ so as to  diagonalize
$W_{ij}$. The equations (\ref{expanz}), (\ref{expanv}) become \bea
{\tilde z} &=&-\frac{Q^2}{(r^2+y^2)^2}\left[1-
\frac{2}{(r^2+y^2)}\left\{W-\frac{6 r^2 W_{\bf n}}{r^2+y^2}
\right\}
\right]\nonumber\\
V_\phi&=&-Q^2\frac{r^2}{2 (r^2+y^2)^2 }\left[ 1
+24\frac{r^2}{(r^2+y^2)^2} W_{\bf n}-\frac{4W}{(r^2+y^2)}-
\frac{8 W_{\bf n}}{(r^2+y^2)}
\right]\\
V_r&=&-Q^2\frac{1}{2}\frac{4{\tilde W}r\sin 2\phi}{(r^2+y^2)^3}\nonumber
\eea

The analysis of the asymptotic form of the metric involves terms which are
decaying like $1/v^{2}$ where $v$ is the radius of the $S^3$ in $AdS$.
At this order we have the gauge field, which is what we are interested in,
and in addition we have some scalar modes of dimension $\Delta =2$. One
of these
modes is excited when $\tilde W \not =0$.
At this order we require that $AdS\times S$ metric is
deformed in a way prescribed by the gauged supergravity.
Making a change of variables which is parameterized by various functions of
${\tilde\phi}$ and ${\tilde u}$:
\bea
u={\tilde u}+\frac{f_1}{{\tilde v}^2}+\frac{f_2}{{\tilde v}^4},\qquad
v={\tilde v}+\frac{g_1}{{\tilde v}}+\frac{g_2}{{\tilde v}^3},\qquad
\phi={\tilde\phi}+\frac{h_1}{{\tilde v}^2}+\frac{h_2}{{\tilde v}^4},
%%\tau={\tilde\tau}+\frac{e_1}{{\tilde v}^2}+\frac{e_2}{{\tilde v}^4}
\eea
and requiring that the only mixing between
sphere and AdS in the first two orders in $1/{\tilde v}^2$ is given by a gauge field,
we arrive at the metric
\bea
ds^2&=&\left[
1+\frac{(3u^2-2Q)(Q^2-2W)+6(Q-u^2){\tilde W}\cos 2\phi}{6Q^2v^2}\right.\nonumber\\
&+&\left.\frac{((2Q-3u^2)(Q^2-2W)-6(Q-u^2){\tilde W}\cos 2\phi)^2}{48Q^4v^4}
-\frac{2g_2}{v^4}\right]\nonumber\\
&\times&\left\{-\left(v^2+Q+\frac{Q^2-2W}{3v^2}\right)dt^2+Q\frac{dv^2}{v^2}
\left(1-\frac{Q}{v^2}\right)+v^2 d{\tilde\Omega}_3^2\right\}
\nonumber\\
&+&g_{uu}du^2+2 g_{u\phi}duD\phi+g_{\phi\phi}D\phi^2+g_{\Omega\Omega}d\Omega_3^2
\eea
where expressions for the metric components on the sphere are rather complicated, and
\bea
D\phi=d\phi+dt-\frac{2W-Q^2}{Qv^2}dt
\eea
So we see that the angular momentum or the mass is equal to
\be \l{ang}
M = J =
 { 2 \pi^2  \over 16 \pi G_{N}^5}  { 2 W  - Q^2  \over Q^2} = { 1 \over
 16 \pi^2 l_p^8 } ( W Q^2 - { Q^4 \over 2} )
\ee
Notice that the term proportional to $W Q^2 $ is proportional
 the energy of the fermions
and the term proportional to $Q^4 $ is subtracting the ground state energy of
$N$ fermions. Using (\ref{qexp}) and (\ref{wexp}) we get (\ref{energyf}).

For completeness we also give the
expressions for the coefficients involved in the reparameterization
\bea
f_1&=&\frac{u^3(Q^2-2W)}{4Q^2}+\frac{u(Q-u^2){\tilde W}\cos 2\phi}{2Q^2},\quad
h_1=-\frac{{\tilde W}\sin 2\phi}{Q},\quad h_2=0\nonumber\\
g_1&=&\frac{Q^2-2W}{6Q}-\frac{u^2(Q^2-2W)}{4Q^2}-
\frac{(Q-u^2){\tilde W}\cos 2\phi}{2Q^2}.
\nonumber
\eea

\section{1/2 BPS chiral primaries in M theory.}
\label{AppMderiv}
\renewcommand{\theequation}{F.\arabic{equation}}
\setcounter{equation}{0}

%=====================================================================

\subsection{Reduction on the spheres.}
\label{AppE1}

Let us now consider supersymmetric solutions of eleven dimensional
supergravity which have $SO(6)\times SO(3)$ symmetry
\bea
ds_{11}^2&=&e^{2\la}\left(\frac{1}{m^2}d\Omega_5^2+e^{2A}d{\tilde\Omega}_2^2+
ds_4^2\right)\\
G_{(4)}&=&G_{\mu_1\mu_2\mu_3\mu_4}dx^{\mu_1}\wedge
dx^{\mu_2}\wedge d x^{\mu_3}\wedge
dx^{\mu_4}+\d_{\mu_1}B_{\mu_2}dx^{\mu_1}\wedge dx^{\mu_2} \wedge
d^2{\tilde\Omega} \eea where $d\Omega_5^2$ and $d\tilde
\Omega_2^2$ are unit radius metrics on the corresponding spheres
and $\mu_i$ are indices in the remaining 4 dimensions. It is
convenient to introduce the following notation for the coordinates
\bea \Omega_5: \ \theta^a,\qquad {\tilde\Omega}_2:\
\theta^\alpha,\qquad ds_4: \ x^\mu \eea and to choose eleven
dimensional gamma matrices as
\bea \Gamma^a=\rho^a\otimes
\gamma^7,\quad \Gamma_\alpha=1\otimes\sigma_\alpha\otimes
\gamma^5,\qquad \Gamma_\mu=1\otimes 1\otimes \gamma_\mu,\nonumber
\eea \bea \gamma_7=(\sigma_1\cdot\sigma_2)\otimes (-i\gamma_5)=
\sigma_3\otimes\gamma_5\qquad (\gamma_7)^2=+1 \eea To find
supersymmetric configurations we will solve the equation for
Killing spinor \bea\label{11DKill}
\nabla_m\eta+\frac{1}{288}\left[{\Gamma_m}^{npqr}-
8\delta_m^n{\Gamma}^{pqr}\right]G_{npqr}\eta=0 \eea Following
\cite{Gauntlett} we first perform a reduction on $S^5$ by
decomposing the spinor as
\bea \eta =\psi(\theta^a)\otimes e^{\la/2}\xi
\eea and perform further reduction on $S^2$ later on. The spinor
on the sphere $S^5$ satisfies the equation \bea\label{SpinorOnS5}
D_a\psi=\frac{i}{2}\rho_a\psi, \eea where $D_a$ is a reduced
covariant derivative on a unit sphere, which is related to $\nabla_a$
as \bea \nabla_a=m D_a-\frac{1}{2}{\gamma^\mu}_a\d_\mu \la \eea
Then (\ref{11DKill}) reduces to equations for seven dimensional
spinor $\xi$ \bea &&\left[\gamma^\mu\d_\mu \la+
\frac{1}{144}e^{-3\la}{{\tilde\Gamma}}^{npqr}G_{npqr}+im\gamma_7\right]
\xi=0\nonumber\\
&&\left[\nabla_\mu-\frac{im}{2}\gamma_\mu\gamma_7-\frac{1}{24}e^{-3\la}
{{\tilde\Gamma}}^{npq}G_{\mu npq}\right]\xi=0,\\
&&\left[{\tilde\nabla}_\alpha-\frac{1}{2}{\gamma^\mu}_\alpha\d_\mu A
-\frac{im}{2}\Gamma_\alpha\gamma_7-\frac{1}{8}e^{-3\la}
{{\tilde\Gamma}}^{\beta \mu\nu}G_{\alpha\beta \mu\nu}\right]\xi=0\nonumber
\eea
We now perform the reduction on $S^2$ by introducing two component spinors
$\chi_+$
and $\chi_-$ which obey the equations
\bea\label{SpinorOnS2}
{\tilde\nabla}_\alpha \chi_\pm=\pm\frac{i}{2}{\tilde e}_\alpha^{\hat\alpha}
{\hat\gamma}_{\hat\alpha}\chi_\pm=
\pm\frac{i}{2}e^{-A}\gamma_{\alpha}\gamma_5 \chi_\pm
\eea
and expanding the spinor $\xi$ over basis on the sphere:
\bea\label{DecompSpinXi}
\xi=\chi_+\otimes \eps_++\chi_-\otimes \eps_-.
\eea
Covariance under $SU(2)$ transformations ensures that we can take
\bea
\chi_-={\hat\gamma}\chi_+,\qquad {\hat\gamma}\equiv i\gamma_7\gamma_5
\eea
without loss of generality. Introducing an operator $P$ which acts on the spinors
$\chi_\pm$ as $P \chi_{\pm}=\pm \chi_\pm$, we can simplify the equations for $\xi$:
\bea\label{KillEqnS5}
&&\left[\gamma^\mu\d_\mu \la+
\frac{1}{144}e^{-3\la}{{\tilde\Gamma}}^{npqr}G_{npqr}+im\gamma_7\right]
\xi=0\\
\label{KillEqnS2}
&&\left[ie^{-A} \gamma^5 P+\gamma^\mu\d_\mu A-im\gamma_7-
\frac{1}{4}e^{-3\la-2A}
{\hat\gamma}\gamma^{\mu\nu}\d_\mu B_\nu\right]
\xi=0\\
\label{KillEqnR4}
&&\left[\nabla_\mu-\frac{im}{2}\gamma_\mu\gamma_7-\frac{1}{4}e^{-3\la-2A}
\d_{[\mu}B_{\nu]}{\gamma}^{\nu}{\hat\gamma}-
\frac{1}{24}e^{-3\la}{\gamma}^{\nu\la\sigma}G_{\mu\nu\la\sigma}
\right]\xi=0
\eea

Before proceeding with analysis of the equations for the spinor,
we observe that the equation for the field strength
\bea
d(~^\star_{11}G_{(4)})=0.
\eea
implies that
\bea\label{GformI1}
G_{\nu_1\nu_2\nu_3\nu_4}=I_1 e^{-3\la-2A}\eps_{\nu_1\nu_2\nu_3\nu_4}
\eea
for some constant $I_1$. This equation tells us that there is a
flux $I_1$ over noncompact four dimensional space. For our purposes we
are not interested in solutions with fluxes over noncompact spaces, so
we set $I_1=0$. Looking back at the equations (\ref{KillEqnS5})--(\ref{KillEqnR4})
we observe that they can be written as two decoupled systems: one for
$\eps_-+\gamma_5\eps_+$ and one for $\eps_--\gamma_5\eps_+$. We will solve these
systems separately by imposing a relation between $\eps_+$ and $\eps_-$
\bea
\eps_-=-a\gamma_5\eps_+
\eea
with $a=\pm 1$.
With this assumption spinor $\xi$ can be expressed in terms of $\eps_+$
and $\chi_+$:
\bea
\xi=(1-a\gamma_5{\hat\gamma})\cdot \chi_+\otimes\eps_+,
\eea
and equations
(\ref{KillEqnS5})--(\ref{KillEqnR4}) can be rewritten in terms of four
dimensional spinor $\eps\equiv\eps_+$:
\bea\label{4dKillEqnS5}
&&\left[\gamma^\mu\d_\mu \la
+\frac{a}{12}e^{-3\la-2A}\gamma_5\gamma^{\mu\nu}\d_\mu B_\nu
+am\right]\eps=0\\
\label{4dKillEqnS2}
&&\left[ie^{-A} \gamma^5+\gamma^\mu\d_\mu A-
\frac{a}{4}e^{-3\la-2A}
{\gamma}_5\gamma^{\mu\nu}\d_\mu B_\nu-am\right]
\eps=0\\
\label{4dKillEqnR4}
&&\left[\nabla_\mu-\frac{am}{2}\gamma_\mu-\frac{a}{8}e^{-3\la-2A}
F_{\mu\nu}\gamma^\nu\gamma_5
\right]\eps=0
\eea
Here $F_{\mu\nu}$ is a field strength of the gauge field $B_\mu$.

From now on we will take $a=1$, and solutions with $a=-1$ can be obtained from our
geometries by changing signs of $B_\mu$ and $m$. Alternatively, solution with $a=-1$
can be obtained from one with $a=1$ by reversing the sign of all four coordinates.

\subsection{Using spinor bilinears to fix the form of the metric.}

Let us now construct bilinears out of four dimensional spinor
$\eps$ \footnote{Notice that while some of our bilinears are equal
to those constructed in \cite{Gauntlett} for $S^5$ reduction (for
example, ${\bar\xi}\gamma_7\xi=-2i(\chi_+^\dagger
\chi_+){\bar\eps}\eps$), bilinears $f_2$ and $L_\mu$ use the split
between $S^2$ and four dimensional space in a nontrivial way. For
example, $f_2=\frac{1}{2}{\bar\xi}{\hat\gamma}\xi$ uses chirality
matrix on the sphere so it is not defined for a generic six
dimensional base. On the other hand, a bilinear $i{\bar\xi}\xi$
exists in general, but it has nontrivial dependence of $S^2$, so
it does not respect $S^2\times M$ split.
}:
\bea
f_1={\bar\eps}\eps,\quad f_2={\bar\eps}\Gamma_5\eps,\quad
K_\mu=-2{\bar\eps}\gamma_\mu\eps,\quad L_\mu=2m{\bar\eps}\gamma_\mu\Gamma_5\eps,
\quad Y_{\mu\nu}={\bar\eps}\gamma_{\mu\nu}\eps.
\eea
There are also bilinears involving $\eps^t$ instead of ${\bar\eps}$, we will consider
them later. Taking derivatives of the bilinears, we get
\bea\label{BilMscal}
&&\nabla_\mu f_1=0,\qquad \nabla_\mu f_2=L_\mu-3\d_\mu\la f_2\\
\label{BilMvect}
&&\nabla_{\nu} {K}_{\mu}=
-2mY_{\mu\nu}
+\frac{e^{-3\la-2A}}{2}F_{\mu\nu}f_2
\eea
We will not need the expression for $\nabla_\mu L_\nu$ and $\nabla_\mu Y_{\nu\la}$.
Notice that due to the first relation in (\ref{BilMscal}) we can choose
a normalization
\bea\label{NormOfSpin}
f_1=-i
\eea
we also write a real bilinear $f_2$ as
\bea
f_2=-\sinh\zeta
\eea

Let us now use the relations between spinor bilinears to
restrict the form of the metric on four dimensional base space.
First we rewrite the second equation in (\ref{BilMscal}) as
\bea
e^{-3\la}\nabla_\mu (e^{3\la}\sinh\zeta)=-L_\mu
\eea
Following \cite{Gauntlett}, we can define a coordinate
\bea
y=-e^{3\la}\sinh\zeta
\eea
and the remaining three coordinates to be orthogonal to $y$. In this coordinate
system vector $L_\mu$ has only one nontrivial component:
\bea\label{DefVectL}
{L}_\mu dx^\mu=e^{-3\la}dy
\eea

Equation (\ref{BilMvect}) implies that $K^\mu$ is a Killing vector,
and we will choose the coordinate $t$ along the vector ${K}^\mu$
(we take ${K}^t=2m$).
One can also repeat the arguments of \cite{Gauntlett} to show that the Lie
derivatives ${\cal L}_{K}$ of $e^{\la}$ and four form field
strength vanish, and thus ${K}^\mu$ generates an isometry of the
entire solution, not just the four dimensional metric.

Let us now use the Fierz identities to determine some of the metric
components. As in the type IIB case we find:
\bea
&&K_\mu L^\mu=0,\qquad \frac{1}{4m^2}L^2=-\frac{1}{4} K^2=
({\bar\epsilon}\Gamma_5\epsilon)^2-({\bar\epsilon} \epsilon)^2=\cosh^2\zeta
\eea
This fixes $g_{t\mu}$ and $g_{yy}$ components of the metric:
\bea\label{4DMetric}
ds_4^2=-\frac{1}{m^2}\cosh^2\zeta (dt+V_idx^i)^2+
\frac{e^{-6\la}}{4m^2\cosh^2\zeta}\left(dy^2+g_{ij}dx^i dx^j\right)
\eea
where $i,j$ take values $1$ and $2$. For later convenience we introduce
two functions:
\bea
g_0=\frac{1}{m}\cosh\zeta,\qquad h=\frac{e^{-3\la}}{2m\cosh\zeta}
\eea
and choose nonzero components of the vielbein to be
\bea\label{4DVierb}
e^{\hat 0}_\mu dx^\mu=g_0(dt+V_idx^i),\quad e_j^{\hat i}=h{\hat e}_j^{\hat i},
\quad e_y^{\hat 3}=h,
\eea

To conclude consideration of the metric, we also find the expression for the
warp factor $e^{2A}$. We begin with writing a linear combination of
(\ref{4dKillEqnS5}) and (\ref{4dKillEqnS2}) which does not contain a
vector field $B_\mu$:
\bea\label{APlus3La0}
{\not \d}(A+3\la)\eps=(-ie^{-A}\gamma_5-2m)\eps
\eea
Using this equation we compute
\bea\label{APlus3La1}
&&2\d_\mu(A+3\la)f_1=
{\bar\eps}\{ {\not\d}(A+3\la),\gamma_\mu\}\eps
=-\frac{i}{m}e^{-A}L_\mu
\eea
Using normalization condition (\ref{NormOfSpin}) as well as expression
for ${L}_\mu$ (\ref{DefVectL}), this equation can be rewritten as
\bea
2m d(A+3\la)=e^{-A-3\la}dy
\eea
and can be easily integrated:
\bea\label{MtheA}
e^{A}=\frac{y}{2m}e^{-3\la}
\eea

\subsection{Restricting the form of the spinor.}

Let us now discuss various projections which should be imposed on the
Killing spinor. We begin with substituting (\ref{MtheA}) into equation
(\ref{APlus3La0}):
\bea
(\cosh\zeta e^{-A}\Gamma_3+ie^{-A}\Gamma_5+2m)\eps=0
\eea
To simplify this projector we first rewrite it as
\bea\label{ProjPrelEps}
\left(e^{-\Gamma^3\zeta}+i\Gamma_3\Gamma_5\right)\eps=0
\eea
After introducing ${\tilde\eps}$:
\bea\label{EpsThrEpsTld}
\eps=e^{\frac{\zeta}{2}\Gamma_3}{\tilde\eps}
\eea
we arrive at a very simple condition:
\bea\label{ProjMth1}
i\Gamma_3\Gamma_5{\tilde\eps}=-{\tilde\eps}
\eea
Now we recall expression for nonzero components of $K$ and $L$:
\bea
2m=K^t=\frac{2}{g_0}\eps^\dagger\eps,\qquad
e^{-3\la}=L_y=2mh {\bar\eps}\Gamma_3\Gamma_5\eps
\eea
and take a linear combination of these relations:
\bea
\eps^\dagger(1+\Gamma^5\Gamma^0\Gamma^3)\eps=0
\eea
This equation implies the projection
\bea\label{ProjSpinMt1}
(1+\Gamma^5\Gamma^0\Gamma^3)\eps=0
\eea
which can also be written as
\bea\label{ProjSpinMt2}
(1-i\Gamma^1\Gamma^2)\eps=0
\eea
To summarize, we have shown that the Killing spinor $\eps$ can be expressed as
(\ref{EpsThrEpsTld}) where ${\tilde\eps}$ is annihilated by two independent
projectors:
\bea
(i\Gamma_3\Gamma_5+1){\tilde\eps}=0,\qquad
(1+\Gamma^5\Gamma^0\Gamma^3){\tilde\eps}=0
\eea
This implies that ${\tilde\eps}$ has only one independent component. Normalization
condition (\ref{NormOfSpin}) implies that
\bea
{\tilde\eps}^\dagger\Gamma^0{\tilde\eps}=-i
\eea
and for one component spinor this translates into relation
${\tilde\eps}=e^{i\phi}{\tilde\eps}_0$ with some constant spinor ${\tilde\eps}_0$. We
can then make a local rotation of the vielbein to make the phase $\phi$ independent of
$x_1,x_2,y$ (but the phase can still be a function of time). From now on we will work
in the frame where ${\tilde\eps}$ does not depend on $x_1,x_2,y$.

\subsection{Equations for the field strength.}

Let us now look at the equations for the four--form field strength:
\bea
dG_{(4)}=0,\qquad d(~^\star_{11}G_{(4)})=0.
\eea
As we already discussed, we set $G_{\nu_1\nu_2\nu_3\nu_4}=0$, then
\bea
G_{(4)}=\d_{\mu_1}B_{\mu_2}dx^{\mu_1}\wedge dx^{\mu_2}
\wedge d^2{\tilde\Omega}
\eea
and we have only one nontrivial equation for the field strength:
\bea\label{FldStrEqn}
d(e^{3\la-2A}~^\star_4 dB)=0
\eea
Let us apply the $3+1$ split to this equation: we use indices
$\alpha,\beta=1,\dots 3$ for coordinates $y,x_i$ and write the
index $t$ explicitly. Then it is convenient to split a gauge field as
\bea
B_\mu dx^\mu=B_t dt+B_\alpha dx^\alpha=B_t(dt+V_idx^i)+(B_\alpha-B_t V_\alpha)
dx^\alpha\equiv B_t(dt+V_idx^i)+{\hat B}\nonumber
\eea

We begin with components of (\ref{FldStrEqn}) along spacial directions and
rewriting them in terms of three dimensional quantities:
\bea
d~^\star_3\left[g_0e^{3\la-2A}\left\{d{\hat B}+B_t dV\right\}\right]=0
\eea
This means that locally we can introduce a dual potential $\Phi$:
\bea\label{DefPhiM}
d{\hat B}+B_t dV=g^{-1}_0e^{-3\la+2A}~^\star_3d\Phi
\eea
We will choose to describe the gauge field $B_\mu$ by specifying $B_t$
and $\Phi$.

The time component of (\ref{FldStrEqn}) leads to the equation:
\bea
d\left[V\wedge d\Phi+g_0 g^{tt}e^{3\la-2A}~^\star_3 dB_t\right]=0
\eea

{\bf Evaluation of $B_t$.}

Let us use bilinears constructed from Killing spinors to express $B_t$
and $\Phi$ in terms of warp factors $e^{2\la}$ and $e^{2\la+2A}$. We
begin with $B_t$:
\bea\label{StartDBt}
\d_\mu B_t=\frac{1}{2m}F_{\mu\nu}K^\nu=-\frac{1}{m} F_{\mu\nu}
{\bar\eps}\gamma^\nu\eps=-\frac{1}{4m}{\bar\eps}[\gamma_\mu,{\not F}]\eps
\eea
To proceed it is convenient to add equations (\ref{4dKillEqnS5})
and (\ref{4dKillEqnS2}):
\bea
\left[ie^{-A}+\gamma_5{\not\d}(A+\la)\right]\eps=
\frac{1}{12}e^{-3\la-2A}{\not F}\eps
\eea
and use the resulting relation to exclude ${\not F}$ from (\ref{StartDBt}):
\bea
\d_\mu B_t=-\frac{3}{m}
e^{3\la+2A}{\bar\eps}[\gamma_\mu,ie^{-A}+\gamma_5{\not\d}(A+\la)]\eps
=-\frac{3}{m}e^{\la}f_2\d_\mu e^{2A+2\la}=-6 e^{A+\la}\d_\mu e^{2A+2\la}\nonumber
\eea
Integrating this equation we find the expression for $B_t$:
\bea\label{MthBt}
B_t=-4e^{3A+3\la}.
\eea

{\bf Evaluation of $\Phi$.}

Next we consider
\bea\label{StartDetPhi}
\eps_{\mu\nu\alpha\beta}F^{\alpha\beta}K^\nu=
-i{\bar\eps}\Gamma_5\{\gamma_\mu,{\not F}\}\eps
\eea
Here we used
\bea
\{\gamma_\mu,\gamma_{\alpha\beta}\}=-2i\eps_{\mu\alpha\beta\nu}\gamma_5
\gamma^\nu,
\qquad \eps_{0123}=\sqrt{-g}
\eea
Now we will use the equation (\ref{4dKillEqnS5}) and its conjugate to
eliminate
${\not F}$ from (\ref{StartDetPhi}):
\bea
\eps_{\mu\nu\alpha\beta}F^{\alpha\beta}K^\nu&=&24e^{3\la+2A}
i{\bar\eps}\left(\gamma_5
\gamma_\mu\gamma_5({\not\d}\la+m)+(-{\not\d}\la+m)\gamma_\mu
\right)\eps\nonumber\\
&=&-48e^{3\la+2A}
i{\bar\eps}\eps \d_\mu\la=-48e^{3\la+2A}\d_\mu\la
\eea
where at the last step we used normalization (\ref{NormOfSpin}).
Notice that the expression in the left hand side of (\ref{StartDetPhi}) is
related to the derivative of $\Phi$:
\bea
\eps_{\mu\nu\alpha\beta}F^{\alpha\beta}K^\nu=
4m e^{2A-3\la}\d_\mu\Phi
\eea
so we finally find the relation between $\la$ and $\Phi$:
\bea
\d_\mu\Phi=-\frac{12}{m} e^{6\la} \d_\mu\la
\eea
Integrating this equation, we arrive at the relation
\bea\label{SlnPhi}
\Phi=-\frac{2}{m}e^{6\la}
\eea

\subsection{Evaluating two dimensional metric.}

To fix the form of the vielbein ${\hat e}^{\hat i}_j$ we construct
a one form as a spinor bilinear\footnote{As in the type IIB case
our conventions are such that $\Gamma^2$ is antisymmetric, so the
factor of $\Gamma_2$ is necessary to ensure invariance under
local Lorentz transformations.}
\bea
\omega=\eps^t\Gamma_2\gamma_\mu\eps~ dx^\mu=\eps^t\Gamma_2\Gamma_a\eps~ h{\hat e}^a_\mu
dx^\mu=
\eps^t\eps~ h(i{\hat e}^1_\mu+{\hat e}^2_\mu)dx^\mu
\eea
Let us compute the derivative of this form:
\bea
d\omega&=&\eps^t\Gamma_2\left(-m\gamma_{\mu\nu}+
\frac{1}{4}e^{-3\la-2A}F_{\mu\sigma}{\gamma_\nu}^\sigma \Gamma_5\right)
\eps~ dx^\mu\wedge dx^\nu
\eea
Using commutator
\bea
[\gamma_{\mu\nu},{\not F}]=2({\gamma_\mu}^\alpha F_{\nu\alpha}-
{\gamma_\nu}^\alpha F_{\mu\alpha})
\eea
as well as equation (\ref{4dKillEqnS5})  and its transpose
\bea
\eps^t\Gamma^2\left\{-{\not\d}\la+m-\frac{e^{-3\la-2A}}{24}
\Gamma_5{\not F}\right\}=0
\eea
we evaluate
\bea
d\omega&=&\eps^t\Gamma_2\left(-m\gamma_{\mu\nu}+
3m \gamma_{\mu\nu}+3\gamma_{\mu}\d_\nu\la
\right)
\eps~ dx^\mu\wedge dx^\nu
\eea
Using projection conditions for $\eps$ one can see that
\bea
\eps^t\Gamma_2\gamma_{\mu\nu}\eps~ dx^\mu\wedge dx^\nu=
2\eps^t\Gamma_2\gamma_{\mu}\Gamma_2\eps~ h(i{\hat e}^1_\nu+{\hat e}^2_\nu)
dx^\mu\wedge dx^\nu
\eea
This leads to the expression for the derivative of $\omega$
\bea\label{DerivZ}
d\omega&=&\left(-m\frac{\epsilon^t\gamma_{\mu}\epsilon}{\epsilon^t
\epsilon}dx^\mu
-3d\la\right)\wedge \omega
\eea
Now we use the expressions for bilinears which can be derived using
(\ref{ProjMth1}) and (\ref{ProjSpinMt1}):
\bea
\eps^t\eps=\cosh\zeta{\tilde \epsilon}^t{\tilde\eps},&&
\eps^t\Gamma_0\eps=-i{\tilde\eps}^t{\tilde\epsilon},\quad
\eps^t\Gamma_3\eps=\sinh\zeta{\tilde\eps}^t{\tilde\eps},\nonumber\\
&&\eps^t\Gamma_1\eps=\eps^t\Gamma_2\eps=0
\eea
This simplifies (\ref{DerivZ}):
\bea\label{DerivZFin}
d\omega&=&\left(i(dt+V)+\frac{ye^{-6\la}}{2\cosh^2\zeta}dy
-3d\la\right)\wedge \omega
\eea
as well as expression for $\omega$
\bea
\omega={\tilde\eps}^t{\tilde\eps}~ h\cosh\zeta (i{\hat e}^1_\mu+{\hat e}^2_\mu)dx^\mu
\eea
We now recall that we fixed a gauge for local
 Lorentz rotations in such a way that
${\tilde\epsilon}^t{\tilde\eps}$
does not depend on $x^i$ or $y$, so $y$ component of
(\ref{DerivZFin}) becomes
\bea
\d_y(h\cosh\zeta {\hat e}^i_\mu)=\frac{ye^{-6\la}}{2\cosh\zeta}h{\hat e}^i_\mu
\eea
This relation guarantees that ${\hat e}^1$ and ${\hat e}^2$ have
the same $y$ dependence, i.e.
\bea
{\hat e}^1=e^{D/2}(f_1 dx^1+f_2 dx^2),\qquad {\hat e}^1=e^{D/2}(f_3 dx^1+f_4 dx^2)
\eea
where $f_i$ do not depend on $y$. Then we can always use reparameterizations in
$x_1$--$x_2$ plane to simplify the vielbein:
\bea
{\hat e}^1=e^{D/2} dx^1,\qquad {\hat e}^1=e^{D/2}dx^2
\eea
With this choice of vielbein we can simplify equation (\ref{DerivZFin}):
\bea
&&\left\{d\log(e^{D/2}e^{-3\la})+\d_t\log({\tilde\eps}^t{\tilde\eps}) dt\right\}
\wedge (idx^1+dx^2)\nonumber\\
&&\qquad=
\left(i(dt+V)+\frac{ye^{-6\la}}{2\cosh^2\zeta}dy
-3d\la\right)\wedge (idx^1+dx^2)
\eea
Writing this in components, we find three equations:
\bea\label{DerivDyEqn}
&&\d_y{D}=\frac{y e^{-6\la}}{1+y^2e^{-6\la}}\\
&&\d_t({\tilde\eps}^t{\tilde\eps})=i{\tilde\eps}^t{\tilde\eps}\\
&&\frac{1}{2}\d_i{D} dx^i\wedge (idx^1+dx^2)=iV\wedge (idx^1+dx^2)
\eea
Second equation specifies time dependence of the Killing spinor
(${\tilde\eps}\sim e^{\frac{it}{2}}$) and taking real and imaginary parts of
the last equation we find an expression for $V_i$ in terms of ${D}$:
\bea\label{VthroughD}
V_i=\frac{1}{2}\eps_{ij}\d_j{D}
\eea

\subsection{Field strength for the vector $V_i$ and Toda equation.}

Let us now look at the vector $V_i$ appearing in the metric. As in the type IIB case
we will determine this vector by looking at the equation (\ref{BilMvect}) for
$\nabla K$. In the right hand side of that equation there is an antisymmetric tensor
$Y_{\mu\nu}$ and we begin with evaluating its components in the orthonormal frame.
Using projections
(\ref{ProjSpinMt1}), (\ref{ProjSpinMt2}) one can show that
$Y_{{\hat 0}{\hat 1}}=Y_{{\hat 0}{\hat 2}}=Y_{{\hat 3}{\hat 1}}=
Y_{{\hat 3}{\hat 2}}=0$: for example,
\bea
Y_{{\hat 0}{\hat 1}}=\eps^\dagger\Gamma^0\Gamma_{01}\eps=
\eps^\dagger\Gamma_{1}\eps=i\eps^\dagger\Gamma_{2}\eps=
-i\eps^\dagger\Gamma_{2}\eps
\eea
So in the orthonormal frame $Y$ has only two nonzero components
\bea
Y_{{\hat 0}{\hat 3}}&=&{\hat\eps}\Gamma_{03}\eps=
{\bar\eps}\Gamma_{5}\eps=-\sinh\zeta\nonumber\\
Y_{{\hat 1}{\hat 2}}&=&{\bar\eps}\Gamma_{12}\eps=-1\nonumber
\eea
so we find
\bea
Y=Y_{\mu\nu}dx^\mu\wedge dx^\nu=-\frac{e^{-3\la}}{2m^2}
\sinh\zeta (dt+V)\wedge dy-h^2 e^D dx^i\wedge dx^j
\eea
Using this expression, we can rewrite antisymmetric part of (\ref{BilMvect}) as
\bea
2m d\left[g_0^2(dt+V)\right]&=&-2\frac{e^{-3\la}}{m}\sinh\zeta (dt+V)\wedge dy-
4mh^2 e^D dx^1\wedge dx^2\nonumber\\
&&+e^{-3\la-2A}\sinh\zeta d\left[B_t(dt+V)+{\hat B}\right] \eea
This equation splits into two equations, and we look at only one
of them:
\bea g_0^2 dV&=&-2 h^2 e^D dx^i\wedge dx^j
+\frac{e^{-3\la-2A}}{2m}\sinh\zeta\left[B_t dV+d{\hat B}\right]
\eea Using definition of $\Phi$ (\ref{DefPhiM}): \bea d{\hat
B}+B_t dV=g_0^{-1}e^{-3\la+2A}~_3^\star d\Phi \eea as well as
duality relation \bea -2 h^2 e^D dx^1\wedge dx^2=-2 h~ _3^\star dy
\eea we find the final expression for the Hodge dual of $dV$: \bea
_3^\star d V=-\frac{m e^{-3\la}}{\cosh^3\zeta}
dy-\frac{1}{2m}g_0^{-3}ye^{-9\la}d\Phi
=-\frac{e^{-3\la}}{m\cosh^3\zeta} d(ye^{-6\la}) \eea Substituting
the value of $V_i$ from (\ref{VthroughD}), we find the equation
for ${D}$: \bea\label{DerivDxEqn} \Delta
{D}&=&-e^{D}\frac{1}{(1+y^2e^{-6\la})^2} \d_y(ye^{-6\la}) \eea
Combining this with (\ref{DerivDyEqn}), one can show that ${D}$
satisfies three dimensional Toda equation:
\bea\label{TodaEquation} \Delta {D}+\d_y^2 e^{D}=0 \eea

To summarize, we have shown that all supersymmetric solutions of M theory with
$SO(6)\times SO(3)$ isometry can be parameterized in terms of a single function $D$
which satisfies (\ref{TodaEquation}):
\bea \l{s2s5_solution}
ds_{11}^2&=&\frac{e^{2\la}}{m^2}\left(d\Omega_5^2+\frac{y^2 e^{-6\la}}{4}d{\tilde\Omega}_2^2
-\cosh^2\zeta (dt+V_idx^i)^2+
\frac{e^{-6\la}}{4\cosh^2\zeta}\left(dy^2+e^D d{\bf x}^2\right)
\right)\nonumber\\
G_{(4)}&=&\left[dB_t\wedge
(dt+V)+\frac{y^2e^{-6\la}}{4m\cosh\zeta}~ ^\star_3
d\Phi\right]\wedge d^2{\tilde\Omega},\qquad \sinh\zeta=-ye^{-3\la}
\eea
\bea
V_i=\frac{1}{2}\eps_{ij}\d_j D,\quad B_t=-\frac{y^3
e^{-6\la}}{2m^3},\quad \Phi=-\frac{2}{m}e^{6\la}, \quad
e^{-6\la}=\frac{\d_y D}{y(1-y\d_y D)}\nonumber \eea

\subsection{Solutions for $AdS_4\times S^7$ and $AdS_7 \times S^4$}
\label{AppAnCntM}

The solutions which we just constructed could have different asymptotics at large values of
$x_1,x_2,y$, but we will be mostly interested in solutions with either $AdS_7\times S^4$ or
$AdS_4\times S^7$ asymptotics. In particular, we can start with any solution which asymptotes
to $AdS_7\times S^4$ and make an analytic continuation to the $AdS_4\times S^7$ case. To derive
this analytic continuation it is useful to recall that $AdS_7\times S^4$ with radius
of $S^4$ equal to $\frac{1}{2m}$ corresponds to the solution
\bea
e^{D}=\frac{m^2 r^2}{1+m^2 r^2},\qquad
\begin{array}{c}
x_1=(1+m^2 r^2)\cos\theta\cos\phi,\\
x_2=(1+m^2 r^2)\cos\theta\sin\phi
\end{array},\qquad
y=m^2 r^2\sin\theta
\eea
Here $r$ is a radial coordinate on AdS, and $\theta$ is an angle on $S^4$. The analytic
continuation to $AdS_4\times S^7$ solution is given by
\bea
m=im',\qquad mr=i\sin\theta',\qquad \sin\theta=im'r'
\eea
This leads to the expressions for the parameters of analytically continued solution:
\bea
e^{D}=-\frac{\sin^2\theta'}{\cos^2\theta'},\qquad
\begin{array}{c}
x_1=\sqrt{1+(m'r')^2}\cos^2\theta'\cos\phi,\\
x_2=\sqrt{1+(m'r')^2}\cos^2\theta'\sin\phi
\end{array},\qquad
y=-im' r'\sin^2\theta' \eea While this solution describes
$AdS_4\times S^7$, it is not the most convenient description of
this space since $e^D$ diverges at $\theta'=\frac{\pi}{2}$. We can
use the conformal transformation in the $x_1$--$x_2$ plane to make
$e^D$ regular. Namely we construct a complex combination \bea
z=x_1+ix_2=\sqrt{1+(m'r')^2}\cos^2\theta' e^{i\phi} \eea and
introduce $z'$ as $z=(z')^2$. Then we can rewrite the two
dimensional metric as \bea e^D dz d{\bar z}=4 e^D z'{\bar z'} dz'
d{\bar z}'\equiv e^{D'}dz' d{\bar z}',\qquad
e^{D'}=-4\sqrt{1+(m'r')^2}\sin^2\theta' \eea Then writing
$z'=x'_1+ix'_2$ we find the final form of $AdS_4\times S^7$
solution: \bea &&
x'_1=(1+(m'r')^2)^{1/4}\cos\theta'\cos\frac{\phi}{2},\qquad
x'_2=(1+(m'r')^2)^{1/4}\cos\theta'\sin\frac{\phi}{2}\nonumber\\
&&e^{D'}=-4\sqrt{1+(m'r')^2}\sin^2\theta',\qquad y=-im'
r'\sin^2\theta' \eea

\subsection{Analytic continuation to $AdS_5 \times S^2$ solutions }
\label{AppAnCntM2}

Let us briefly comment on the analytic continuation to the
$AdS_5\times S^2$ reduction. It is straightforward to find an
analytic continuation of the solutions which we constructed and
find the geometries dual to ${\cal N}=2$ superconformal theories
(see (\ref{AnalCont52In})--(\ref{AnalCont52Out})), however there
are two potential subtleties. First, these new solutions have
space--like Killing direction $\chi$, so we can introduce a
topological mixing between $\chi$ and $S^2$. This leads to a
slight generalization of the ansatz which we have considered, and
in this subsection we will show that in the presence of such
mixing, a Killing spinor which has a nontrivial charge under
translations in $\chi$ cannot be a doublet of $SU(2)$.

The second subtlety is related to the flux over four
dimensional base. In the case of $S_5\times S^2$
reduction we have argued that $G_{\mu_1\mu_2\mu_3\mu_4}=0$ based on the absence of flux
over non--compact base. In the case of $AdS_5\times S^2$ we can have compact four
dimensional base, so this argument does not apply. However we will show that starting from
a solution with $I_1=0$ one cannot switch on $I_1$ in a regular manner.

In the appendix \ref{AppE1} we started with a metric which has no
mixing between $S^2$ and four dimensional base. However there
seems to be more general ansatz which is consistent with $SU(2)$
symmetry: assuming that the base has a space--like Killing
direction $\chi$, we can introduce a mixing of the form
$f(d\chi+A_\alpha dx^\alpha)^2$ where $x^\alpha$ are coordinates
on $S^2$ and $A_\alpha$ is a gauge field on the sphere with a
field strength proportional to the volume form
($F_{\alpha\beta}=\mbox{const}~ \eps_{\alpha\beta}$). Let us show
that if a Killing spinor which has a nontrivial charge under
translations in $\chi$, then the supercharges cannot form doublet
of $SU(2)$.

We recall that after reducing (\ref{11DKill}) on $S^5$ we find an equation for a spinor in six
dimensions:
\bea\label{WspinorEqn}
\left[\nabla_\mu-W_\mu\right]\xi=0.
\eea
Let us now assume that the metric in six dimensional space has a form
\bea
ds^2=f^2(d\chi+V_\mu dx^\mu)^2+\eta_{MN}e^M_\mu e^N_\nu dx^\mu dx^\nu.
\eea
then in the combination
\bea
\nabla_\mu-V_\mu \nabla_\chi={\hat\nabla}_\mu-V_\mu \d_\chi\nonumber
\eea
information about $V_\mu$ and $f$ disappears from ${\hat\nabla}_\mu$, in other words,
${\hat\nabla}_\mu$ is constructed from spin connection which uses only the reduced
five dimensional metric. We now split the five coordinates $x^\mu$ into the sphere
$S^2$ and the rest, and write the combination of (\ref{WspinorEqn}) with indices on
the sphere:
\bea
\left[{\hat\nabla}_\alpha -V_\alpha \d_\chi+W_\alpha-V_\alpha W_\chi\right]\xi=0
\eea
As in equation (\ref{DecompSpinXi}) we can split the spinor into the spinor on the
sphere and spinor on three dimensional base, and we arrive at the equation for a spinor $\eta$
on a unit sphere $S^2$:
\bea\label{EqnSpinWarped}
\left[{\nabla}_\alpha -V_\alpha \d_\chi\right]\eta=a\gamma_\alpha \eta+
b\eps_{\alpha\beta}\gamma_\beta \eta
\eea
Let us determine the constants $a,b$. First we notice that by making a rotation
$\eta\rightarrow e^{\mu\gamma_1\gamma_2}\eta$ with $\tan 2\mu=-\frac{b}{a}$ we can set $b=0$.
Then taking the commutator of different components of (\ref{EqnSpinWarped}), we get a relation
\bea
\left[-\gamma_{12}-F_{12}\d_\chi\right]\eta=2  a^2\gamma_{12}\eta
\eea
If $F_{12}=0$, this leads to $a=\pm \frac{i}{2}$ and equation
(\ref{EqnSpinWarped}) reduces to (\ref{SpinorOnS2}).
However if $\eta \sim e^{i\chi/2}$, then
\bea
F_{12}=\pm 2(2a^2+1),\qquad \gamma_{12}\eta=\mp i\eta
\eea
Let us take the upper signs in these relations. Then multiplying (\ref{EqnSpinWarped}) by
$(\gamma_{12}+i)$, we find that $a=0$ (recall that we already made a rotation to set $b=0$).
To summarize, we see that we should have either $F_{12}=0$, or $F_{12}=\pm 2$, and in the
latter case for each choice of the sign we have only one chiral spinor on the
2--sphere\footnote{This can be thought of as topological twisting: the contribution from spin
connection in the covariant derivative of the spinor cancels the contribution from the gauge
field and the spinor effectively behaves as a scalar on $S^2$.}.

Let us now address an issue of nonzero flux $I_1$. While we have not analyzed a possibility of
constructing solutions with $I_1\ne 0$, we can show that starting from
a solution with $I_1=0$ one cannot switch on $I_1$ in a regular manner.

We take a linear combination of (\ref{KillEqnS5}) and (\ref{KillEqnS2}) which does
not contain ${\not F}$, then using relation (\ref{GformI1}) we find:
\bea\label{CompFluxMink}
\left[{\not \d}(A+3\la)+ie^{-A}\Gamma^5 P+2im\gamma_7+\frac{i}{12}e^{-6\la-2A}
I_1\Gamma_5
\right]\xi=0
\eea
Let us rewrite this equation in terms of $\eps_+$ and $\eps_-$:
\bea\label{ModifEqnAPlLa}
\left[\pm ie^{-A}+\Gamma_5{\not \d}(A+3\la)+\frac{i}{48}I_1e^{-6\la-2A}\right]\eps_{\pm}
=\pm 2m\eps_{\mp}
\eea
If $I_1=0$ then the system of equations (\ref{KillEqnS5})--(\ref{KillEqnR4}) separates for
$\psi_+\equiv \eps_-+\Gamma_5\eps_+$ and $\psi_-\equiv \eps_--\Gamma_5\eps_+$, but
for nonzero flux there is no separation. In particular, equation (\ref{ModifEqnAPlLa})
becomes:
\bea\label{EqnI1Psi}
&&\left[ie^{-A}+\Gamma_5{\not \d}(A+3\la)-2m\Gamma_5\right]\psi_{+}=-
\frac{i}{48}I_1e^{-6\la-2A}\psi_{-}\nonumber\\
&&\left[ie^{-A}+\Gamma_5{\not \d}(A+3\la)+2m\Gamma_5\right]\psi_{-}=-
\frac{i}{48}I_1e^{-6\la-2A}\psi_{+}
\eea
Let us start from some solution with $I_1=0$ (we denote it $\psi^{(0)}$). Such solutions
have
either $\psi^{(0)}_+=0$ or $\psi^{(0)}_-=0$ and we choose a solution with $\psi^{(0)}_+=0$.
Let us view $I_1$ as a perturbation parameter and write equations (\ref{EqnI1Psi}) at the zeroes
and first orders in $I_1$. To do this we write $\psi=\psi^{(0)}+I_1\psi^{(1)}$ and use similar
expansion for the bosonic fields, however we will not need terms which contain corrections to
the bosonic background. Let us write zeroes order for the second equation in (\ref{EqnI1Psi}) and
first order for the first equation:
\bea
&&\left[ie^{-A}+\Gamma_5{\not \d}(A+3\la)+2m\Gamma_5\right]\psi^{(0)}_{-}=0\nonumber\\
&&I_1\left[ie^{-A}+\Gamma_5{\not \d}(A+3\la)-2m\Gamma_5\right]\psi^{(1)}_{+}=-
\frac{iI_1}{48}e^{-6\la-2A}\psi^{(0)}_{-}
\eea
Notice that $\psi^{(0)}_{-}$ is the same spinor as $\eps$ which we have been using in this
appendix. Since bosonic fields are solutions of equations of motion, we observe that
the first equation above can be rewritten as (\ref{ProjPrelEps}), and making similar
manipulations with second equation, we find:
\bea
&&\left[e^{-\Gamma_3\zeta}+i\Gamma_3\Gamma_5\right]\psi^{(0)}_{-}=0\nonumber\\
&&I_1 e^{-A}\left[e^{\Gamma_3\zeta}+i\Gamma_3\Gamma_5\right]\psi^{(1)}_{+}=-
\frac{iI_1}{48}e^{-6\la-2A}\Gamma_3\Gamma_5\psi^{(0)}_{-} \eea
Introducing spinors ${\tilde\eps}$ and ${\tilde\eta}$: \bea
\psi^{(0)}_{-}=e^{\frac{\zeta}{2}\Gamma_3}{\tilde\eps},\qquad
\psi^{(1)}_{+}=e^{-\frac{\zeta}{2}\Gamma_3}{\tilde\eta}
\eea
we find simple equations:
\bea
\left[1+i\Gamma_3\Gamma_5\right]{\tilde\eps}=0,\qquad
I_1 e^{\frac{\zeta}{2}\Gamma_3}\left[1+i\Gamma_3\Gamma_5\right]{\tilde\eta}=-
\frac{iI_1}{48}e^{-6\la-A}e^{-\frac{\zeta}{2}\Gamma_3}\Gamma_3\Gamma_5{\tilde\eps}
\eea
Multiplying the last equation by
$\left[1-i\Gamma_3\Gamma_5\right]e^{-\frac{\zeta}{2}\Gamma_3}$, we
find a relation which does not contain ${\tilde\eta}$: \bea
\frac{iI_1}{48}e^{-6\la-A}\cosh\zeta {\tilde\eps}=0 \eea This
proves that we can't switch on $I_1$ in perturbation theory.

\section{Derivation of the gauged supergravity solution}
\label{AppGauSUGRA}
\renewcommand{\theequation}{G.\arabic{equation}}
\setcounter{equation}{0}

In this appendix we provide the derivation of the gauged supergravity
solution described in section \ref{gsugrasection}.

\subsection{7-d gauged supergravity}
\label{7dsugra}

 In this subsection we list some of the properties of
$N=4$ supergravity in seven dimensions \cite{gaugedsugra} and we
also write the formulas \cite{nastase} for lifting solutions of
gauged supergravity into the solutions of M theory.

Bosonic sector of seven dimensional gauged supergravity
\cite{gaugedsugra} contains metric, $SO(5)$ gauge field ${A_{\mu I}}^J$,
14 scalar degrees of freedom which form a $SL(5,R)/SO(5)$ coset
${V_I}^i$
and five three--forms $C^I_3$. Since we will be looking for
solutions which have $C^I_3=0$, we will suppress the three forms
from the beginning. The equations of motion for the bosonic
fields come from the Lagrangian\footnote{In this section we use parameter
${\tilde m}$ which is natural from the point of view of gauged supergravity. It
is related to $m$ which is used in the rest of the paper as ${\tilde m}=2m$.}
\bea\label{GSGRLag}
2e^{-1}L=
R+\frac{{\tilde m}^2}{2}(T^2-2T_{ij}T^{ij})-\mbox{Tr}(P_\mu P^\mu)-\frac{1}{2}({V_I}^i{V_J}^j
F_{\mu\nu}^{IJ})^2+e^{-1}{\tilde m}^{-1}p_2(A,F)
\eea
where
\bea
T_{ij}={(V^{-1})_i}^I{(V^{-1})_j}^J\delta_{IJ},\qquad
{(V^{-1})_i}^I\nabla_\mu {V_I}^j\equiv (Q_\mu)_{[ij]}+(P_\mu)_{(ij)}
\eea
The fact that we have a coset is encoded in the identification:
\bea
{V_I}^i\sim {O_I}^J {V_J}^i
\eea
where ${O_I}^J$ is an element of $SO(5)$.

The fermionic degrees of freedom consist of gravitino $\psi_\mu$
and spin--$\frac{1}{2}$ fermions $\la_i$ which transform under
spinor representation of $SO(5)$. The supersymmetry
transformations are given by \bea\label{AVarySpinors} \delta
\la_i&=&\left[\frac{{\tilde
m}}{2}(T_{ij}-\frac{1}{5}\delta_{ij}T)\Gamma^j+
\frac{1}{2}\gamma^\mu P_{\mu
ij}\Gamma^j+\frac{1}{16}\gamma^{\mu\nu}
(\Gamma^{kl}\Gamma^i-\frac{1}{5}\Gamma^i\Gamma^{kl}){V_K}^k{V_L}^l
F^{KL}_{\mu\nu}\right]\eps\nonumber\\
\delta\psi_\mu&=&
\left[\nabla_\mu+\frac{{\tilde m}}{20}T\gamma_\mu-\frac{1}{40}({\gamma_\mu}^{\nu\la}-
8\delta_\mu^\nu\gamma^\la) \Gamma^{ij}{V_I}^i{V_J}^jF_{\nu\la}^{IJ}
\right]\eps
\eea
Starting from a solution of gauged supergravity, one can construct a
corresponding solution of supergravity in eleven dimensions. The
general procedure for such lifting was derived in \cite{nastase}, and
here we just cite the result for the case when the three form is
switched off:
\bea
ds_{11}^2&=&\Delta^{1/3}ds_7^2+\frac{1}{{\tilde m}^2}\Delta^{-2/3}T^{-1}_{IJ}
(dY^I+2{\tilde m} (A_\mu)^{IK}Y_K)(dY^J+2{\tilde m} (A_\mu)^{JL}Y_L)\\
\frac{\sqrt{2}}{3}F_{(4)}&=&
\eps_{A_1\dots A_5}\left[
\frac{2}{{\tilde m}\Delta}F^{A_1A_2}(DY)^{A_3}(DY)^{A_4}(T\cdot Y)^{A_5}+
\frac{1}{{\tilde m}}F^{A_1A_2}F^{A_3A_4}Y^{A_5}\right.\\
&+&\left.\frac{1}{3{\tilde m}^3}(DY)^{A_1}(DY)^{A_2}(DY)^{A_3}\left\{
-\frac{1}{\Delta}(DY)^{A_4}(T\cdot Y)^{A_5}+
4D\left(\frac{(T\cdot Y)^{A_4}}{\Delta}\right)Y^{A_5}\right\}
\right]\nonumber
\eea
where $Y^I$ is a five--dimensional unit vector ($Y\cdot Y=1$) and
\bea
\Delta=Y\cdot T\cdot Y,\qquad DY^I=dY^I+2{\tilde m} A^{IJ}Y_J
\eea
We will now use the above formulas to construct a
regular supersymmetric solution of gauged supergravity and to lift it
to the solution of M theory.

\subsection{Constructing a regular solution of gauged supergravity.}
\label{AppSlvGS}

Let us construct a supersymmetric solution of
M theory which has $SO(6)\times SO(3)\times U(1)$ symmetry. This
will not be the most general solution of M theory with such
symmetry, but it will be the most general solution which is
described by 7--dimensional gauged supergravity. The
reduction to gauged supegravity introduces a $7+4$ split of
eleven dimensional theory, and we will require that four
dimensions contain a round sphere $S^2$, while the seven
dimensional base space contains $S^5$ as well as one more
Killing direction $\d_t$. To have $SO(6)\times SO(3)\times U(1)$
symmetry we also require that all functions entering the ansatz
of gauged supergravity are invariant under rotations of $S^5$
and under translations of $t$. These conditions fix the metric
of seven dimensional space and the matter fields of gauged
supergravity to be
\bea\label{AnsGauSUGRA}
ds^2&=&-e^{2n}dt^2+g_{rr} dr^2+e^{2k} d\Omega_5^2\\
{V_I}^i&=&\left[\begin{array}{cc}
e^{-3\chi}g&{\bf 0}_{2\times 3}\\
{\bf 0}_{3\times 2}&e^{2\chi}{\bf 1}_{3\times 3}\\
\end{array}
\right],\qquad
{A_{\mu I}}^J=\left[\begin{array}{ccc}
iA_\mu\sigma_2&{\bf 0}_{2\times 3}\\
{\bf 0}_{3\times 2}&{\bf 0}_{3\times 3}\\
\end{array}
\right],\quad A=h dt\nonumber
\eea
Here $g$ is an element of $SL(2,R)/U(1)$ coset which can be parameterized
in terms of two functions $\rho$ and $\theta$:
\be
g=\exp(i\theta\sigma_2)\exp(-\rho\sigma_3)
\ee
All scalar functions introduced above depend only on one coordinate $r$,
and we can use the remaining reparameterization invariance to fix the
gauge
\bea
g_{rr}=r^6e^{-6k-2n}
\eea
We can now proceed with computing various functions which enter the
equations of gauged supegravity:
\bea\label{SpecifTPQ}
T_{ij}&=&\left[\begin{array}{cc}
e^{6\chi}(\cosh 2\rho+\sigma_3\sinh 2\rho)&{\bf 0}_{2\times 3}\\
{\bf 0}_{3\times 2}&e^{-4\chi}{\bf 1}_{3\times 3}
\end{array}
\right]\nonumber\\
P_{ij}&=&\left[\begin{array}{cc}
-3d\chi-\sigma_3 d\rho+(eA+d\theta)\sinh(2\rho)\sigma_1&{\bf 0}_{2\times 3}\\
{\bf 0}_{3\times 2}&2d\chi {\bf 1}_{3\times 3}
\end{array}
\right]\\
Q_{ij}&=&\left[\begin{array}{cc}
i(eA+d\theta)\cosh(2\rho)\sigma_2&{\bf 0}_{2\times 3}\\
{\bf 0}_{3\times 2}&{\bf 0}_{3\times 3}
\end{array}
\right]\nonumber
\eea
Here $e\equiv 2{\tilde m}$ is the charge of the coset ${V_I}^i$. We
observe that $\theta$ enters only through the combination
$eA+d\theta$, so we can always make a gauge transformation
of $A_\mu$ to make $\theta=0$, and from now on we will
work in this gauge. Notice however that after we
make such a choice, the value of $A_\mu$ at infinity acquires
a physical meaning.

To find supersymmetric solutions we have to solve the equations
\bea
\delta\la_i=0,\qquad \delta\psi_\mu=0
\eea
with variations given by (\ref{AVarySpinors}). First we consider
the equation $\delta\psi_\alpha=0$ where $\alpha$ is an index on
$S^5$. In particular we would need the expression for the
covariant derivative:
\bea
\nabla_\alpha\eps=\d_\alpha\eps+\frac{1}{4}{\tilde\omega}_\alpha \eps
-\frac{1}{2}{\gamma^r}_\alpha\eps \d_r k=
{\tilde\nabla}_\alpha\eps-\frac{1}{2}{\gamma^r}_\alpha\eps \d_r k
\eea
where ${\tilde\nabla}_\alpha$ is a covariant derivative on the sphere of unit radius
and the second term in the last equation comes from the warp factor. Since we have
a symmetry under rotations of $S^5$, the covariantly constant spinor should satisfy
\bea
{\tilde\nabla}_\alpha\eps=\frac{s}{2}e^{-k}\gamma_\alpha{\hat\gamma}\eps
\eea
where ${\hat\gamma}$ is a chirality operator on the sphere and $s=\pm i$
\cite{SphKilling}. We will take $s=-i$. Using above projection, we can reduce
the equations  $\delta\psi_\alpha=0$ to a single relation
\bea
\left[se^{-k}{\hat\gamma}+\frac{{\tilde m}}{2}e^{-4\chi}+\gamma^r\d_r (k+\chi)
\right]\eps=0
\eea
We also notice that since $t$ is a Killing direction, we can choose
the time dependence of $\eps$ to be
\bea
\eps\sim e^{iEt/2}
\eea
To write the remaining equations it is convenient to choose a particular
basis of gamma matrices:
\bea
\gamma^r=e^r_{\hat r}\sigma_3,\quad \gamma^t=ie^t_{\hat t}\sigma_2
\eea
and decompose a spinor $\eps$ in this basis as
\bea
\eps=\left(\begin{array}{c}\eps_+\\
\eps_-\end{array}\right)
\eea
With this decomposition we arrive at the system of equations:
\bea\label{SpinEqnOne}
&&\left[{\tilde m}(e^{-10\chi}-\cosh 2\rho)+
\frac{5}{r^3} e^{-6\chi+n+3k} \d_r\chi\right]\eps_+
+ie^{-12\chi+3k}r^{-3}\d_r h\eps_-=0\nonumber\\
&&\left[{\tilde m}(e^{-10\chi}-\cosh 2\rho)-
\frac{5}{r^3} e^{-6\chi+n+3k} \d_r\chi\right]\eps_-+
ie^{-12\chi+3k}r^{-3}\d_r h\eps_+=0\\
\label{SpinEqnTwo}
&&\left[\frac{1}{2}e^r_{\hat r}\d_r\log\tanh\rho-me^{6\chi}
\right]\eps_+-2i{\tilde m} e^t_{\hat t} A_t\eps_-=0\nonumber\\
&&\left[\frac{1}{2}e^r_{\hat r}\d_r\log\tanh\rho+me^{6\chi}
\right]\eps_--2i{\tilde m} e^t_{\hat t} A_t\eps_+=0\\
\label{SpinEqnThree}
&&\left[\frac{{\tilde m}}{2}e^{-4\chi}+e^r_{\hat r}\d_r (k+\chi)\right]\eps_+
-se^{-k}\eps_-=0\nonumber\\
&&\left[\frac{{\tilde m}}{2}e^{-4\chi}-e^r_{\hat r}\d_r (k+\chi)\right]\eps_-
-se^{-k}\eps_+=0\\
&&\left[\frac{{\tilde m}}{2}e^{-4\chi}+e^r_{\hat r}\d_r (\chi+n)\right]\eps_++
\left[(iE+2i{\tilde m} A_t\cosh 2\rho)e^t_{\hat t}+i
e^{-6\chi+3k}r^{-3}\d_r h
\right]\eps_-=0\nonumber\\
\label{SpinEqnFour}
&&\left[\frac{{\tilde m}}{2}e^{-4\chi}-e^r_{\hat r}\d_r (\chi+n)\right]\eps_-+
\left[-(iE+2i{\tilde m} A_t\cosh 2\rho)e^t_{\hat t}+
ie^{-6\chi+3k}r^{-3}\d_r h
\right]\eps_+=0\nonumber\\
\eea We have written all nontrivial equations except
$\delta\psi_r=0$. The reason for separating this equation is that
unlike (\ref{SpinEqnOne})--(\ref{SpinEqnFour}), it is not
algebraic in $\eps$, but rather gives a differential equation
which involves $\d_r\eps$. So the simplest way to proceed is to
solve the equations (\ref{SpinEqnOne})--(\ref{SpinEqnFour}) first
and to find a bosonic background, and then use the equation
$\delta\psi_r=0$ to determine the radial dependence of $\eps$.
Since we are not interested in constructing $\eps(r)$ we will not
need the equation $\delta\psi_r=0$ for the moment.

Notice that due to our gauge condition, we have
\bea
e^r_{\hat r}=r^{-3}e^{n+3k},
\eea
so derivative $\d_r$ always appear in the combination
\bea
r^{-3}\d_r=\d_R,\qquad R\equiv \frac{r^4}{4}
\eea
To find the solution of the system (\ref{SpinEqnOne})--(\ref{SpinEqnFour}),
we observe that the general systems of the form:
\bea
&(a+b)\eps_++c\eps_-=0,\qquad\qquad & (d+e)\eps_++(f+g)\eps_-=0\\
&(a-b)\eps_-+c\eps_+=0,\qquad\qquad & (d-e)\eps_-+(f-g)\eps_+=0
\eea
implies that
\be
cd=af+bg,\qquad ce=ag+bf,\qquad a^2=b^2+c^2,\qquad
d^2-e^2=f^2-g^2
\ee
By applying these formulas to various pairs from the system
(\ref{SpinEqnOne})--(\ref{SpinEqnFour}) we will find equations
which do not contain $\eps$.
For example, integrability condition for (\ref{SpinEqnOne}) and
(\ref{SpinEqnTwo}) leads to an equation
\bea
-10{\tilde m} e^{-6\chi+n+3k-n}h\d_R\chi=-{\tilde m}e^{-6\chi+3k}\d_R h
\eea
which can be easily integrated:
\bea
e^{\chi}=\left(\frac{h}{h_0}\right)^{1/10}
\eea
Now we combine (\ref{SpinEqnTwo}) and (\ref{SpinEqnThree}):
\bea
e^{4\chi+4k}\d_R(k+\chi)=\frac{is}{2}h_0^{-1}\quad\rightarrow\quad
e^{4\chi+4k}=\frac{is}{2}h_0^{-1} (r^4+c_1)
\eea
where $c_1$ is an integration constant.
To find $n$ we take a determinant of (\ref{SpinEqnThree}):
\bea
\frac{1}{16}e^{2n}[\d_R e^{4k+4\chi}]^2=\frac{{\tilde m}^2}{4}e^{2k}-s^2e^{8\chi}.
\eea
and rewrite it as an expression for $e^{2n}$:
\be
e^{2n}=-\frac{1}{s^2}h_0^{2}[{\tilde m}^2 e^{2k}-4 s^2 e^{8\chi}]=
h_0^{2}[{\tilde m}^2 e^{2k}+4e^{8\chi}]
\ee
Finally we take a determinant of (\ref{SpinEqnOne}) to find an
expression for $\cosh 2\rho$:
\bea
\cosh 2\rho=h_0\d_R[(r+\frac{c_1}{4}) h^{-1}]
\eea
Let us summarize the results we have so far:
\bea\label{SolutOne}
h\equiv h_0 H^{-1},\quad e^\chi=H^{-\frac{1}{10}},\quad
e^k=(r^4+c_1)^{1/4}H^{\frac{1}{10}}(2h_0)^{-1/4},\quad
e^{2n}=(2h_0)^2 fH^{-4/5}
\eea
where we defined
\be\label{DefFunc}
f=1+\frac{{\tilde m}^2}{4}e^{2k-8\chi}=1+\frac{{\tilde m}^2}{4\sqrt{2h_0}}H
\left(r^4+c_1\right)^{1/2}
\ee
Substituting these expressions into the equations
(\ref{SpinEqnOne})--(\ref{SpinEqnFour}) and writing
all possible integrability conditions, we find only two
nontrivial equations:
\bea\label{NontrivEqn}
&&2f{\tilde R}^{1/2}\d^2_R({\tilde R}H)=-\frac{{\tilde m}^2}{\sqrt{2h_0}}
([\d_R({\tilde R}H)]^2-1)\\
\label{EqnForEnrg} &&{\tilde m}-2\left(\frac{E}{2h_0}+{\tilde
m}H^{-1}\d_R({\tilde R}H)\right)- 2{\tilde m}{\tilde R}H\d_R
H^{-1}=0 \eea To simplify these equations we introduced \bea
{\tilde R}\equiv R+\frac{c_1}{4} \eea Notice that equation
(\ref{EqnForEnrg}) reduces to the expression for $E$: \bea
E=-{\tilde m}h_0 \eea and we are left with only one nontrivial
differential equation (\ref{NontrivEqn}).

To summarize, we have solved the equations for Killing spinors and
the solution is given by (\ref{SolutOne}), (\ref{DefFunc}) with
$H$ satisfying the equation (\ref{NontrivEqn}). At this point we
have two integration constants $h_0,c_1$ and we will fix their
values momentarily. Substituting the expressions (\ref{SolutOne})
into (\ref{AnsGauSUGRA}), we find the solution of gauged
supergravity
\bea &&ds^2=-(2h_0)^2
fH^{-4/5}dt^2+\frac{H^{1/5}}{\sqrt{2h_0}} \left[ \frac{r^6
dr^2}{f(r^4+c_1)^{3/2}}+\left(r^4+c_1\right)^{1/2}
d\Omega_5^2\right]\nonumber\\
&&A_t^{(1)}=h_0 H^{-1},\qquad e^{\chi}=H^{-\frac{1}{10}},\qquad
\cosh 2\rho=\d_R({\tilde R}H)
\eea
First we notice that by rescaling $t,r$ and $c_1$ we can set $h_0$
to any value, we will choose this value to be $h_0=\frac{1}{2}$.
Then we see that the metric, matter fields and equation
(\ref{NontrivEqn}) can all be expressed in terms of coordinate
${\tilde R}$, then $c_1$ never appears explicitly. After this
is done we can define a new coordinate ${\tilde r}=(4{\tilde R})^{1/4}$
instead of the coordinate $r$,
which is equivalent to setting $c_1$ to zero. Thus without loss
of generality we can take $c_1=0$.

Now we write the solution of gauged supergravity in its final form:
\bea
&&ds^2=-fH^{-4/5}dt^2+H^{1/5}
\left[
\frac{dr^2}{f}+r^2 d\Omega_5^2\right]\nonumber\\
&&A_t^{(1)}=\frac{1}{2} H^{-1},\qquad
e^{\chi}=H^{-\frac{1}{10}},\qquad \cosh 2\rho=\d_R(RH),\qquad
R\equiv \frac{r^4}{4} \l{gsugraan} \eea where \bea
f=1+\frac{{\tilde m}^2 r^2}{4}H \eea and $H$ satisfies a nonlinear
differential equation: \bea 2f{R}^{1/2}\d^2_R({R}H)=-{\tilde
m}^2([\d_R({R}H)]^2-1) \l{feqn} \eea To formulate the problem
entirely in terms of $R$ one can rewrite the metric as \bea
&&ds^2=-fH^{-4/5}dt^2+H^{1/5} \left[
\frac{dR^2}{8fR^{3/2}}+2\sqrt{R} d\Omega_5^2\right] \eea Using the
formulas from Appendix \ref{7dsugra}, one can easily lift this
geometry to the solution of eleven dimensional supergravity, and
the result is given by (\ref{GaugSGRsln11}). The coordinates
introduced there are related to the vector $Y^I$ by \bea
Y_1=\cos\theta\cos\phi,\quad Y_2=\cos\theta\sin\phi,\
\left(\begin{array}{c} Y_3\\Y_4
\end{array}\right)=
\left(\begin{array}{c}
\sin\theta\cos\psi\cos\zeta\\\sin\theta\cos\psi\sin\zeta
\end{array}\right)
,\ Y_{5}=\sin\theta\sin\psi\nonumber
\eea
and the sphere $S^2$ is parameterized by $(\psi,\zeta)$.

\subsection{The asymptotic behavior and the charge of the regular gauged supegravity
solution} \label{AppCharge}

 In this appendix, we study various approximations
to the non-linear differential equation (\ref{TheEquation}), and
the asymptotic behavior and the charge of the 1/2 BPS regular
solution of M theory from 7d gauged supergravity.

This family of smooth solutions from 7d gauged supergravity are
characterized by a function $H(x)$ which satisfies the non-linear
differential equation
\begin{eqnarray}  \label{TheEquation2}
(2\sqrt{x}+F)F^{\prime\prime}=(1-(F^{\prime})^2)
\end{eqnarray}
where $x=4 m^4 r^4, F(x)=xH(x)$.

We will consider smooth solutions to equation
(\ref{TheEquation2}), which at the origin $x=0$ obeys the boundary
condition:
\begin{eqnarray} \label{bc}
F(x)|_{x=0}=0, \quad \quad F^{\prime}(x)|_{x=0}=C
\end{eqnarray}
These smooth solutions are parametrized by $C$, where $C>1$
because $C$ is equal to the maximal value of $\cosh 2\rho$ at
$x=0$.

At large $x$, the asymptotic solution to (\ref{TheEquation2}) is
\begin{eqnarray}
F(x)=[x^{2}+2Px+(P^{2}-d)]^{1/2}, \quad \quad \quad {\rm for}
\quad x\gg P \label{largex}
\end{eqnarray}
where $P$ and $d$ are two parameters related to the charge and the
"dipole moment" of the solution and are both functions of $C$.
From (\ref{largex}), the asymptotic solution of $H(r)$ when $r
\rightarrow \infty$ is therefore
\begin{equation}
H(r)|_{r \rightarrow \infty} \rightarrow 1+\frac{Q}{r^4}
\end{equation}
so we see that $Q=\frac{P}{4m^4}$ is the charge of the solution.

Another useful approximation is to drop the 1 in equation
(\ref{TheEquation2}), which is valid for $F^{\prime }(x) \gg 1$.
After change of variables
\begin{equation}
x=C^{-2} e^{t}, \quad\quad F(x)={\sqrt{x}}G(t)
\end{equation}
equation (\ref{TheEquation2}) when dropping 1 becomes:
\begin{equation}
2(G+2)\ddot{G}+2\dot{G}^{2}+2G\dot{G}-G=0 \label{scalefree}
\end{equation}
where dot means $\frac{d}{dt}$. The solution $G(t)$ to equation
(\ref{scalefree}) is independent of $C$.

We are able to get analytical expressions for the charge and
"dipole moment" of the solution as functions of $C$ by matching
(\ref{largex}) and (\ref{scalefree}) in the large $C$ limit:
\begin{equation}
Q\approx \frac{1}{8m^4}(\ln C)^{2}, \quad \quad d \approx
(4{m^4}Q)^{2}+O((\ln C)^{2}),  \quad \quad C \rightarrow \infty
\end{equation}

\subsection{The regular gauged SUGRA solution as a  solution of the Toda
equation}
\l{todagauged}

In this appendix, we derive the solution to the Toda equation (\ref{toda}),
corresponding to the regular gauged SUGRA solution in section 3.2, which is a
particular example of our general 1/2 BPS chiral-primary solutions (\ref{4d_general})
in section 3.

The metric and 4-form flux of our gauged SUGRA are written in
(\ref {GaugSGRsln11}). It can be brought into the form of our
general solutions in (\ref{4d_general})
or (\ref{s2s5_solution}) by non-linear coordinate transformations.

By comparing the 11d metric components of (\ref {GaugSGRsln11})
and (\ref{s2s5_solution}), we can easily read off that:
\begin{equation}
e^{2\lambda }=m^{2}r^{2}\Delta ^{1/3}H^{1/5},\quad \cosh ^{2}\zeta
=1+m^{-2}r^{-2}\Delta ^{-1}H^{-3/5}\sin ^{2}\theta ,\quad
y=m^{2}r^{2}\sin \theta
\end{equation}

Using (\ref{gsugraan}), we can rewrite (\ref{feqn}) in terms of the function $\rho$:
\begin{equation}
\partial _{r}\rho =\frac{-2m^{2}r\sinh 2\rho }{f}  \label{rho}
\end{equation}

From the solution in (\ref{GaugSGRsln11}), the metric of the 4d
base space is
\begin{eqnarray}
ds_{4}^{2} &=&\frac{1}{m^{2}r^{2}}\left[ -fH^{-1}d{\tilde t}^{2}+
\frac{dr^{2}}{f}+\frac{%
H^{-1/5}}{4{m}^{2}\Delta }\left\{ e^{4\chi }\cos ^{2}\theta d\theta
^{2}\right. \right.   \nonumber \\
&+&e^{-6\chi -2\rho }\left[ d(\cos \theta \cos \phi )+2mH^{-1}\cos \theta \sin
\phi d{\tilde t}\right] ^{2}  \label{4d_gauge}\\
&+&\left. \left. e^{-6\chi +2\rho }\left[ d(\cos \theta \sin \phi
)-2mH^{-1}\cos \theta \cos \phi d{\tilde t}\right] ^{2}\right\}
\right]  \nonumber
\end{eqnarray}

We will write this in terms of our general solution
\begin{equation}
ds_{4}^{2}=-\frac{\cosh ^{2}\zeta }{m^{2}}(dt +V)^{2}+\frac{e^{-6\lambda }%
}{4m^{2}\cosh ^{2}\zeta }\left[ e^{D}(dx^{2}+x^{2}d\psi ^{2})+dy^{2}\right]
\label{4d_ansatz2}
\end{equation}%
where $x,\psi $ are the polar coordinates in $x_{1}, x_{2}$ plane and $%
D,x,\psi $ are functions of $r,\theta ,\phi $.

By comparing the metrics (\ref{4d_gauge}) and (\ref{4d_ansatz2}%
), we read off
\begin{equation}
dt +V=md{\tilde t}-\frac{m^{4}r^{4}e^{-6\lambda }}{2\cosh ^{2}\zeta }
\left[ \frac{%
s_{2}{\tilde{s}}_{2}}{2}\sinh 2\rho d\theta -(e^{-2\rho }{\tilde{s}}%
^{2}+e^{2\rho }{\tilde{c}}^{2})c^{2}d\phi \right]
\end{equation}
where we introduced the notations in this appendix:
\begin{equation}
c=\cos \theta ,\quad s=\sin \theta ,\quad s_{2}=\sin 2\theta ,\quad {\tilde{c%
}}=\cos \phi ,\quad {\tilde{s}}=\sin \phi ,\quad {\tilde{s}}_{2}=\sin 2\phi
\end{equation}

To solve $D,x,\psi $ in terms of $r,\theta ,\phi $, we need to
bring the 2d metric of the $x_{1}, x_{2}$ plane to the form:
\begin{eqnarray}
&&\frac{e^{-6\lambda }}{4m^{2}\cosh ^{2}\zeta }\left[ e^{D}(dx^{2}+x^{2}d%
\psi ^{2})\right]   \nonumber \\
&&=\frac{1}{m^{2}r^{2}f}dr^{2}-\frac{e^{-6\lambda }}{4m^{2}\cosh ^{2}\zeta }%
(2m^{2}r\sin \theta dr+m^{2}r^{2}\cos \theta d\theta )^{2} \nonumber \\
&&+\frac{H^{2/5}}{4m^{4}r^{2}\Delta }\left\{ (e^{-2\rho }{\tilde{c}}%
^{2}+e^{2\rho }{\tilde{s}}^{2})s^{2}d\theta ^{2}+H^{-1}c^{2}d\theta ^{2}\right.
-\left. s_{2}{\tilde{s}}_{2}\sinh 2\rho d\theta d\phi +(e^{-2\rho }{\tilde{s}%
}^{2}+e^{2\rho }{\tilde{c}}^{2})c^{2}d\phi ^{2}\right\}   \nonumber \\
&&+\frac{1}{\cosh ^{2}\zeta }\frac{m^{6}r^{8}}{4e^{12\lambda }}\left[ \frac{1%
}{2}s_{2}{\tilde{s}}_{2}\sinh 2\rho d\theta -(e^{-2\rho }{\tilde{s}}%
^{2}+e^{2\rho }{\tilde{c}}^{2})c^{2}d\phi \right] ^{2}
\label{2dmetric}
\end{eqnarray}

After solving a set of nonlinear differential equations from (\ref{2dmetric}) and using
(\ref{rho}), we get the solution of $D,x,\psi $:
\begin{equation}
e^{D}=\frac{m^{2}r^{2}f}{\left( \widetilde{F}\right) ^{2}},\quad x=\sqrt{%
e^{-2\rho }{\cos }^{2}{\phi }+e^{2\rho }\sin ^{2}\phi
}\widetilde{F}\cos \theta ,\quad \psi =\mathrm{arc}\tan (e^{2\rho
}\tan \phi ) \label{gauge_exp}
\end{equation}%
where $\widetilde{F}$ is a function of $r$, which satisfies
\begin{equation}
\partial _{r}\widetilde{F}(r)=\frac{2m^{2}r\widetilde{F}(r)\cosh 2\rho }{f}
\label{F_tilde}
\end{equation}
Using (\ref{rho}), we see that the solution to (\ref{F_tilde}) is
\begin{equation}
 \tilde F(r) = \sqrt{
 \sinh 2 \rho(r=0) \over
 \sinh 2 \rho(r)}
\end{equation}
where the numerator $\sinh 2\rho(r=0)$ is a normalization factor
which is related to the charge of the solution.

The expressions for $x,\psi $ in (\ref{gauge_exp}) can be
conveniently written in terms of a complex coordinate $w$ in
$x_{1}, x_{2}$ plane:
\begin{equation}
w=xe^{i\psi }=\left( e^{i\phi }\cosh \rho -e^{-i\phi }\sinh \rho
\right) \widetilde{F}\cos \theta
\end{equation}

The solution (\ref{gauge_exp}) to the Toda equation can reduce to the
solution (\ref{ads7_exp}) corresponding to $AdS_{7}\times S^{4}$ in
the case $\rho =0$. For $AdS_{7}\times S^{4}$ we can get its expression from (\ref{gauge_exp}),
(\ref{F_tilde}):
\begin{equation}
e^{D}=\frac{m^{2}r^{2}}{1+m^{2}r^{2}},\quad \quad
w=(1+m^{2}r^{2})\cos \theta e^{i\phi },\quad \quad
\widetilde{F}=f=1+m^{2}r^{2}  \label{Ads_7}
\end{equation}

In the case of $AdS_{7}\times S^{4}$, the region on the $x_{1},
x_{2}$ plane where the $S^{5}$ shrinks is a perfectly round disk.
Outside the disk, the $S^{2}$ shrinks instead. The effect of
turning on the charged scalar $\rho $ in the gauged SUGRA solution
is to deform the
perfectly round disk to an elliptic disk. In this case, the region on the $%
x_{1}, x_{2}$ plane where the $S^{5}$ shrinks is described by \bea
x^{2}=\left( a^{2}\sin ^{2}\phi +b^{2}{\cos }^{2}{\phi }\right)
\cos ^{2}\theta \leq \left( a^{2}\sin ^{2}\phi +b^{2}{\cos
}^{2}{\phi } \right) =\left( a^{-2}\sin ^{2}\psi +b^{-2}{\cos
}^{2}{\psi }\right) ^{-1}
 \nonumber \eea where
$a=\widetilde{F}e^{\rho }\mid _{r=0}$,
$b=\widetilde{F}e^{-\rho}\mid _{r=0}.$ The boundary of this region
is an  ellipse  with long-axis $a$ and short-axis $b$. Since ${a
\over b}=e^{2\rho(r=0)}$, the ellipticity of the ellipse is a
function of the charge of the solution. The deformation of the
round disk to an elliptic disk is related to the breaking of the
gauge group $SO(2)$ by turning on $\rho $ in ansatz
(\ref{SO3Ans}).

\end{document}